\documentclass[fleqn,usenatbib]{mnras}

\usepackage{newtxtext,newtxmath}

\usepackage[T1]{fontenc}
\usepackage{ae,aecompl}

\usepackage{graphicx}	
\usepackage{amsmath}	
\usepackage{amssymb}	

\usepackage{siunitx}
\DeclareSIUnit
	\microns{\micro\metre}
\DeclareSIUnit
	\parsec{pc}

\usepackage[caption=false]{subfig}
\usepackage{booktabs}

\usepackage[normalem]{ulem}

\usepackage{natbib}
\usepackage{hyperref}
\usepackage{cleveref}

\newcommand{\ssg}{S$^4$G}
\newcommand{\sersic}{S\'{e}rsic\,}
\newcommand{\msol}{\, \mathrm{M_{\odot}}}
\newcommand{\pc}{\, \mathrm{pc}}

\newcommand{\rbar}{R_{\mathrm{bar}}}
\newcommand{\rbulge}{R_{\mathrm{eff, bulge}}}
\newcommand{\rmax}{R_{\mathrm{max}}}
\newcommand{\riso}{R_{\mathrm{25.5}}}
\newcommand{\Mth}{\mathrm{M_{th}}}
\newcommand{\rbr}{R_{\mathrm{br}}}
\newcommand{\hin}{h_{\mathrm{in}}}
\newcommand{\hout}{h_{\mathrm{out}}}
\newcommand{\hdisk}{h_{\mathrm{disk}}}
\newcommand{\e}[1]{\times 10^{#1}}
\newcommand{\rev}[1]{{#1}}
\newcommand{\revv}[1]{{#1}}

\hypersetup{draft}

\title[Spiral arm contrasts and other galaxy properties]{How do spiral arm contrasts relate to bars, disk breaks and other fundamental galaxy properties?}

\author[Bittner et al.]
{Adrian Bittner,$^{1,2}$\thanks{E-mail: A.Bittner@physik.uni-muenchen.de}
Dimitri A. Gadotti,$^{1}$
Bruce G. Elmegreen,$^{3}$
E. Athanassoula,$^{4}$\newauthor
Debra M. Elmegreen,$^{5}$
Albert Bosma,$^{4}$
Juan-Carlos Mu\~{n}oz-Mateos$^{1}$
\\
$^{1}$European Southern Observatory, Casilla 19001, Santiago 19, Chile													\\
$^{2}$University Observatory Munich, Scheinerstr. 1, 81679 Munich, Germany												\\
$^{3}$IBM Research Division, T.J. Watson Research Center, Yorktown Heights, NY 10598, USA								\\
$^{4}$Aix Marseille Univ, CNRS, LAM, Laboratoire d'Astrophysique de Marseille, Marseille, France						\\
$^{5}$Vassar College, Dept. of Physics and Astronomy, Poughkeepsie, NY 12604, USA										
}

\date{Accepted XXX. Received YYY; in original form ZZZ}

\pubyear{2017}

\begin{document}
\label{firstpage}
\pagerange{\pageref{firstpage}--\pageref{lastpage}}
\maketitle

\begin{abstract}
%
	We investigate how the properties of spiral arms relate to other fundamental
	galaxy properties, including bars and disc breaks.
%
	We use previously published measurements of those properties, \revv{and} our
	own measurements of arm and bar contrasts for a large sample of galaxies,
	using $3.6 \micron$ images from the Spitzer Survey of Stellar Structure in
	Galaxies (\ssg). 
%
	Flocculent galaxies are clearly distinguished from other spiral arm classes,
	especially by their lower stellar mass and surface density.  Multi-armed and
	grand-design galaxies are similar in most of their fundamental parameters,
	excluding some bar properties and the bulge-to-total ratio.  Based on these
	results, we revisit the sequence of spiral arm classes, and discuss
	classical bulges as a necessary condition for standing spiral wave modes in
	grand-design galaxies. 
	\revv{We find} a strong correlation between bulge-to-total \rev{ratio} and
	bar contrast, and a weaker correlation between arm and bar contrasts. Barred
	and unbarred galaxies exhibit similar arm contrasts, but the highest arm
	contrasts are found exclusively in barred galaxies. Interestingly, the bar
	contrast, and its increase from flocculent to grand-design galaxies, is
	systematically more significant than that of the arm contrast. 
	We corroborate previous findings concerning a connection between bars and
	disc breaks. In particular, \rev{in grand-design galaxies the bar contrast}
	correlates with the \revv{normalised} disc break radius. This does not hold
	for other spiral arm classes or the arm contrast. 
%
	\revv{Our} measurements of arm and bar contrast \rev{and} radial contrast
	profiles \rev{are} publicly available. 
\end{abstract}



\section{Introduction}
\label{sec:introduction}
Spiral arms in disc galaxies show different geometrical properties, levels of
symmetry and amplitudes. They can be classified visually in three distinct
classes: flocculent, multi-armed and grand-design.  Modern theories on the
formation of spiral arms indicate different underlying physics for these three
arm classes. 
Flocculent galaxies show short and patchy spiral arms without underlying density
waves in the old stellar component \rev{\citep{elmegreen1984,buta2015}}. This
indicates that the spiral arms of flocculent galaxies are mainly triggered by
local gravitational instabilities concerning old stars
\citep{julian1966,toomre_kalnajs1991}, gas \citep{goldreich1965} and the
resulting formation of new stars
\citep[e.g.][]{seiden1979,elmegreen1981,elmegreen1984}.  \rev{In addition,
	\citet{ee2011} find that only 15\% of optically flocculent galaxies show a
weak two-armed spiral pattern in near-IR observations. }
In contrast, grand-design galaxies show spiral arms with high symmetry on large
scales. These spiral arms may be caused by spiral density waves as initially
suggested by \citet[][]{lindblad1959}.  Such density waves can be driven by bars
or satellites \citep[e.g.][]{lia1980}, or triggered by companions
\citep[e.g.][]{dobbs2010}, or form self-consistently as in the density wave
theory of \citet{linshu1964}, or the swing amplification mechanism
(\citeauthor{toomre1981} \citeyear{toomre1981}, see also
\citeauthor{goldreich1965} \citeyear{goldreich1965}).  An alternative theory for
grand-design spirals is the manifold theory \citep[e.g.][and references
therein]{romero-gomez2006,lia2012,harsoula2016} but this has so far been fully
worked out only for grand-design spirals due to bars.  For both 
theories, gas is compressed and shocked in the high density regions of these
spirals.  As a direct result of this process gas densities and star formation
rates are enhanced turning the spiral arms bright in optical observations
\citep{roberts1969}.  
Multi-armed galaxies are believed to be an intermediate case between
grand-design and flocculent galaxies. \citet{elmegreen1984,elmegreen1995} find
that the stellar spirals of multi-armed galaxies are regular in the central
parts, \rev{but become} more and more irregular at larger radii. 
For a general review of the observational properties of spirals and particularly
of the theories that have strived to explain their properties and evolution see
\citet{lia1984} and \citet{dobbs-baba2014}. 

A second fundamental structural component of disc galaxies is bars. It is
well-known that the majority of disc galaxies ($\sim$ 65\%) in the local
universe have a bar
\citep[e.g.][]{eskridge2000,knapen2000,whyte2002,menendez2007,marinova2007}.
Some bar properties depend strongly on the Hubble type of the galaxy.
Early-type disc galaxies show the tendency to have long, high amplitude bars
with flat radial surface brightness profiles. The amplitude of the corresponding
spiral arms is radially decreasing. In contrast, late-type galaxies have short,
low amplitude bars with an exponential profile and increasing arm amplitudes
\citep{ee1985}. 
\citet{lia2002} and \citet{combes1993} have conducted N-body simulations of
barred galaxies and conclude that flat bars emerge in galaxies with quickly
increasing inner rotation curves, whereas exponential bars correlate with slowly
rising rotation curves. The main difference between these two cases is the
component dominating in the bar region. Therefore \citet{lia2002} introduce the
nomenclature ``MH'' for the halo-dominated and ``MD'' for the disc-dominated
case.  Further studies (\citealt{lia2009}; see also \citealt{gadotti2007})
conclude that the angular momentum redistribution in galaxies with flat bars is
more efficient than in those with exponential bars.  This could be connected to
the findings of \citet{ee1985} that exponential bars mainly occur in
multiple-arm and flocculent galaxies, whereas flat bars primarily emerge in
symmetric, 2-armed spirals.  Indeed, \citet{lia2002} showed with the help of
N-body simulations that weaker bars should have exponential profiles, while
stronger ones have flatter profiles.  Another result of this observation is that
strong bars show the tendency to have strong arms as well \citep{ee1985,ann1987,
elmegreen2007,salo2010} although other studies have showed the opposite
\citep[e.g.][]{seigar1998,seigar2003,durbala2009}, while \citet{buta2009} find
only a weak trend with a correlation coefficient of 0.3.  One possible physical
explanation of this correlation is that the spiral structure is driven by the
bar at the same pattern speed, or is linked to the bar via energy and angular
momentum exchange at dynamical resonances \citep{tagger1987}.

A third important property of disc galaxies is their radial surface brightness
profile. Typically disc galaxies have been described with a single exponential
profile (Type I) by e.g. \citet{freeman1970}. However, \citet{freeman1970} and
\citet{kruit1979} also showed that some of the discs have a break in their
radial surface brightness profile in the outer parts of the discs. Recent
observations of face-on as well as edge-on galaxies
\citep{pohlen2002,erwin2005,pohlen2006,erwin2008,gutierrez2011,maltby2012,jc2013,
comeron2012,martin-navarro2012} indicate that in many cases the slope of the
radial surface brightness profile changes abruptly.  Therefore it is better to
model these galaxies with two different exponential profiles - one for the inner
and another for the outer part of the disc.  The radius at which the change of
the slope occurs is usually called the break radius. 
If the disc inner scale length is larger than the outer scale length, the galaxy
has a down-bending (Type II) profile. An up-bending (Type III) profile refers to
a disc with steeper inner and shallower outer profile
\citep[e.g.][]{pohlen2002,erwin2005,elmegreen2006,pohlen2006}.

Recent studies indicate that an essential fraction of all disc galaxies requires
modelling with a double exponential profile.  The exact fraction of galaxies in
each of these types is, however, still highly debated.  \citet{pohlen2006}
investigated the light profiles of late-type galaxies in the optical band. They
conclude that $\sim$ 60\% of the galaxies have Type II and $\sim$ 30\% Type III
profiles. Only $\sim$ 10\% showed a single exponential profile. A similar study
of early type disc galaxies indicates that the fraction of Type I, II and III
profiles is 27\%, 42\% and 24\%, respectively \citep{erwin2008}. It is
remarkable that the remaining 7\% of the galaxies exhibit a combination of Type
II and Type III discs. \citet{hunter2006} studied the disc profiles in very
late-type galaxies.  They find that 22\% of the galaxies have a down-bending and
8\% an up-bending profile. The remaining galaxies in their sample are well
fitted with a single exponential profile. 

The physical origin of these disc breaks is not fully understood. In the last
decades several theories about the origin of disc breaks were developed.
\citet{kruit1987} proposed that the down-bending break is a result of the
formation process of the galaxy itself. In this scenario the break is located at
the radius of the maximum angular momentum of the original spheroidal cloud. 
Another theory is related to the density of the gas in the disc. Once this gas
density falls below a certain threshold, stars cannot form with the same
efficiency anymore and thus a break should develop at this point in the galactic
disc \citep{kennicutt1989}.  Nevertheless, other processes such as turbulent
compressions of the gas could trigger star formation in the outer parts of the
galaxies \rev{despite the} low gas densities. \citet{elmegreen2006} suggest that
these different boundary conditions of star formation could introduce a double
exponential disc profile in the galactic disc as well.
Further suggestions of \citet{elmegreen1994} and \citet{schaye2004} connect the
disc break to the transition between the cool and warm phases of the
interstellar medium. 
Finally, non-axisymmetric features such as bars could play a major role in the
development of Type II disc breaks. A number of N-body simulations
\citep[e.g.][]{sellwood1980, lia2003,debattista2006} indicate that bars can
strongly redistribute the angular momentum in galaxies. Therefore bars also
influence the distribution of material in the disc and could trigger the
formation of a double exponential profile.  \citet{foyle2008} simulated the
development of single exponential discs neglecting cosmological influences such
as interactions and accretion. They found that these exponential discs easily
develop a down-bending disc break. It is remarkable that the inner disc profiles
strongly evolve whereas the outer disc profiles remain in a similar state. This
is probably due to the difference between the dynamical time scale of the inner
and outer disc. Thus the density profiles in the outer parts could provide clues
about the properties of the original disc profile. 

In this paper we make use of a large dataset obtained by the Spitzer Survey of
Stellar Structure in Galaxies (\ssg) to investigate how the properties of
galaxies, and of their bars and bulges, relate to the properties of their spiral
arms. Our selection of suitable subsamples of \ssg\, are described in Sect.
\ref{sec:data_sample}.  

We use three different approaches:
Firstly, we connect fundamental physical properties of the galaxies and their
structural components with their classical morphological arm classification
(Sect.  \ref{sec:general_properties}).
Secondly, in order to quantify the amplitude of the spiral arms and bar, 
the strength of the arm and bar \rev{is parametrized} by measuring
the arm-interarm as well as bar-interbar contrasts. As a result, we provide
these measurements for a large subsample of S$^4$G \rev{galaxies}. 
Section \ref{sec:contrast_measurements} describes the measurement of
these contrasts and connects them to some fundamental parameters of the
galaxies.
Thirdly (Sect. \ref{sec:disk_breaks}), we explore differences in the break of
Type II discs according to the arm classes, as well as the measured arm and bar
contrasts.  Furthermore features in the contrast profiles
\rev{are connected} to the break radius.
Finally our conclusions \rev{are presented} in Sect.
\ref{sec:conclusions}.

\section{Sample Selection and Data}
\label{sec:data_sample}
\subsection{Sample selection}
\label{subsec:sample_selection}
\begin{figure*}
	\subfloat{
		\includegraphics[width=0.5\hsize]{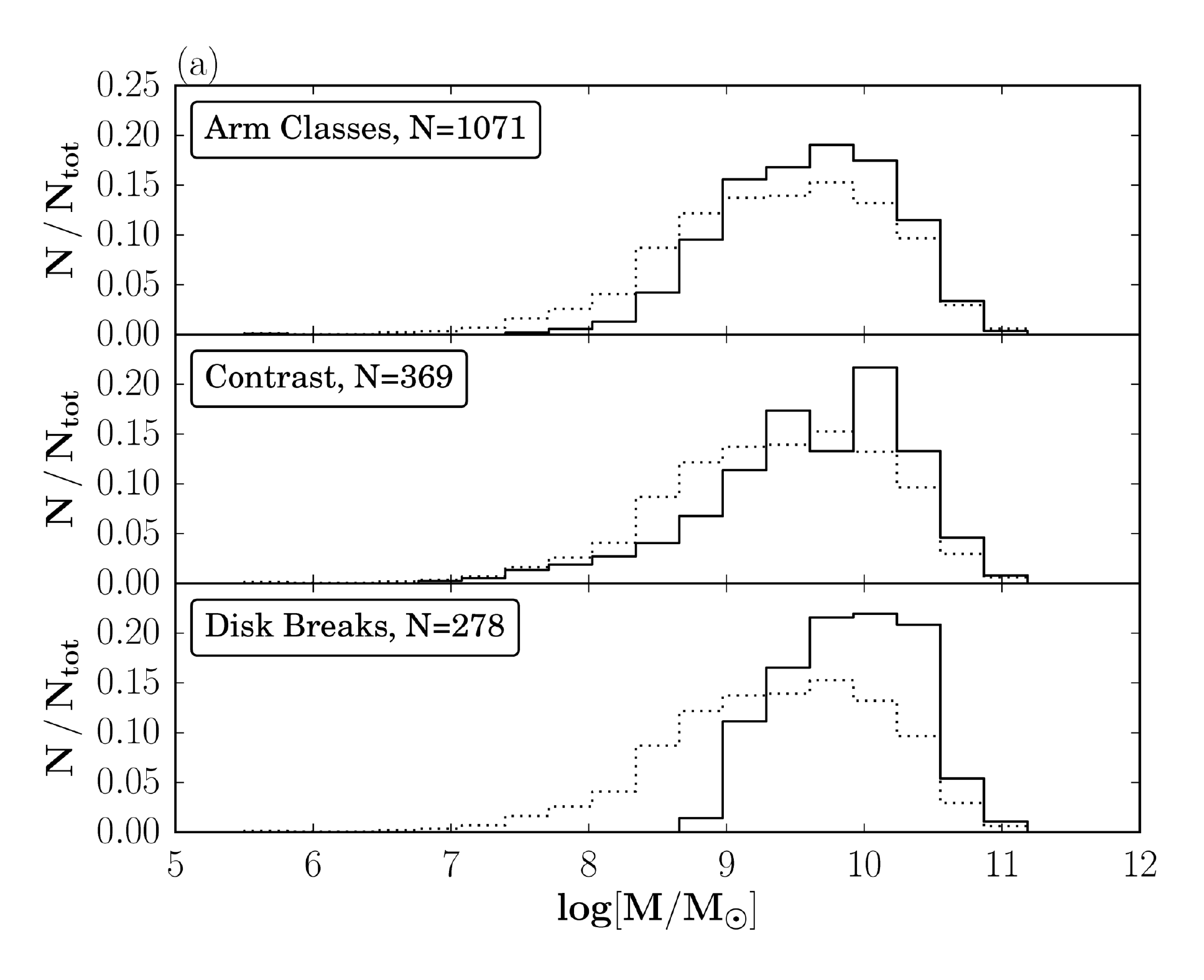}
		\label{subfig:massdist_sample}
	}
	\subfloat{
		\includegraphics[width=0.5\hsize]{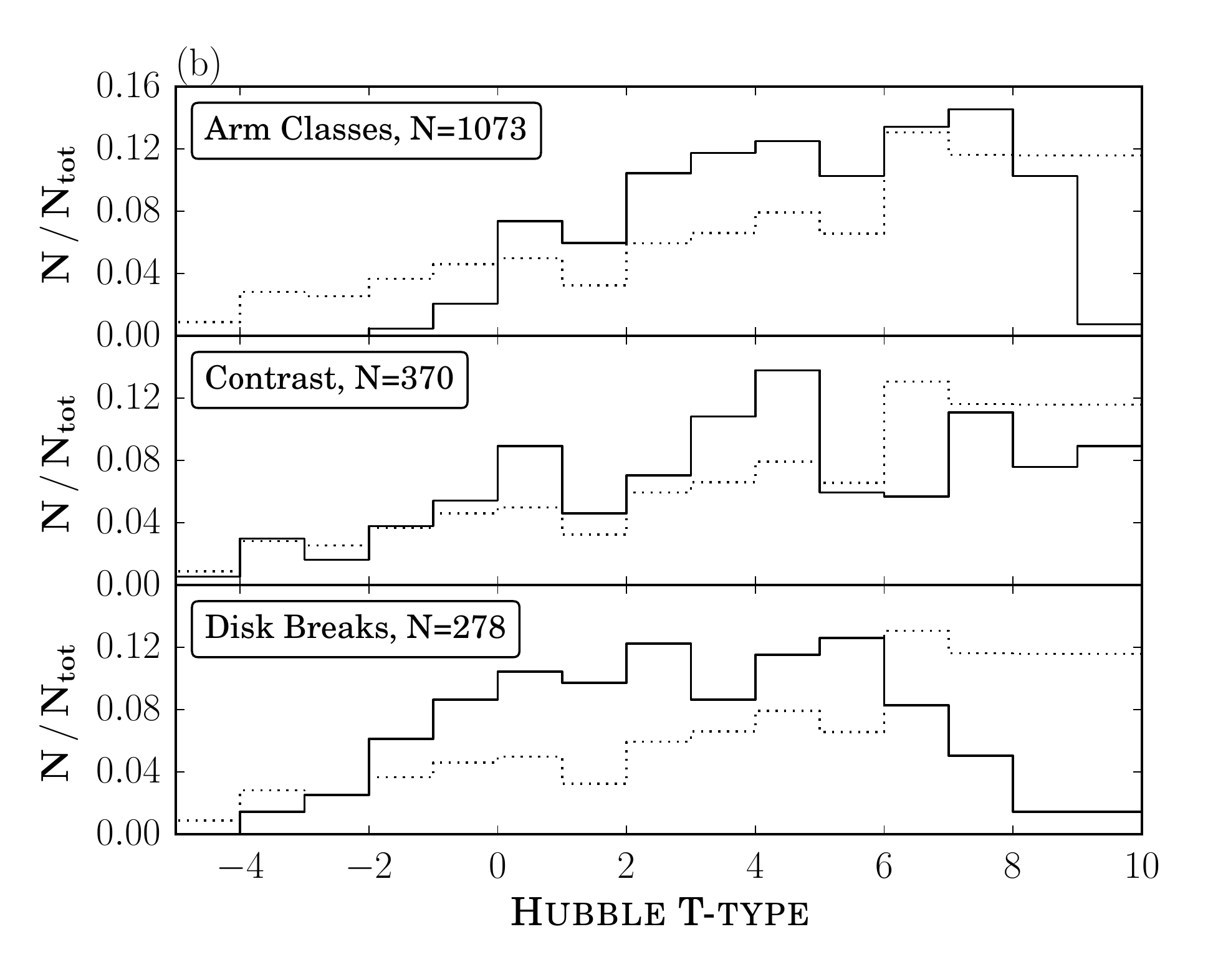}
		\label{subfig:tdist_sample}
	}
	\caption{
		Distributions of (a) total stellar masses and (b) \rev{mid-IR} Hubble
		T-types.  The dotted line displays the distribution of the full \ssg\
		sample, whereas the solid lines refer to our chosen subsamples.
		\rev{The arm class subsample is slightly shifted towards later
			Hubble-types whereas the disc break subsample has higher masses and
		earlier Hubble-types with respect to the parent sample. } The
		\rev{contrast subsample is} similar to the parent
		sample. Different numbers of galaxies in the left panel arises
		from the fact that stellar masses are not provided for all galaxies. 
	}
\end{figure*}
The parent sample of our study is the Spitzer Survey of Stellar Structure in
Galaxies (\ssg) \citep{sheth2010}. It is a volume- (d < \SI{40}{\mega\parsec}),
magnitude- ($m_{B_{\mathrm{corr}}} < 15.5$) and size-limited ($D_{25}$ >
\ang{;1;}) survey of 2352 galaxies. The galactic latitude is constrained to $
|b| > \ang{30;;}$, and the distance determined from HI redshifts. Each galaxy
is observed at \num{3.6} and \SI{4.5}{\microns} and mapped to $1.5 \times
D_{25}$. In azimuthally averaged surface brightness profiles a depth of
$\mu_{3.6\mu m}(AB)(1\sigma) \sim 27 \mathrm{\, mag} \mathrm{\, arcsec}^{-2}$
is reached, which corresponds to a stellar mass surface density of $\sim 1
\msol \pc^{-2}$. 

The galaxies are processed uniformly through the five \ssg\, pipelines.
Pipeline 1 (P1) produces science ready images from the two observations of each
galaxy \citep{regan2013}.  In pipeline 2 (P2), masks for fore- and background
objects are created. Pipeline 3 (P3) measures the local sky level, the surface
brightness profiles, ellipticities and position angles based on IRAF ellipse
fits \citep{jc2015}.  Pipeline 4 (P4) decomposes the two-dimensional stellar
distributions into subcomponents using GALFIT version 3 \citep{salo2015}.
Finally, pipeline 5 (P5) produces maps of the stellar mass distribution within
each galaxy \citep{querejeta2015}. 

In our study we \textbf{(i)} connect fundamental physical parameters with the
spiral arm class (Sect. \ref{sec:general_properties}), \textbf{(ii)} measure
the arm-interarm as well as bar-interbar contrast (Sect.
\ref{sec:contrast_measurements}) and \textbf{(iii)} investigate the nature of
down-bending disc breaks (Sect. \ref{sec:disk_breaks}).  Therefore it is
necessary to define three different subsamples that are suitable for the
corresponding investigations. The selection of these three subsamples is
described below. 

Section \ref{sec:general_properties} compares galaxy properties to their
spiral arm class as defined by \citet{buta2015}.  Therefore we constrain the
parent sample to the galaxies classified in that paper and we are left with a
subsample of 1074 galaxies.  Some plots in that section may contain a smaller
number of objects. For example, when analysing bulge properties, bulgeless
galaxies, as defined from the decompositions of \citet{salo2015}, are of course
not included.

In Sect. \ref{sec:contrast_measurements} we present the measurement of the
arm-interarm and bar-interbar contrasts. This measurement requires additional
constraints to our parent sample.  To avoid excessive measurement errors caused
by projection effects, the sample \rev{is limited} to galaxies with inclinations lower
than $\ang{50;;}$. When the determination of the coordinates of the galaxy
centres, as provided by pipeline 3, is uncertain, problems arise in the
transformation of the images to polar coordinates. Therefore those
galaxies \rev{are removed} from the sample. Since the bar-interbar contrast \rev{is measured} from the
bulge effective radius $\rbulge$ up to the bar semi-major axis $\rbar$ and the
arm-interarm contrast from the bar semi-major axis up to the radius $\rmax$ at
which the background noise gets dominant, only galaxies that fulfill
the condition $\rbulge < \rbar < \rmax$ \rev{are chosen}.  Furthermore, we visually inspect all
remaining objects to check the images for contamination by e.g. foreground
stars, stray light and companion galaxies, excluding such cases from the
sample. Finally, we exclude galaxies that are overly disturbed, have a
non-axisymmetric disc or peculiar spiral structure, since these galaxies are
not suitable for the measurement of the arm-interarm contrast. 
\rev{Thus the final subsample consists of 288 galaxies. }

For a discussion of Type II disc breaks in Sect. \ref{sec:disk_breaks} we use
both the data provided by two-dimensional multicomponent decompositions of
\citet{kim2014} and azimuthally averaged radial profiles by \citet{jc2013}.
\citet{kim2014} chose barred galaxies with Hubble types from S0 to Sdm and
excluded disturbed, faint and irregular galaxies. They also excluded images
which are contaminated by foreground stars or stray light and constrained the
inclination to $i < \ang{60;;}$. Their subsample contains 144 galaxies.
\citet{jc2013} uses galaxies with Hubble T-types in the numerical range $-3 \le
T \le 7$ and explicitly excludes Sdm and Sm galaxies due to their patchy and
asymmetric morphology. Furthermore they constrain the inclination to $i <
\ang{60;;}$ and the total stellar mass to $M > 2 \e{9} \msol$. They end up with
a subsample of 218 galaxies.  In order to obtain a larger sample for our
investigations of disc breaks, both of their subsamples \rev{are used} together. Since
a number of galaxies are common to both subsamples, we obtain a sample of 278
galaxies. 

To check how our samples compare with the full \ssg\ sample we plot the
distributions of total stellar masses (see Fig.  \ref{subfig:massdist_sample})
and \rev{mid-IR} Hubble T-types \rev{from \citet{buta2015}} (see Fig.  \ref{subfig:tdist_sample}). 
The subsamples intended for use in the discussion of
spiral arm classes (Sect.  \ref{sec:general_properties}) as well as the
measurements of the contrast profiles (Sect. \ref{sec:contrast_measurements})
show only a small shift towards higher masses, compared  to the full \ssg\
sample. 
Only the subsample that is used for the investigation of disc breaks (Sect.
\ref{sec:disk_breaks}) differs significantly. It lacks low-mass galaxies below
$M \simeq 10^9 \msol$ and shows a higher fraction of galaxies with masses
around $M \simeq 10^{10} \msol$. Considering the distribution of the Hubble
T-types,
\rev{the arm class subsample is slightly shifted towards late-types whereas 
the disc break subsample has earlier Hubble-types. In particular, the latter has
fewer } Hubble-types later than $T = 6$. These findings are consistent
with the sample selection criteria of \citet{kim2014} and \citet{jc2013} and
will be addressed below. 

\subsection{Data}
\label{subsection:data}
In this subsection we describe the datasets \rev{used} for our investigations.
Regardless of which subsample is used below, the data are based on the same
sources.
In this study we connect fundamental physical parameters of the galaxies with
their spiral arm class. The latter are taken from a classical morphological
analysis by \citet{buta2015}. 
General properties of the galaxies in the samples are adapted from the
two-dimensional multi-component decompositions conducted in pipeline 4 of the
\ssg\ survey \citep{salo2015}.  Some of these decompositions provide two
exponential disc profiles for the galaxies. Since it is not clear whether they
refer to a disc break or replace another subcomponent with different underlying
physics, such galaxies \rev{are not used} in this study.  Furthermore, the
decompositions can contain an unresolved point source in the centre of the
galaxies. We point out that this component does not necessarily refer
to a structural component with distinct underlying physics (e.g. active
galactic nuclei). In the following we will refer to this unresolved point
source as a nucleus and also consider the nucleus-to-total ratio in our
results.  The decompositions also provide measurements of the nucleus, bulge,
bar and disc-to-total ratio.  These ratios refer to the fraction of the flux of
the corresponding component to the total flux of all galaxy components
together. 
\rev{Due to different inclinations and position angles of the galaxies, it is
necessary to deproject the bar semi-major axis. Treating the bar as a one dimensional 
line, the deprojected bar length $r_{\mathrm{real}}$ is given by
\begin{eqnarray}
	r_{\mathrm{real}} = r_{\mathrm{obs}} \left( \sin^2\alpha \sec^2i  +  \cos^2\alpha \right)^{1/2}
	\label{eqn:deprojection}
\end{eqnarray}
with the observed bar length $r_{\mathrm{obs}}$, inclination $i$ and position 
angle $\alpha$ \citep{gadotti2007,martin1995}. }
Furthermore, \ssg \ provides the isophotal radius at a surface brightness level
of $25.5 \mathrm{\, AB} \mathrm{\, mag} \mathrm{\, arcsec}^{-2}$. In the
following we will refer to this radius as $\riso$.
The bar \sersic index is taken from \citet{kim2014} whereas disc breaks as well
as the inner and outer scale lengths come from both \citet{kim2014} and
\citet{jc2013}.
Whenever possible, distances are based on the mean redshift-independent
distances from the NASA Extragalactic Database (NED). If these are not
available, distances are calculated from the radial velocity corrected for
Virgo-centric infall as provided by the LEDA database. 
The mass represents the total stellar mass of the galaxy including all of its
components.  It is based on the absolute magnitude of the galaxy and converted
to mass using the calibration of \citet{eskew2012}. Details of this calibration
are also discussed in \citet{jc2013}.

\subsection{Plots}
\label{subsection:plots}
The plots in this paper are mostly designed in a uniform way.  They are split
in three normalized panels with the upper panel referring to flocculents
(``F''), the central panel referring to multi-armed (``M'') and the lower panel
referring to grand-design galaxies (``G''). This is also stated in the legends
of the plots.  In addition, the legend displays the total number of galaxies
within each panel (``n'') and their portion of the total number of galaxies
with this spiral arm class in \citet{buta2015} (only in Sect.
\ref{sec:general_properties}).  The width of the bins corresponds to
approximately 20\% of the median of the particular distributions.  
\rev{Since a Student t-test is a widely accepted method to significantly distinguish 
distributions, we provide tables with the results of this statistical test for all
presented plots (see Tables \ref{tab:ttest_general_properties} to \ref{tab:ttest_disk_breaks}).}

\section{General galaxy properties}
\label{sec:general_properties}
In the following, we compare some fundamental physical parameters between
galaxies with different arm classes. These comparisons are shown in Figs.
\ref{fig:tdist_class} to \ref{fig:l-bar_massbinned}. In addition, 
the results of Student t-tests for all presented plots \rev{are provided} in Table
\ref{tab:ttest_general_properties} and \ref{tab:ttest_general_properties_2}. 

Since flocculent galaxies tend to be of lower mass and smaller size than
multi-armed or grand-design galaxies, as discussed below, it is not
\rev{straightforward} to compare these spiral arm classes to each other
\citep[see][]{bosma1999}.  Thus, when we put forward differences between 
grand-design and flocculent galaxies, we are fully aware that these could be due to,
or at least linked to, differences in the stellar masses of their host
galaxies. To minimise this effect, this analysis \rev{was performed} anew, dropping
all galaxies beyond type Scd (type 6), since the later Hubble types (7 - 10)
include mainly small galaxies, with a rotation velocity less than $\sim 140
\mathrm{\, km\, s}^{-1}$ \citep[see][]{bosma2016}.  Comparing the results
obtained with and without this constraint, \rev{no significant differences are found}.  
Thus we chose to conduct the following investigations for the
full sample as presented in Sect. \ref{sec:data_sample}. 

\subsection{Results}
\label{subsec:general_results}
\begin{table}
	\centering
	\begin{tabular}{llcl}
\toprule
\toprule
Name									& Arm Cl.		&	T-statistic			&	P-value								\\
\midrule
Hubble types	 						& 	F - M	 	& 	$2.9\e{1}$	 		& 	$8.8\e{-123}$ 	 					\\ 
(Fig. \ref{fig:tdist_class})			& 	F - G	 	& 	$2.6\e{1}$	 		& 	$7.1\e{-75}$ 	 					\\ 
										& 	M - G	 	& 	$6.2	 $	 		& 	$1.5\e{-9}$ 	 					\vspace{0.1cm}\\ 

Stellar Mass							& 	F - M	 	& 	$-2.6\e{1}$	 		&$ 1.2\e{-109}$  						\\ 
(Fig. \ref{fig:mass_distribution})		& 	F - G	 	& 	$-1.9\e{1}$	 		&$ 3.3\e{-53 }$	 						\\ 
										& 	M - G	 	& 	$9.1\e{-1}$  		&$ 3.7\e{-1 } $   						\vspace{0.1cm}\\ 

Surface density							& 	F - M	 	& 	$-1.9\e{1}$			&$ 	2.3\e{-65 }$	 					\\ 
(Fig. \ref{fig:surf_dens})				& 	F - G	 	& 	$-1.2\e{1}$			&$ 	1.7\e{-28 }$	 					\\ 
										& 	M - G	 	& 	$5.8\e{-1}$ 		&$  5.7\e{-1 } $   						\vspace{0.1cm}\\ 

Bulge eff. radius						& 	F - M	 	& 	$4.7     $    		&$ 2.3\e{-5 } $   						\\ 
(Fig. \ref{fig:bulge_re_r25})   		& 	F - G	 	& 	$4.3     $    		&$ 6.1\e{-5 } $   						\\ 
                    					& 	M - G	 	& 	$-2.9\e{-1}$  		&$ 7.7\e{-1 } $	  						\vspace{0.1cm}\\ 

Bulge \sersic index						& 	F - M	 	& 	$-3.8$				& $	2.2\e{-4} $	 						\\ 
(Fig. \ref{fig:bulge_sersic})           & 	F - G	 	& 	$-5.4$				& $	2.9\e{-7} $	 						\\ 
										& 	M - G	 	& 	$-2.6$				& $	1.2\e{-2} $	 						\vspace{0.1cm}\\ 

Bulge-to-total ratio					& 	F - M	 	& 	$-4.2     $ 		&$  4.8\e{-5 } $   						\\
(Fig. \ref{fig:bulge-total})			& 	F - G	 	& 	$-8.4     $			&$  7.9\e{-14 }$	 					\\ 
										& 	M - G	 	& 	$-5.6     $			&$ 	1.2\e{-7 }$	 						\vspace{0.1cm}\\ 

Nucleus-to-total ratio					& 	F - M	 	& 	$-4.6     $			&$ 	6.1\e{-6 }$	 						\\ 
(Fig. \ref{fig:nucleus-total})			& 	F - G	 	& 	$-4.8     $			&$ 	5.1\e{-6 }$	 						\\ 
                    					& 	M - G	 	& 	$-1.8     $			&$ 	8.3\e{-2 }$	 						\vspace{0.1cm}\\ 

Bar axial ratio							& 	F - M	 	& 	$-1.7     $   		& $   9.9\e{-2 } $  					\\
(Fig. \ref{subfig:bar_ar})   			& 	F - G	 	& 	$-2.2     $   		& $   2.7\e{-2 } $  					\\
										& 	M - G	 	& 	$-5.5\e{-1}$  		& $   5.8\e{-1 } $  					\vspace{0.1cm}\\

Bar semi-major axis						& 	F - M	 	& 	$2.3				$ &$ 	2.0\e{-2}$	 	 				\\ 	
(Fig. \ref{subfig:bar_rout})			& 	F - G	 	& 	$-4.4\times10^{-1}	$ &$ 	6.6\e{-1}$ 	 					\\ 	
										& 	M - G	 	& 	$-2.0   			$ &$ 	4.4\e{-2}$ 	 					\vspace{0.1cm}\\ 	

Bar-to-total ratio						& 	F - M	 	& 	$-3.8\e{-1}$		&$ 	7.0\e{-1 }$	 						\\ 
(Fig. \ref{subfig:bar-total})			& 	F - G	 	& 	$-5.6     $			&$ 	9.3\e{-8 }$	 						\\ 
										& 	M - G	 	& 	$-4.9     $			&$ 	1.9\e{-6 }$	 						\vspace{0.1cm}\\ 

Bar \sersic index						& 	F - M	 	& 	$3.3     $  		&$ 1.4\e{-3 } $   						\\ 
(Fig. \ref{subfig:bar_sersic}) 			& 	F - G	 	& 	$3.7     $    		&$ 6.1\e{-4 } $   						\\ 
                						& 	M - G	 	& 	$-2.2\e{-1}$		&$ 	8.3\e{-1 }$							\\
\bottomrule
	\end{tabular}
	\caption{
		Overview of the results of the Student t-tests for plots presented in
		Sec. \ref{sec:general_properties}. 
	}
	\label{tab:ttest_general_properties}
\end{table}

\begin{table}
	\centering
	\begin{tabular}{lllcl}
\toprule
\toprule
Name								&		Mass 											& Arm Cl.		&	T-statistic		&	P-value					\\
\midrule
$\rbar / h_{\mathrm{disk}}$ 		&					$M > M_{\mathrm{th}}$				&	F - M	 	& 	$-1.6     	$ 	& 	$1.2\e{-1} 		$ 		\\ 
(Fig. \ref{fig:l-bar_hr-disk_mass})	&														& 	F - G	 	& 	$-3.1     	$ 	& 	$5.7\e{-3} 		$ 		\\ 
									&														& 	M - G	 	& 	$-2.3     	$ 	& 	$2.6\e{-2} 		$ 		\vspace{0.1cm}\\ 

									&					$M < M_{\mathrm{th}}$			 	& 	F - M	 	& 	$2.5\e{-1} 	$ 	& 	$8.0\e{-1} 		$ 		\\ 
									&														& 	F - G	 	& 	$8.8\e{-1}	$ 	& 	$3.8\e{-1} 		$ 		\\ 
									&														& 	M - G	 	& 	$4.9\e{-1} 	$ 	& 	$6.2\e{-1} 		$ 		\vspace{0.1cm}\\ 

									&	$M_{\mathrm{l}} \leftrightarrow M_{\mathrm{h}}$		&   F - F	 	&	$1.6     	$ 	& 	$	1.4\e{-1}	$ 		\\
									&														&	M - M	 	&	$-5.1\e{-1}	$ 	& 	$	6.1\e{-1}	$ 		\\
									&														&	G - G	 	&	$-3.1     	$ 	& 	$	2.4\e{-3}	$ 		\vspace{0.1cm}\\

$\rbar / R_{\mathrm{25}}$ 			&					$M > M_{\mathrm{th}}$				& 	F - M	 	& 	$-2.1$	 		& 	$4.8\e{-2}$	 			\\ 
(Fig. \ref{fig:l-bar_r25_mass})&															& 	F - G	 	& 	$-3.8$	 		& 	$5.2\e{-4}$	 			\\ 
									&														& 	M - G	 	& 	$-2.5$	 		& 	$1.6\e{-2}$	 			\vspace{0.1cm}\\ 
                                                                                            
									&					$M < M_{\mathrm{th}}$			 	& 	F - M	 	& 	$1.9$    		& 	$6.1\e{-2}$	 			\\ 						
									&														& 	F - G	 	& 	$9.7\e{-1}$		& 	$3.3\e{-1}$	 			\\ 						
									&														& 	M - G	 	& 	$-4.4\e{-1}$	& 	$6.6\e{-1}$	 			\vspace{0.1cm}\\ 						
                                                                                            
									&	$M_{\mathrm{l}} \leftrightarrow M_{\mathrm{h}}$		& 	F - F	 	& 	$3.8$ 			& 	$1.6\e{-3}$	  		 	\\ 
									&														& 	M - M	 	& 	$1.1\e{-1}$		& 	$9.2\e{-1}$ 	 	 	\\ 
									&														& 	G - G	 	& 	$-1.6$			& 	$1.1\e{-1}$ 	 	 	\\ 	
\bottomrule
	\end{tabular}
	\caption{
		Overview of the results of the Student's t-tests for plots of the bar
		semi-major axis in units of the disk scale length $\hdisk$ (see Fig.
		\ref{fig:l-bar_hr-disk_mass}) and isophotal radius $\riso$ (see Fig.
		\ref{fig:l-bar_r25_mass}).  These plots are splitted into two stellar
		mass bins. The separation is made at $\Mth = 10^{10.25} \msol$ with
		$M_{\mathrm{l}}$ and $M_{\mathrm{h}}$ refering to lower and higher
		total stellar masses.
		}
	\label{tab:ttest_general_properties_2}
\end{table}
\begin{figure}
	\centering
	\includegraphics[width=\hsize]{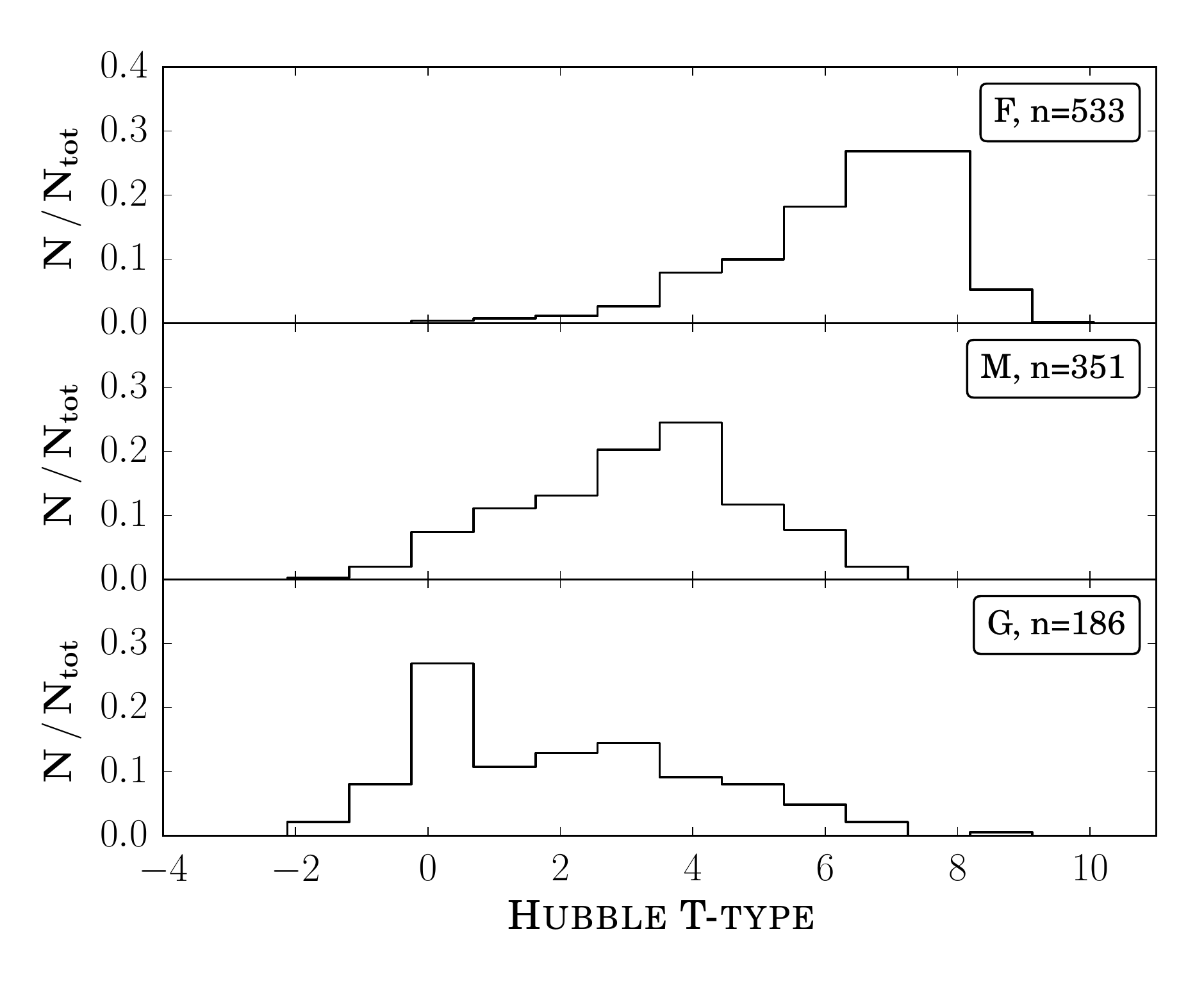}
	\caption{
		Distributions of the \rev{mid-IR} Hubble T-type for galaxies separated by arm class.
		Flocculent galaxies have later, grand-design galaxies earlier Hubble
		types.  Multi-armed galaxies are an intermediate case. 
	}
	\label{fig:tdist_class}
\end{figure}
The distributions of \rev{mid-IR} Hubble T-types separated by arm class \rev{are} displayed in Fig.
\ref{fig:tdist_class}. It clearly shows that grand-design galaxies \rev{tend to} have earlier
Hubble types, whereas \rev{most} flocculents \rev{tend to have later Hubble types}. Multi-armed galaxies
are an intermediate case. However, a Student's t-test indicates a significant
difference between all three distributions. 

\begin{figure*}
	\subfloat
	{
		\includegraphics[width=0.5\hsize]{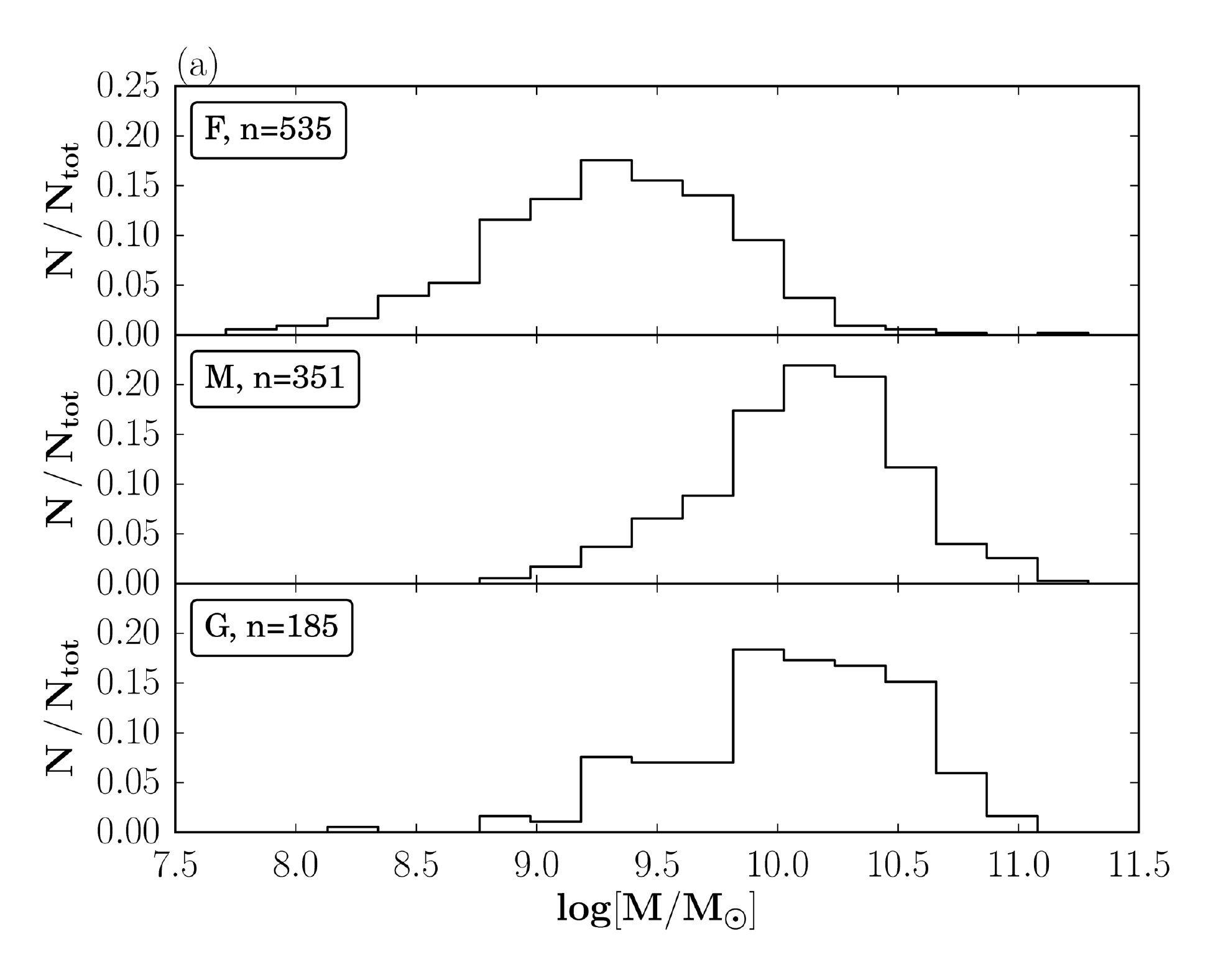}
		\label{fig:mass_distribution}
	}
	\subfloat
	{
		\includegraphics[width=0.5\hsize]{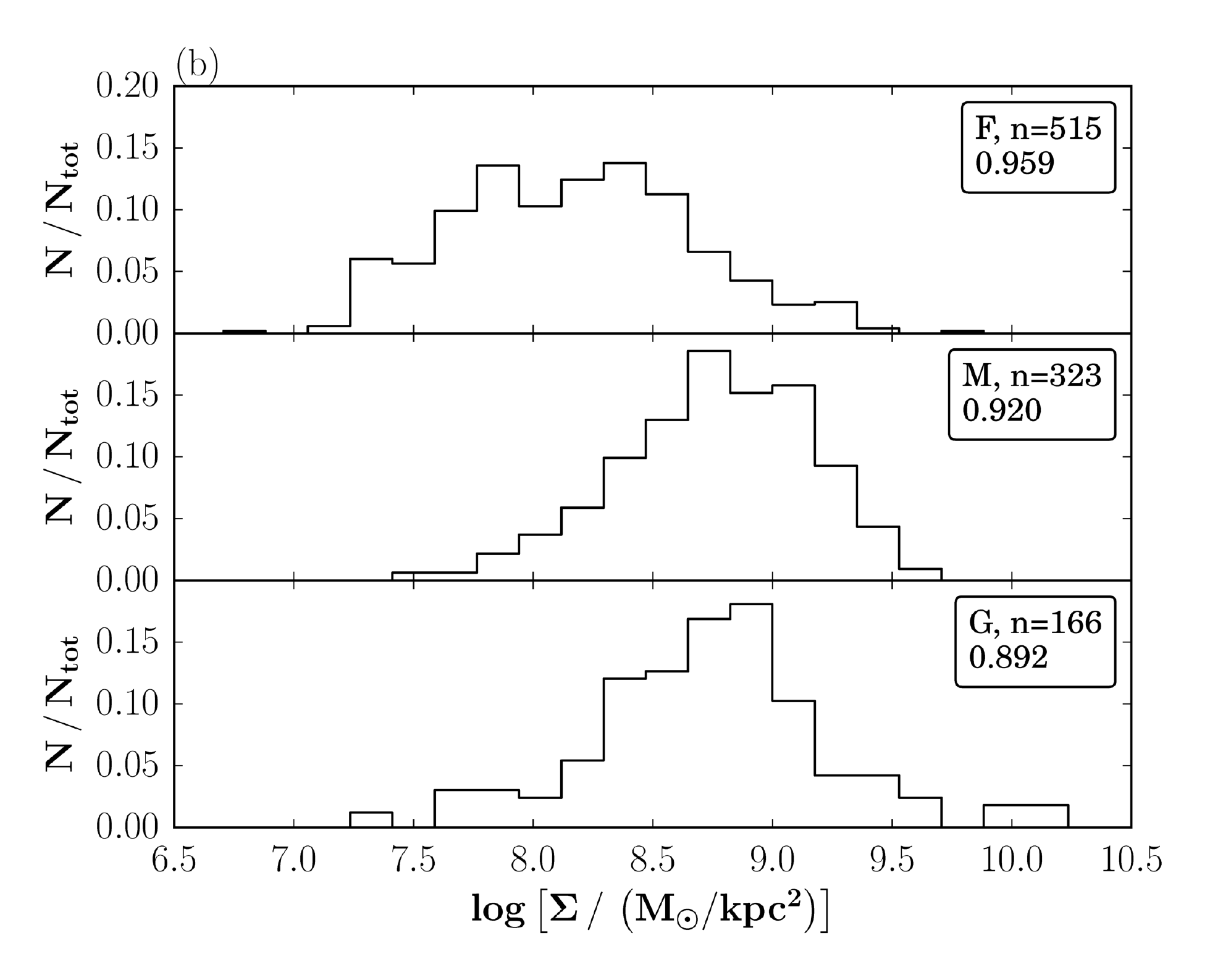}
		\label{fig:surf_dens}
	}
	\caption{
		Distributions of (a) total stellar mass and (b) total stellar mass
		surface density separated by arm class.  Flocculent galaxies have
		significantly lower masses and surface densities whereas the
		distributions of multi-armed and grand-design galaxies are similar.
		This represents a fundamental difference between the arm classes.  
	}
	\label{fig:mass_surfdens}
\end{figure*}

The distributions of stellar mass and stellar mass surface density of the
galaxies are presented in Fig. \ref{fig:mass_surfdens}.  In order to better
illustrate the results, in these figures the width of the bins is not exactly
20\% of the median of the distributions.  The distributions of the total stellar
mass show a high similarity between multi-armed and grand-design galaxies,
whereas flocculent galaxies have significantly lower masses. 

The galaxy stellar mass surface density is given by
\begin{eqnarray}
	\Sigma = \dfrac{M_{\mathrm{st}}}{h_{\mathrm{disc}}^2 \pi}
	\label{eqn:surfdens}
\end{eqnarray}
with the total stellar mass $M_{\mathrm{st}}$ of the galaxy and the disc
exponential scale length $h_{\mathrm{disc}}$.  The distributions of the stellar
mass surface density clearly show that flocculent galaxies have a lower galaxy
surface density \rev{whereas} multi-armed and grand-design galaxies
are very similar.  
\rev{When using the isophotal radius $\riso$ instead of the disc exponential scale length $\hdisk$
for the calculation of the surface density, all distributions were shifted towards
lower values while the differences between the spiral arm classes remained similar. }

\begin{figure*}
	\subfloat
	{
		\includegraphics[width=0.5\hsize]{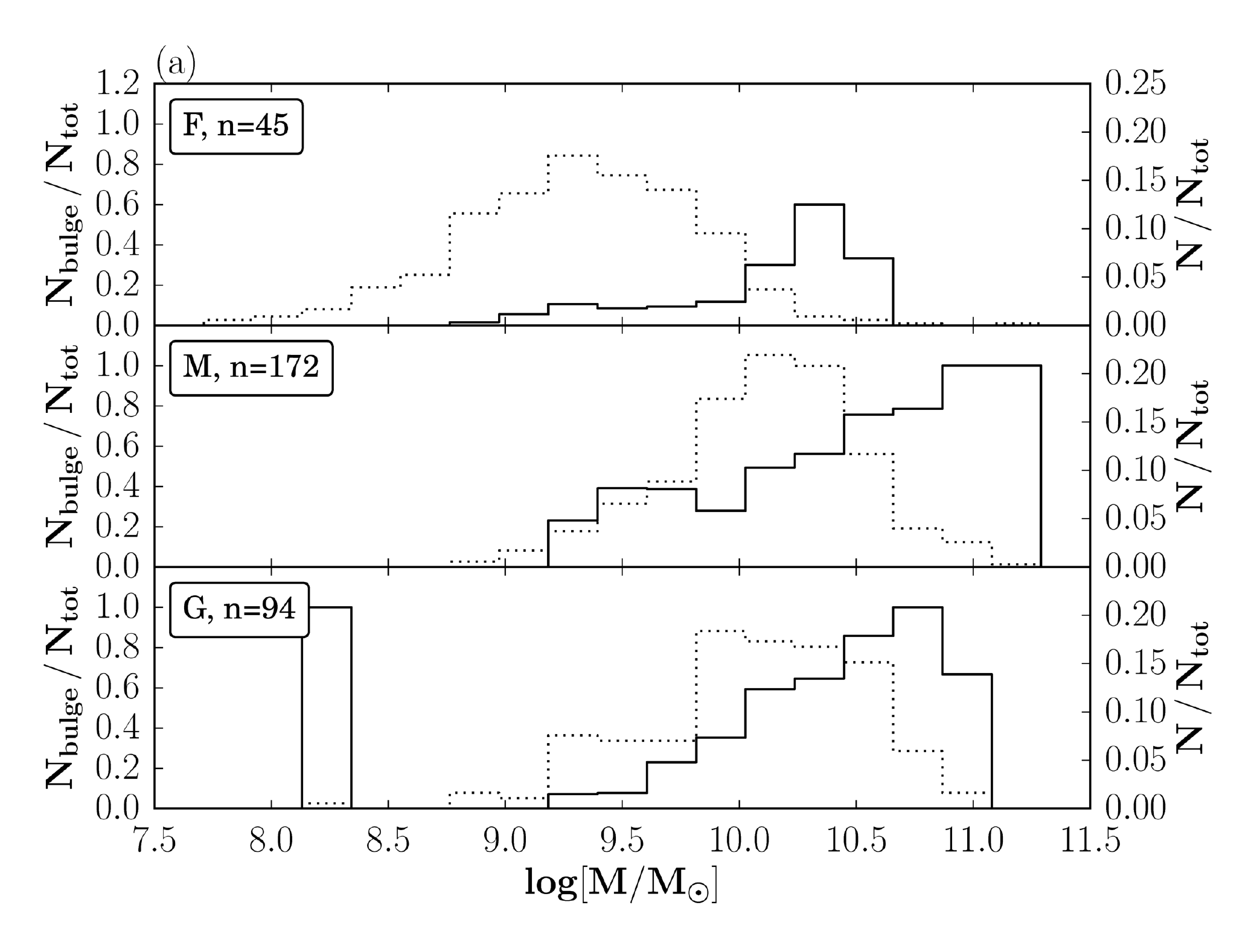}
		\label{fig:bulge_frequency}
	}
	\subfloat
	{
		\includegraphics[width=0.5\hsize]{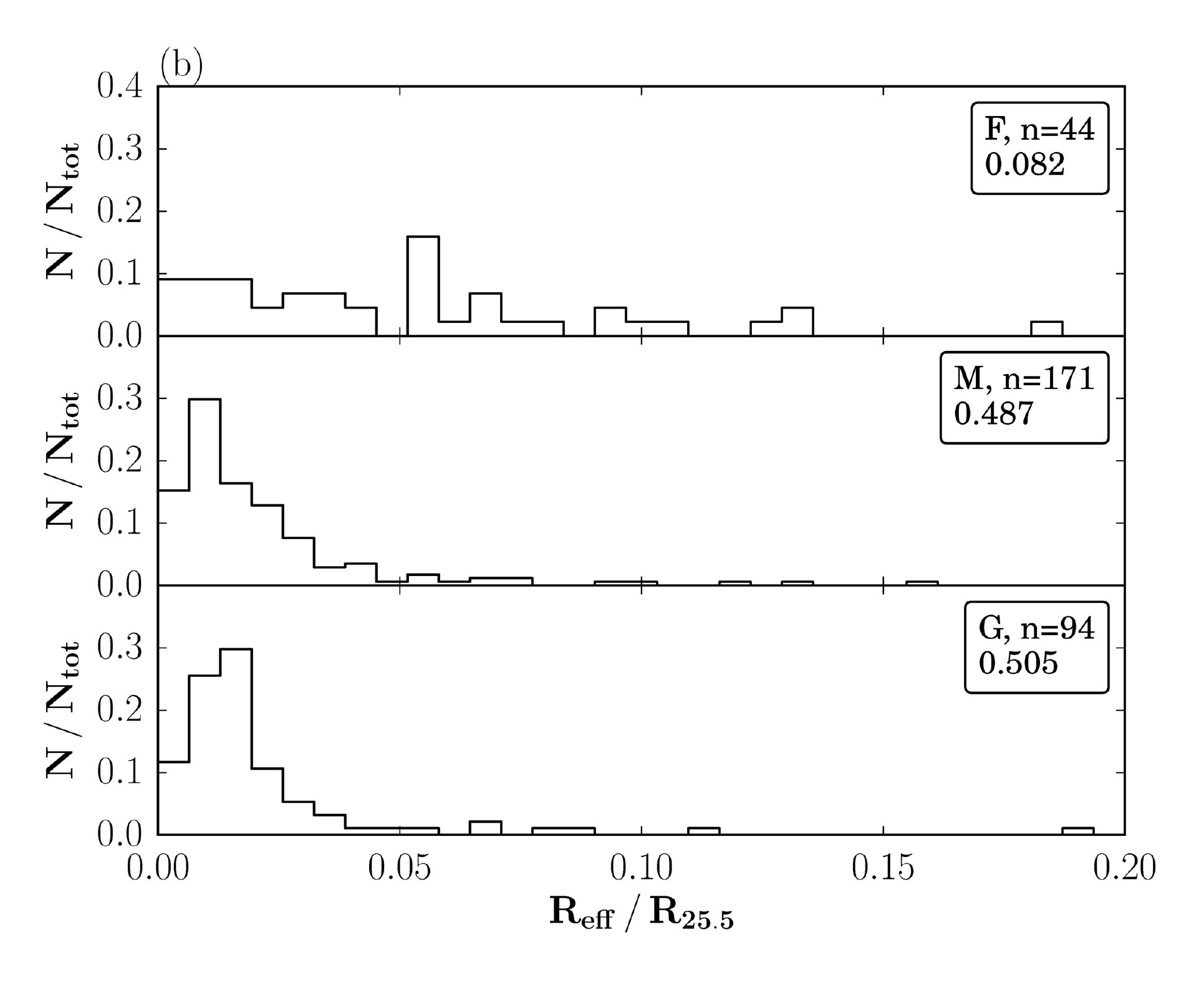}
		\label{fig:bulge_re_r25}
	} \\
	\subfloat
	{
		\includegraphics[width=0.5\hsize]{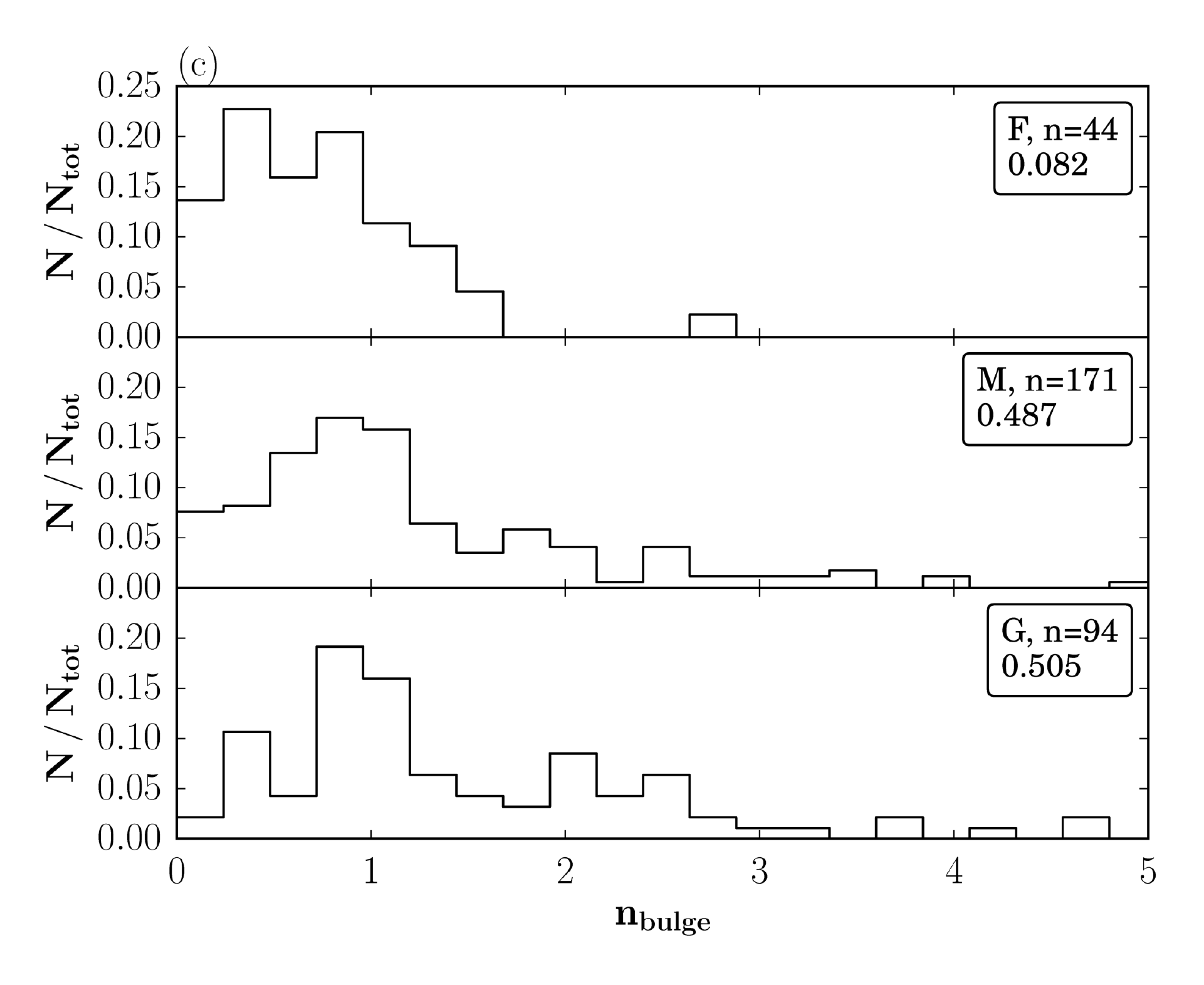}
		\label{fig:bulge_sersic}
	}
	\subfloat
	{
		\includegraphics[width=0.5\hsize]{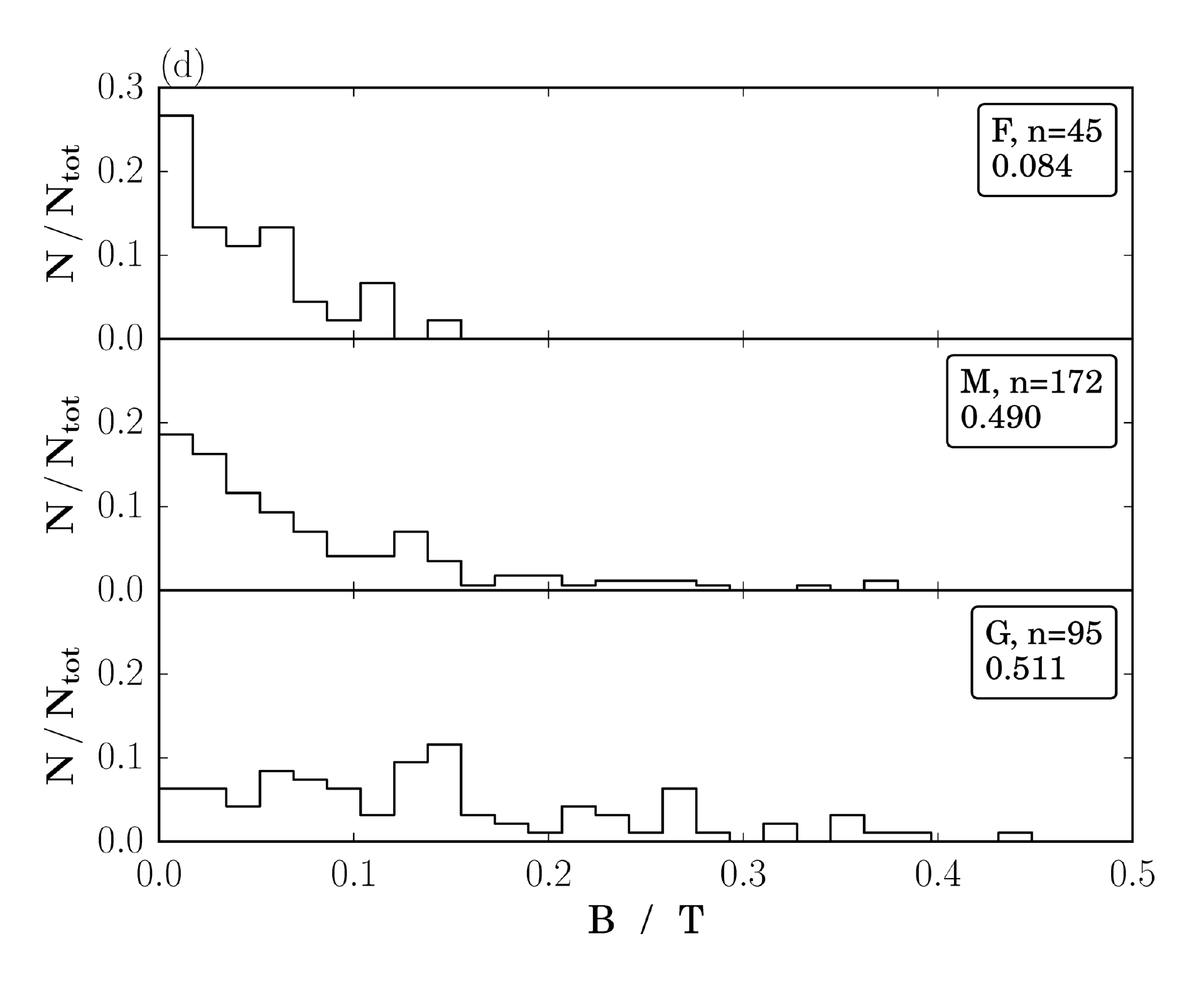}
		\label{fig:bulge-total}
	}
	\caption{
		(a) Fraction of galaxies that requires a bulge component in the
		decompositions as a function of the total stellar mass (solid lines,
		left ordinate).  The dotted lines (right ordinate) represent the
		distributions of the total stellar mass.  The fraction of galaxies with
		bulges increases with mass.  Galaxies with masses below $10^9 \msol$
		are mostly bulgeless.
		(b) Distributions of the bulge effective radius in units of the
		isophotal radius $\riso$. Flocculents show larger normalized bulge
		radii compared to multi-armed and grand-design galaxies, which show
		similar distributions. 
		(c) Distributions of the bulge \sersic index. The \sersic indices of
		flocculents are mostly lower than 2 whereas the distributions of
		multi-armed and grand-design galaxies are similar and extend to higher
		values. 
		(d) Distributions of the bulge-to-total luminosity ratio.  This ratio
		increases from flocculent to grand-design galaxies.
	}
	\label{fig:bulge}
\end{figure*}

\begin{figure}
	\includegraphics[width=\hsize]{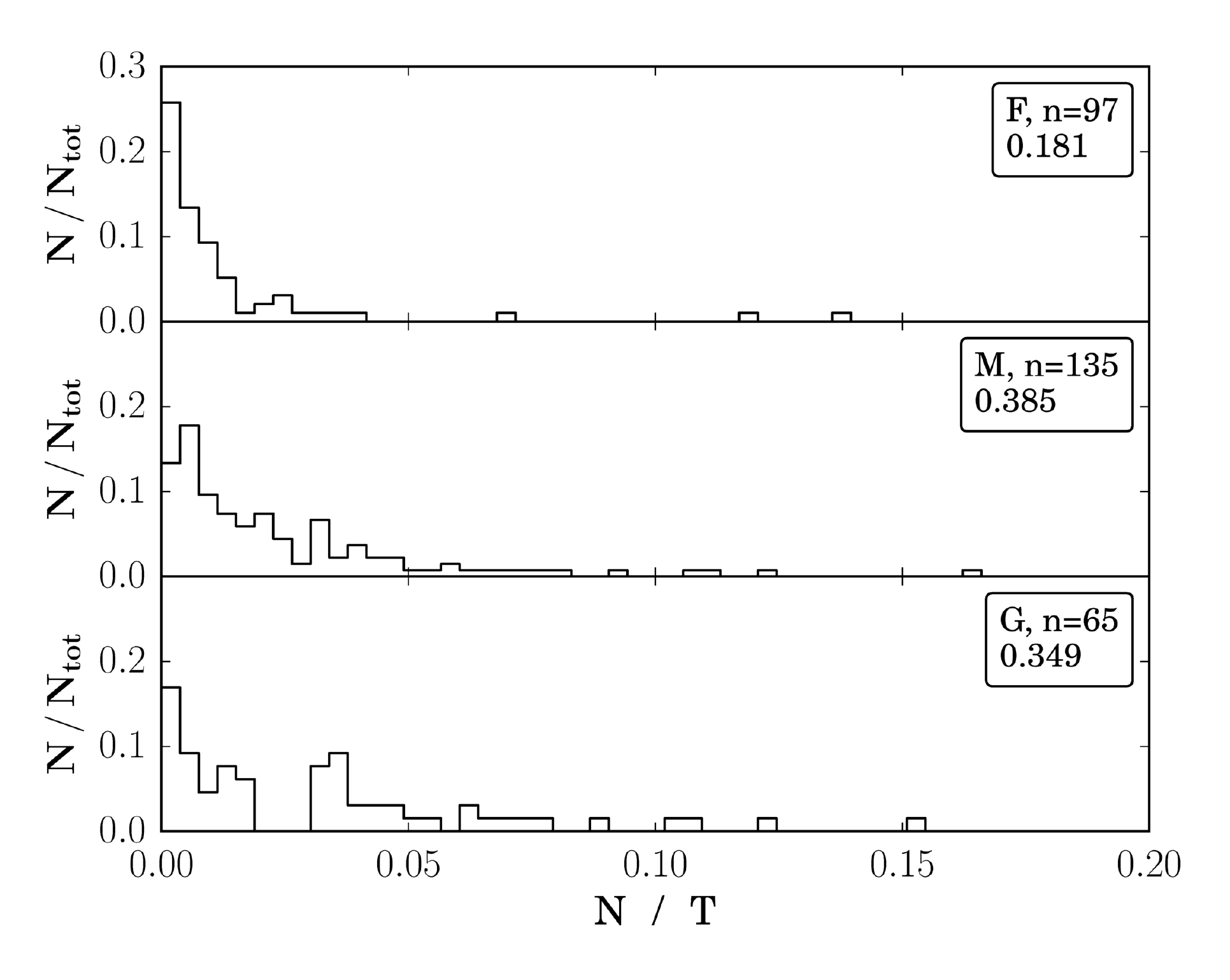}
	\caption{
		Distributions of the nucleus-to-total luminosity ratio.  Flocculent
		galaxies show, on average, a lower nucleus-to-total ratio as compared
		to multi-armed and grand-design galaxies. 
	}
	\label{fig:nucleus-total}
\end{figure}

Moreover, the portion of models which require a bulge component as a
function of the total stellar mass is presented in Fig. \ref{fig:bulge_frequency}. 
The fraction of bulge components in the decompositions increases with
mass \rev{and,} regarding multi-armed and grand-design galaxies, this fraction seems to
reach 100\% for high mass galaxies around $10^{11} \msol$.  However, one should
keep in mind that the number of galaxies with those high masses is very low in
the chosen subsample. The fraction of
flocculent galaxies that require a bulge component in the decompositions also
increases with stellar mass \rev{but never reaches more than 60\%}. We point out that galaxies with stellar masses
below $10^9 \msol$ are mostly bulgeless. 
In order to analyse the properties of the bulge in greater detail, the
bulge effective radius $\rbulge$ \rev{is plotted} in units of the isophotal radius $\riso$
separated by arm class (see Fig. \ref{fig:bulge_re_r25}).  The bulges of
flocculent galaxies are larger relative to the size of the galaxy itself.  The
distributions for multi-armed and grand-design galaxies are similar and
indicate that these galaxies have more compact bulge components. 
Furthermore, nearly all of the bulge \sersic \rev{indices} of flocculent galaxies are
within a range from 0 to 2 (see Fig. \ref{fig:bulge_sersic}) \rev{whereas the} \sersic
\rev{indices} of multi-armed and grand-design galaxies are very similar to each other
and extend to higher values. 
In addition, the bulge-to-total luminosity ratio increases from
flocculent to grand-design galaxies (see Fig. \ref{fig:bulge-total}). 
Finally, the unresolved point source component in the
decompositions (``nucleus'') \rev{is also considered}.  The fraction of decompositions that require this
component in the models is approximately twice as high for multi-armed ($\sim
39\%$) and grand-design galaxies ($\sim 35\%$) than for flocculent galaxies
($\sim 18\%$).  In addition, the nucleus-to-total luminosity ratio of
flocculents is, on average, lower than for the other spiral arm classes (see
Fig.  \ref{fig:nucleus-total}).

\begin{figure*}
	\subfloat
	{
		\includegraphics[width=0.5\hsize]{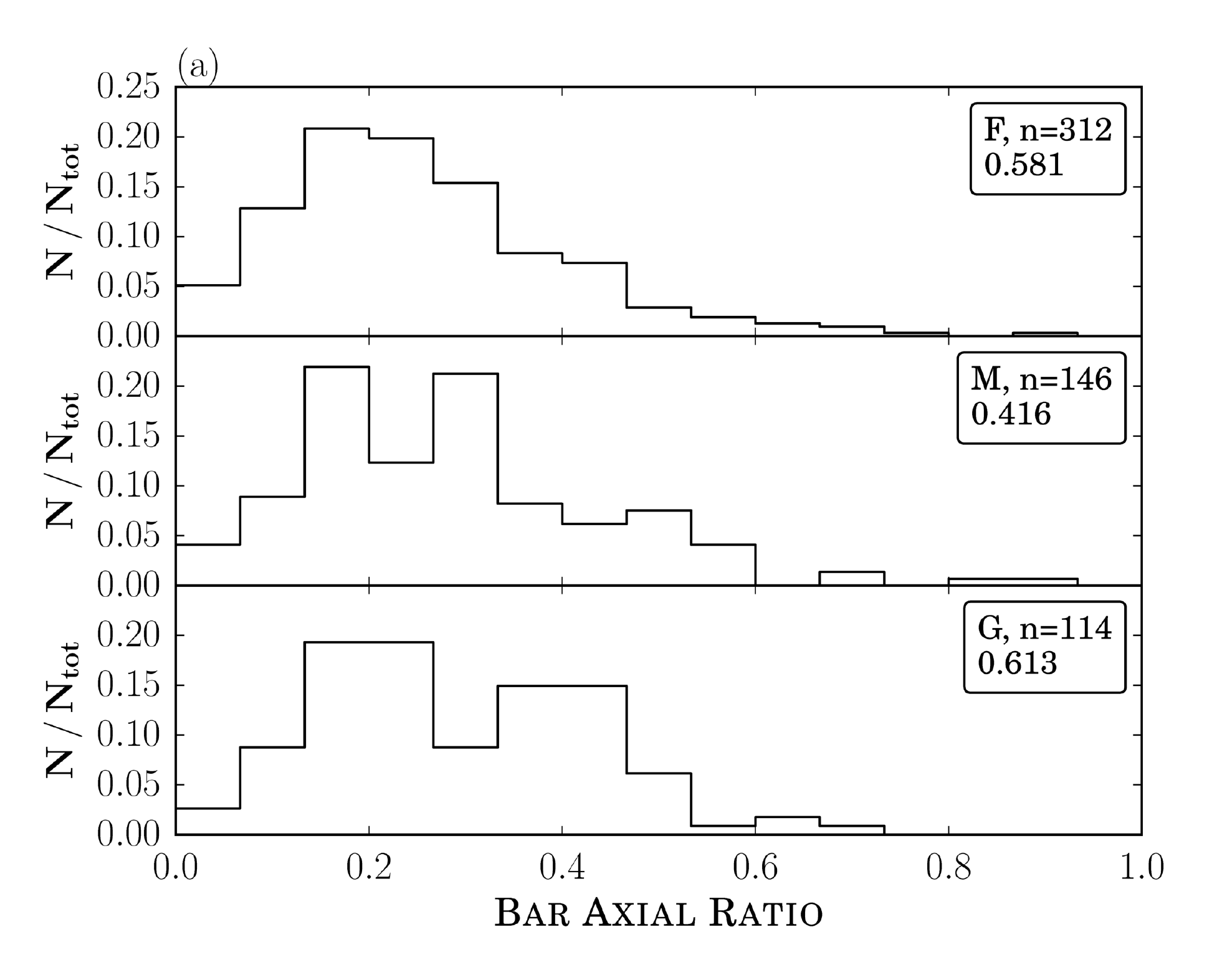}
		\label{subfig:bar_ar}
	}
	\subfloat
	{
		\includegraphics[width=0.5\hsize]{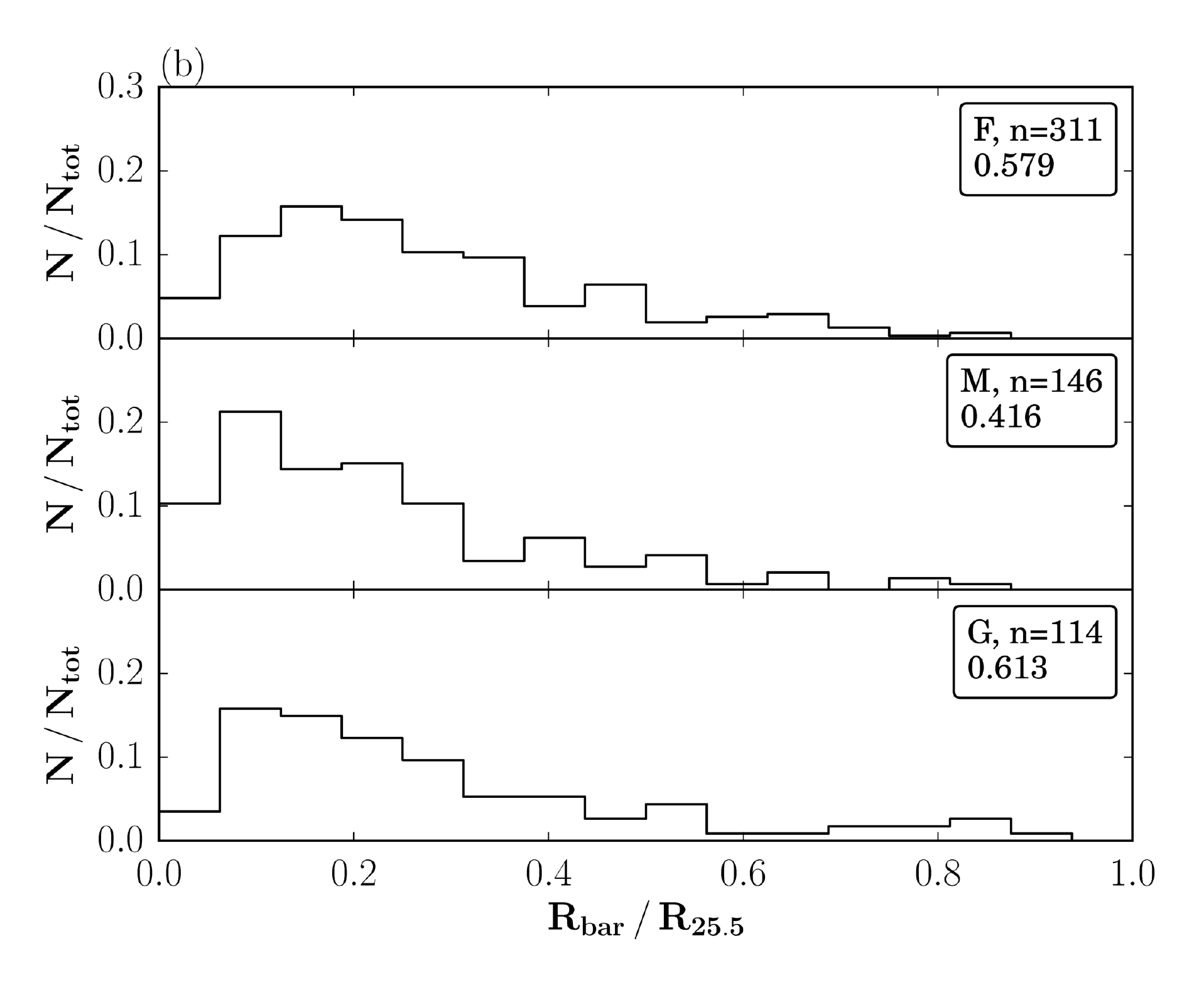}
		\label{subfig:bar_rout}
	}	\\	
	\subfloat
	{
		\includegraphics[width=0.5\hsize]{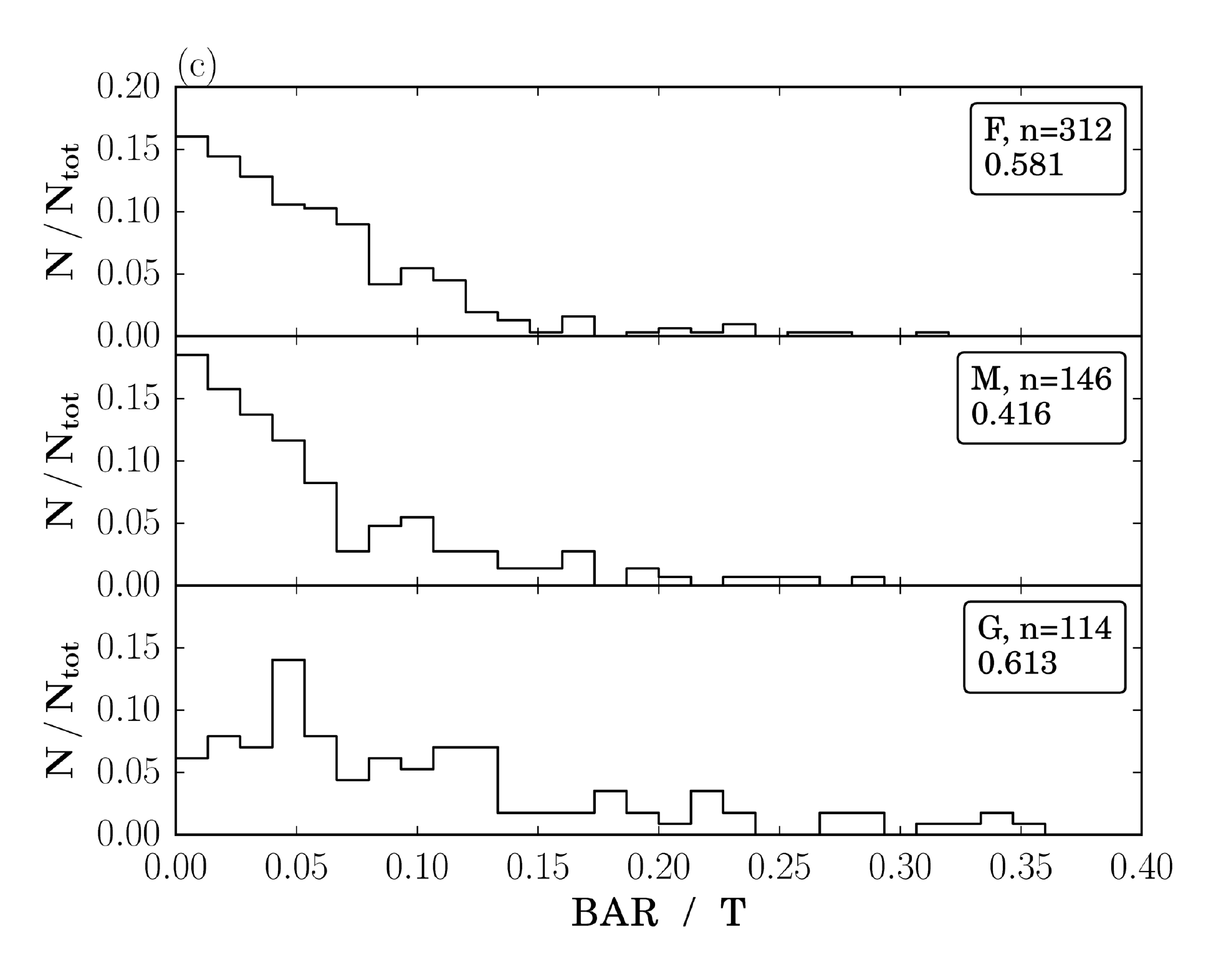}
		\label{subfig:bar-total}
	}	
	\subfloat
	{
		\includegraphics[width=0.5\hsize]{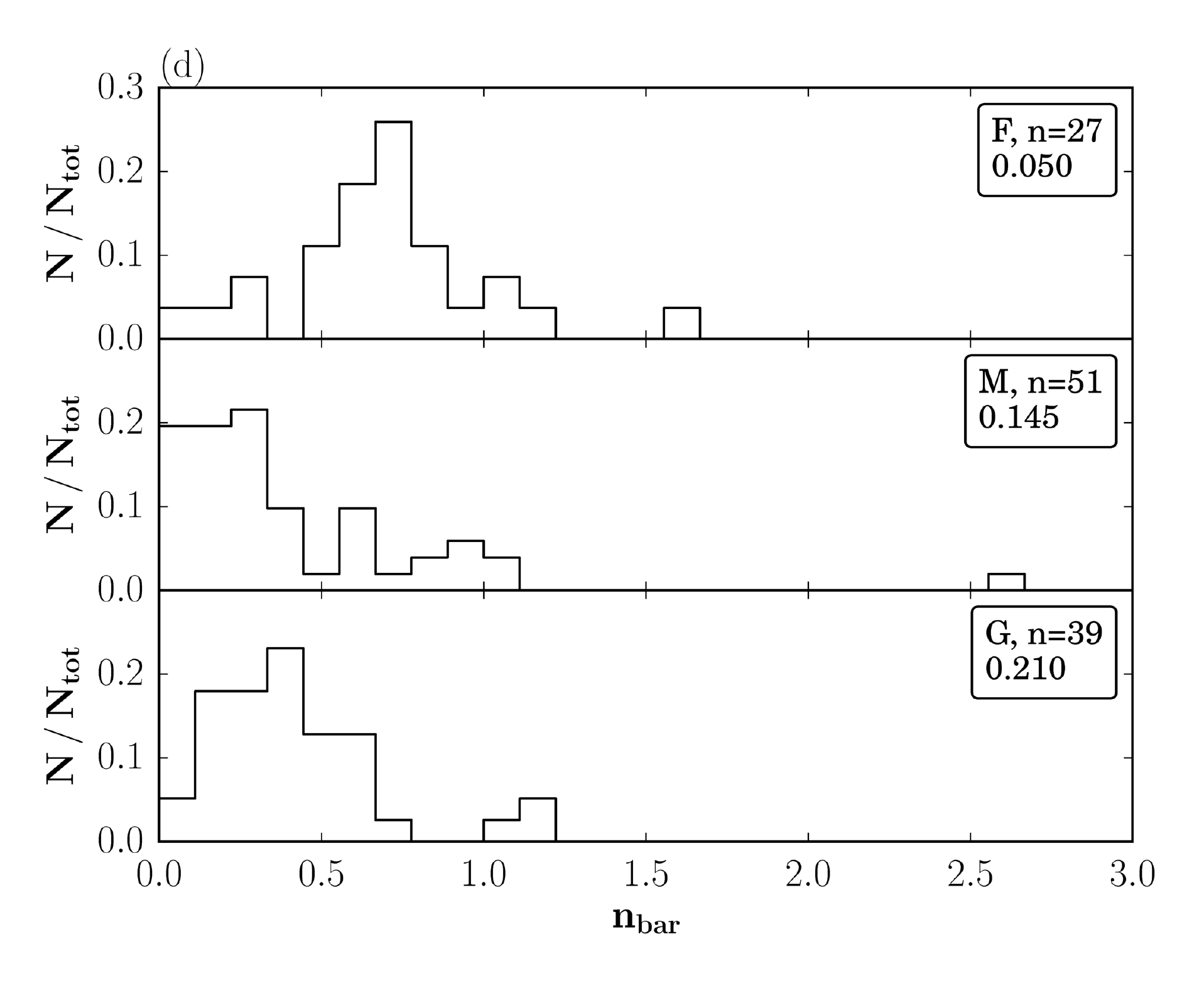}
		\label{subfig:bar_sersic}
	}
	\caption{
		\rev{Distributions of the bar axial ratio (a) and bar semi-major axis in
		units of the isophotal radius (b). The distributions of the different
		spiral arm classes are indistinguishable. }
		(c) Distributions of the bar-to-total luminosity ratio. The distribution
		of grand-design galaxies shows higher average values. Flocculents and
		multi-armed galaxies have similar distributions. 
		(d) Distributions of the bar \sersic index.  Flocculent galaxies have
		exponential bars whereas multi-armed and grand-design galaxies have
		bars with a flatter radial profile. 
	}
	\label{fig:bar}
\end{figure*}

\rev{Figs. \ref{subfig:bar_ar} and
\ref{subfig:bar_rout} show
the bar axial ratio and bar semi-major axis in units of the isophotal radius
$\riso$.} A Student t-test does not indicate a significant
difference of the distributions. 
In addition, the bar-to-total luminosity ratio (see Fig.
\ref{subfig:bar-total}) of grand-design galaxies is higher compared to
multi-armed and flocculent galaxies \rev{which have similar distributions.}
The \sersic index of the bar, as provided by \citet{kim2014}, is higher for
flocculent galaxies (see Fig. \ref{subfig:bar_sersic}). Thus, the radial
profile of their bars is more similar to an exponential profile even if their
bar \sersic index is not exactly unity. In contrast, multi-armed and
grand-design galaxies have bar \sersic \rev{indices} around $\sim 0.5$ and therefore
a flatter profile. 

\begin{figure*}
	\subfloat{
		\includegraphics[width=0.5\hsize]{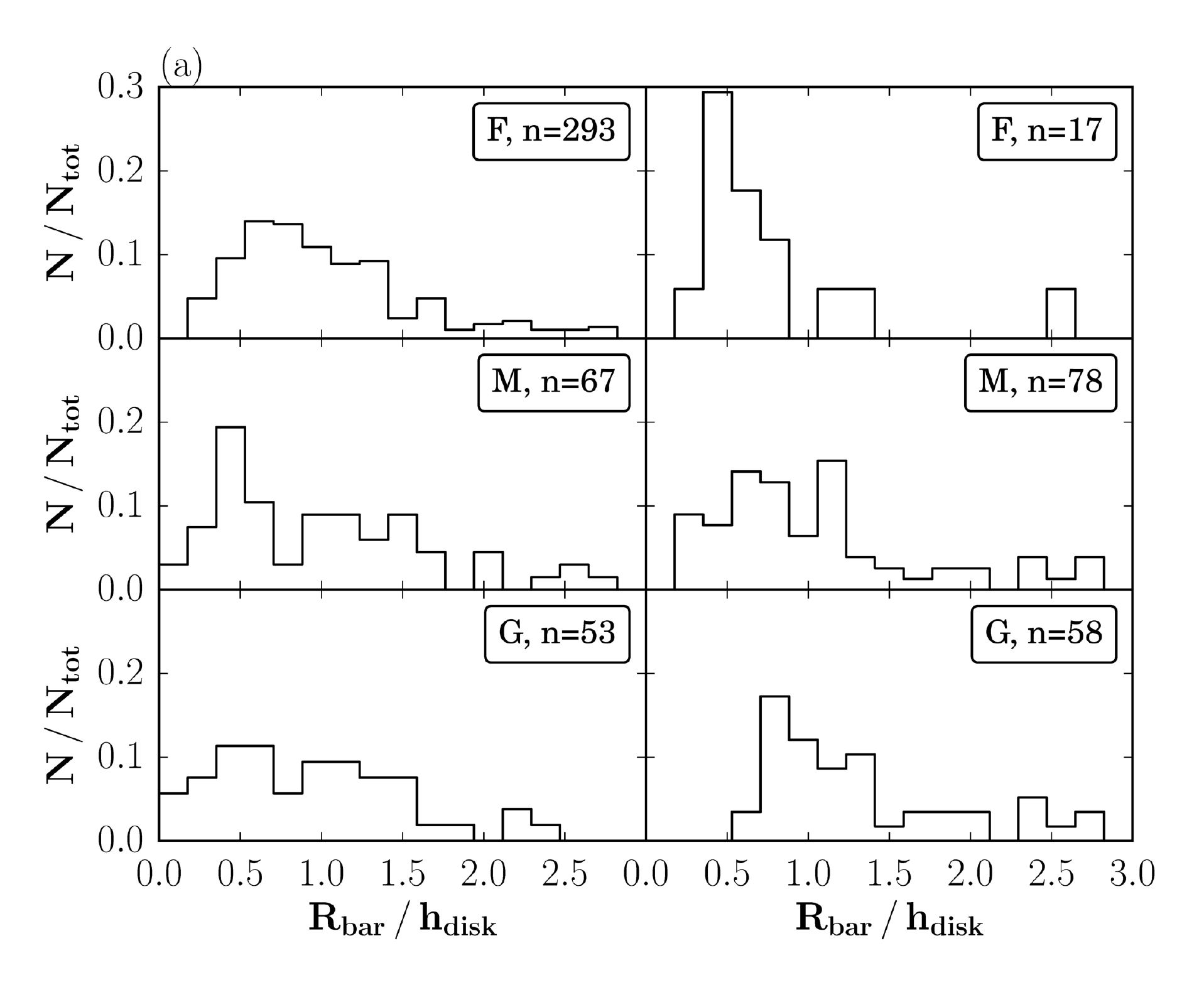}
		\label{fig:l-bar_hr-disk_mass}
	}
	\subfloat{
		\includegraphics[width=0.5\hsize]{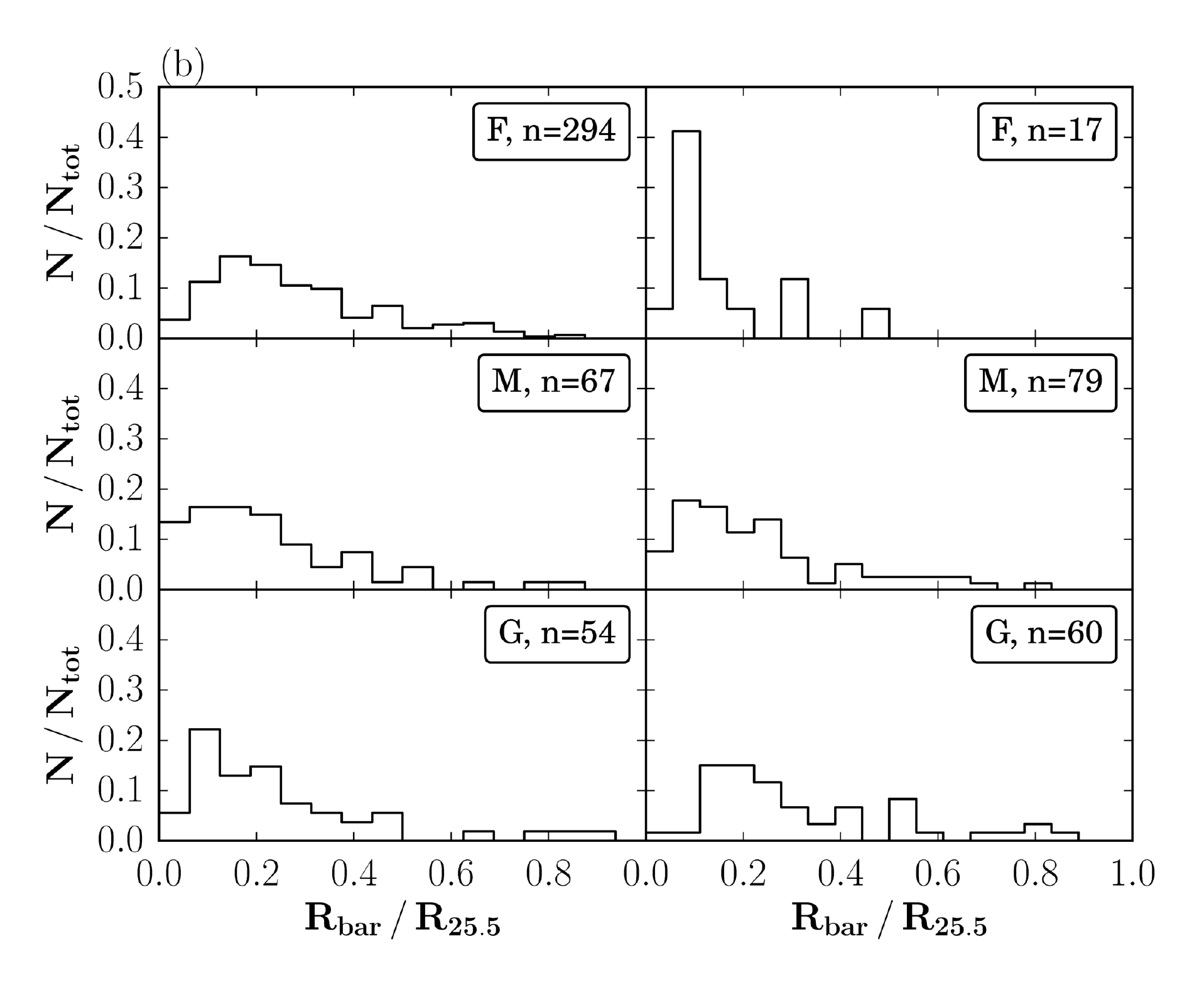}
		\label{fig:l-bar_r25_mass}
	}
	\caption{
		(a) Distributions of the bar semi-major axis in units of the disc
		exponential scale length and (b) isophotal radius $\riso$ separated by
		spiral arm class.  The plots are split into two mass bins at $\Mth =
		10^{10.25} \msol$.  The left columns of the panels refer to low stellar
		masses, whereas the right columns correspond to high stellar masses,
		respectively.
	}
	\label{fig:l-bar_massbinned}
\end{figure*}

Fig. \ref{fig:l-bar_hr-disk_mass} shows the bar semi-major axis in units of the
disc exponential scale length. In this plot we split the sample into two mass
bins in order to investigate this parameter for a possible dependency \rev{on} the
total stellar mass.  With the purpose of having a comparable number of galaxies
in both bins, this separation in mass is made at $\Mth = 10^{10.25} \msol$.
\rev{In the low mass range a Student's t-test does not indicate a significant difference 
between the distributions. 
For galaxies with masses above $\Mth = 10^{10.25} \msol$, grand-design galaxies
have, on average, longer bars as compared to flocculent galaxies.
Moreover, a comparison of the distributions of the low and high mass range
indicates that the distributions of flocculent and multi-armed galaxies are
similar, whereas more massive grand-design galaxies have longer bars as compared to less massive grand-design galaxies.
The same comparison is made for the bar semi-major axis in units of the
isophotal radius $\riso$ \rev{and} displayed in Fig.
\ref{fig:l-bar_r25_mass}. Considering the distributions in the low and high mass
ranges respectively, our findings corroborate the previous results. Comparing 
the distributions of the low and high mass ranges, however, we now find a significant difference 
for flocculent but not for grand-design galaxies. In fact, more massive flocculents tend to have shorter bars
as compared to less massive flocculents, when normalizing the bar radius by $\riso$.}
However, \rev{the number of flocculents in the high mass range is small.}

\subsection{Discussion}
\label{subsec:general_discussion}
\subsubsection{General Remarks}
\label{subsubsec:general_remarks}
\rev{Figure \ref{fig:mass_surfdens} presents} the distributions of total stellar
mass and stellar mass surface density of the galaxies.  The
stellar masses and surface densities of flocculent galaxies are nearly one
order of magnitude lower than those of multi-armed and grand-design galaxies. 
\rev{In fact, this is not surprising since it is well-known that late-type 
galaxies have lower luminosities. }

\rev{Figure \ref{fig:bulge} examines bulge properties of galaxies separated by arm class. Multi-armed
and grand-design galaxies have higher bulge \sersic indices} (see Fig. \ref{fig:bulge_sersic}). In contrast, all bulge \sersic
indices of flocculent galaxies are lower than $n = 2$.  It is widely accepted
that most classical bulges (arguably built through violent processes such as
mergers of individual galaxies, or of clumps in a protogalaxy) are best
described by a \sersic function with $n > 2$ whereas disc-like bulges (i.e.,
those built through secularly evolving instabilities in the major disc), as
well as boxy-peanut bulges/barlenses, tend to have $n < 2$
\citep[e.g.][]{fisher2008,gadotti2009}. 
\rev{Using a single criterion to distinguish between classical and disc-like bulges
is prone to uncertainties. 
Firstly, there is no clear physical justification to use a \sersic index of 2 to
distinguish between classical and disc-like bulges. Secondly, the error in the
determination of the \sersic index is typically of the order of 0.5 which is large compared to a
threshold value of 2 and the observed range from $\sim 0.5$ to $\sim 6$. Thus, this criterion is
prone to uncertainties but still widely accepted as a good first order
approximation. It is difficult to access the impact of using a single criterion
and this would require a full dedicated study 
\citep[but see discussions in e.g.][]{gadotti2008,gadotti2009,fisher2016,neumann2017}. }
Interpreting our results in terms of the \sersic index, all
flocculents have disc-like bulges \rev{whereas} multi-armed and grand-design galaxies have
mainly classical bulges.  However, only a small fraction ($\sim$ 8\%) of
flocculents has a bulge at all whereas approximately 50\% of multi-armed and
grand-design galaxies are modelled with a bulge component in the
decompositions.  These clear distinctions in the bulge properties indicate
different formation processes of the bulges and therewith evolutionary paths of
the galaxies themselves. 

Figure \ref{fig:bulge_re_r25} shows that the bulges in flocculents are less
compact than bulges in multi-armed or grand-design galaxies.  Furthermore, the bulge-to-total luminosity ratio \rev{is investigated in Fig.
\ref{fig:bulge-total}}. This ratio increases from flocculent to grand-design
galaxies.  All these differences in the bulges of flocculent galaxies as
compared to bulges in multi-armed and grand-design galaxies suggest that bulges
in flocculent galaxies are predominantly disc-like bulges. Conversely,
multi-armed and grand-design galaxies appear to have a higher fraction of
classical bulges. The nucleus-to-total ratio also
shows interesting characteristics.  Flocculents have significantly lower
nucleus-to-total ratios \rev{compared to} the distributions of multi-armed and grand-design
galaxies, \rev{which} are similar. In addition, the fraction of decompositions that require
a nuclear point source component is twice as high for multi-armed and
grand-design galaxies as for flocculents.  This is another indication that the
formation scenario of the central regions of flocculent galaxies differ from
those of galaxies \rev{with} the other spiral arm classes. 

\rev{In Fig. \ref{fig:bulge_frequency} we investigate the occurrence of bulges 
as function of the total
stellar mass. The} results indicate that
bulges do not exist in galaxies with total stellar masses below $\sim 10^{8.8}
\msol$ for flocculents and $\sim 10^{9.2} \msol$ for multi-armed and
grand-design galaxies. The fraction of galaxies with bulges increases with mass
for all arm classes. Flocculents reach a maximum bulge fraction of $\sim 60\%$
around $10^{10.4} \msol$ whereas the fraction of galaxies with bulges in
multi-armed and grand-design galaxies seems to converge towards 100\% as the
stellar mass increases. 

\rev{The properties of bars separated by arm class are examined in Fig.
\ref{fig:bar}, which shows the distributions of bar axial ratio,
semi-major axis, bar-to-total ratio and \sersic index. 
The three arm classes show similar distributions of bar axial ratio and semi-major axis, but less so
of \sersic index. } 
Flocculents tend to show a bar \sersic index that is close to unity,
implying that the radial profile of their bars is (nearly) exponential.  This
is in contrast to the flat bars of multi-armed and grand-design galaxies. This
result is in agreement with previous studies on the radial profiles of bars
\citep[e.g.][]{ee1985} and of simulated bars \citep{lia2002}.  However, it is
remarkable that the fraction of galaxies that require a bar component in the
decompositions is similar regardless of the spiral arm class.

Previous findings from \citet{ee1985} and \citet{regan1997} indicate that
early-type galaxies tend to have longer bars relative to the disc scale length.
We check these results by plotting the bar semi-major axis in units of the disc
exponential scale length (see Fig. \ref{fig:l-bar_hr-disk_mass}) \rev{and} investigate this parameter for a possible dependency of the total stellar
mass of the galaxy. 
Considering masses above $10^{10.25} \msol$, grand-design galaxies
have significantly longer bars compared to flocculent galaxies. This result does not hold true for the low mass range. 
In Fig. \ref{fig:l-bar_r25_mass} the bar semi-major axis \rev{is plotted} in units of the
isophotal radius $\riso$, to explore a different normalization parameter for
the bar size. This plot \rev{roughly} supports our findings with the disc scale length as
normalization parameter.

\rev{
Thus, grand-design galaxies have significantly brighter (see Fig. 
\ref{subfig:bar-total}) and, in the high mass range, longer bars.
Moreover, considering the bar length relative to the disc scale length, 
flocculent and multi-armed galaxies have similar bar sizes for both low and 
high stellar masses (see Fig. \ref{fig:l-bar_hr-disk_mass}). 
Theoretical results indicate that bars grow longer and stronger as galaxies
evolve \citep[see][for a review]{lia2013_book}, which is consistent with the
observational results in \citet{kim2016}.  This could indicate that massive
grand-design galaxies are older or evolved faster than their low-mass
counterparts \citep[see also][]{elmegreen2007}.  This effect could also be
connected to differences in the gas content.  Grand-design spirals are mainly
early-types (see Fig. \ref{fig:tdist_class}) and thus have on average less gas.
Indeed, \citet*{lia2013} showed that an increased gas fraction leads to weaker
and shorter bars and discussed this both in terms of the angular momentum
exchange within the galaxy and of an increased central concentration.   
Note also that this is consistent with the more prominent appearance of bars in
grand-design spirals \citep{ee1985,ee2011} and with the theoretical results
that bars can drive spirals \citep[e.g.][]{lia1980}. We emphasize that we can
corroborate these studies only for galaxies with masses above $10^{10.25}
\msol$. 
}

Taking into account the bar length in units of the isophotal radius, 
different results \rev{are found}. Firstly, the relative bar length of grand-design galaxies
does not depend on their mass. Secondly, a Student's t-test indicates that
massive flocculents have shorter bars. Since this plot contains only 17 high
mass but 295 low mass flocculents, the reliability of this result \rev{is questionable}. 
However, the result may be significant, since the
low-density discs of massive flocculent galaxies presumably cannot form a
density wave or a strong bar. 

\citet{gadotti2011} determines the normalized bar length of galaxies with
masses above $10^{10} \msol$ and $b/a \ge 0.9$ based on SDSS data.  He finds a
median bar length of 1.5 disc scale lengths which is larger than $0.92 \pm
0.50$ disc scale lengths measured in this study.  However, he states that his
decompositions probably miss small bars with sizes around \num{2} to
\SI{3}{\kilo\parsec} due to the spatial resolution of the observations. This
could explain the higher values of his measurement.  In addition, he finds that
no bar is longer than $\sim 3$ disc scale lengths and nearly all bars are
shorter than $R_{24}$ (the isophotal radius at which the surface brightness in
the r-band is $24 \mathrm{\, mag \, arcsec^{-2}}$).  Our results corroborate
his findings. 

\subsubsection{The sequence of spiral arm classes}
\label{subsubsec:multi-armed}
Previous studies on the different spiral arm classes indicate a clear
difference between flocculent and grand-design galaxies. This distinction
arises not only in the visible structure of the galaxies, but also, as the
results above suggest, in the physical processes that drive their formation.
Since multi-armed galaxies show regions with both regular and irregular
morphologies, they are believed to be an intermediate case
\citep{elmegreen1984,elmegreen1995}.  In the following we discuss the position
of multi-armed galaxies as an intermediate case between flocculent and
grand-design galaxies. 

Figure \ref{fig:tdist_class} shows clearly that flocculents, multi-armed and
grand-design galaxies form a sequence from late to early-type galaxies.
However, \rev{there are also} several similarities between multi-armed and grand-design
galaxies.  
\rev{For multi-armed and grand-design galaxies, the distribution of the total
stellar mass as well as the stellar mass surface density is similar (see Figs.
\ref{fig:mass_surfdens}). For flocculents, however, the medians in the distribution of these
properties are nearly one order of magnitude smaller.}
Considering the \sersic indices of bulge and bar (see Fig.
\ref{subfig:bar_sersic}), striking similarities between
multi-armed and grand-design galaxies \rev{are obvious}. Our results suggest that all
flocculent galaxies have disc-like bulges and exponential bars whereas
multi-armed and grand-design galaxies have mostly classical bulges and bars
with flat luminosity profiles.  Bars with flat profiles could be an indication
that these bars are dynamically evolved \citep[see e.g.][]{kim2015}.  Moreover,
Fig. \ref{fig:bulge-total} indicates that multi-armed galaxies have a
significantly lower bulge-to-total ratio as compared to grand-design galaxies.
This exhibits one striking difference between the two spiral arm classes and is
discussed below. 

\subsubsection{The connection between dense bulges and spiral waves}
\label{subsubsec:spiral_waves}
In Fig. \ref{fig:bulge} we investigated the effect of bulge properties
separately for the three arm classes and found indications that galaxies with
grand-design spirals tend to have classical bulges more often than flocculent
spirals.  A first explanation, put forward by some of the authors of this
paper, is based on the theory of spiral wave modes. 
The spiral arms in galaxies with high symmetry may be caused by spiral density
waves as initially suggested by \citet{lindblad1959} and advanced by
\citet{linshu1964}.  However, \citet{toomre1969} showed that these spiral waves
have an inward group velocity so that they would wrap up. 
As solution to this problem, \citet{lin1970} and \citet{mark1976_2} suggest
that incoming waves are reflected off the central regions of the galaxies.
\rev{This requires a region in the centre with a high Toomre Q parameter}, as provided by a classical bulge,
and produces a weak leading wave moving outwards.  Thereupon this leading wave
is amplified at corotation similar to the swing amplification theory proposed
by \citet{toomre1981}. Thus, a strong trailing wave is produced with inwards as
well as outwards moving components.  This feedback loop results in a growing
standing spiral wave mode
\citep[see][]{mark1976_1,mark1976_3,mark1977,bertin1983,lin-bertin1985,bertin1989_1,bertin1989_2}.
We emphasize that a necessary condition for this process is the existence of a
high-Q region in the centre which reflects the incoming wave before it reaches
the Inner Lindblad Resonance (ILR) and is absorbed. 

This theory is also consistent with recent work of \citet{saha2016} who
simulated galaxies in a sequence of increasing bulge masses. For intermediate
bulge masses they find optimal conditions for a shielding of the ILR by the
bulge. Using this configuration, a strong and persistent two-armed \rev{wave
mode}
arises. 

As we showed in Fig. \ref{fig:bulge}, flocculent galaxies tend to have either
weak or no bulges, extended bulges with low densities, or bulges with an
exponential, disc-like profile, which are therefore not likely high-Q.
Therefore the inner regions of flocculent galaxies \rev{are not expected} to be able
to reflect an incoming spiral wave and prevent it from being absorbed at the
ILR. 
In contrast, most of the grand-design galaxies have dense bulges, many with a
high bulge-to-total ratio so that these bulges are more suitable to reflect the
incoming wave. 
In Sect. \ref{subsubsec:multi-armed} the many similarities of
fundamental parameters between multi-armed and grand-design galaxies \rev{are discussed}.
Considering the significantly lower bulge-to-total ratio of multi-armed
galaxies, their bulges probably provide less shielding of the ILR so that we
would expect, according to the theory reviewed above, a lower degree of
symmetry in their spiral structure. Therefore it could be possible to reconcile
the many similar fundamental parameters of multi-armed and grand-design
galaxies with their difference in Hubble type. 

Thus, our results are consistent with the theory presented above as well as
with the simulations conducted by \citet{saha2016}.  Using measurements of the
pattern speed of the spiral structure, one could estimate the radius of the ILR
and show if the bulges are indeed large enough to prevent the incoming wave
from reaching the ILR. However, these measurements are beyond the scope of this
paper. 

\begin{figure}
	\includegraphics[width=\hsize]{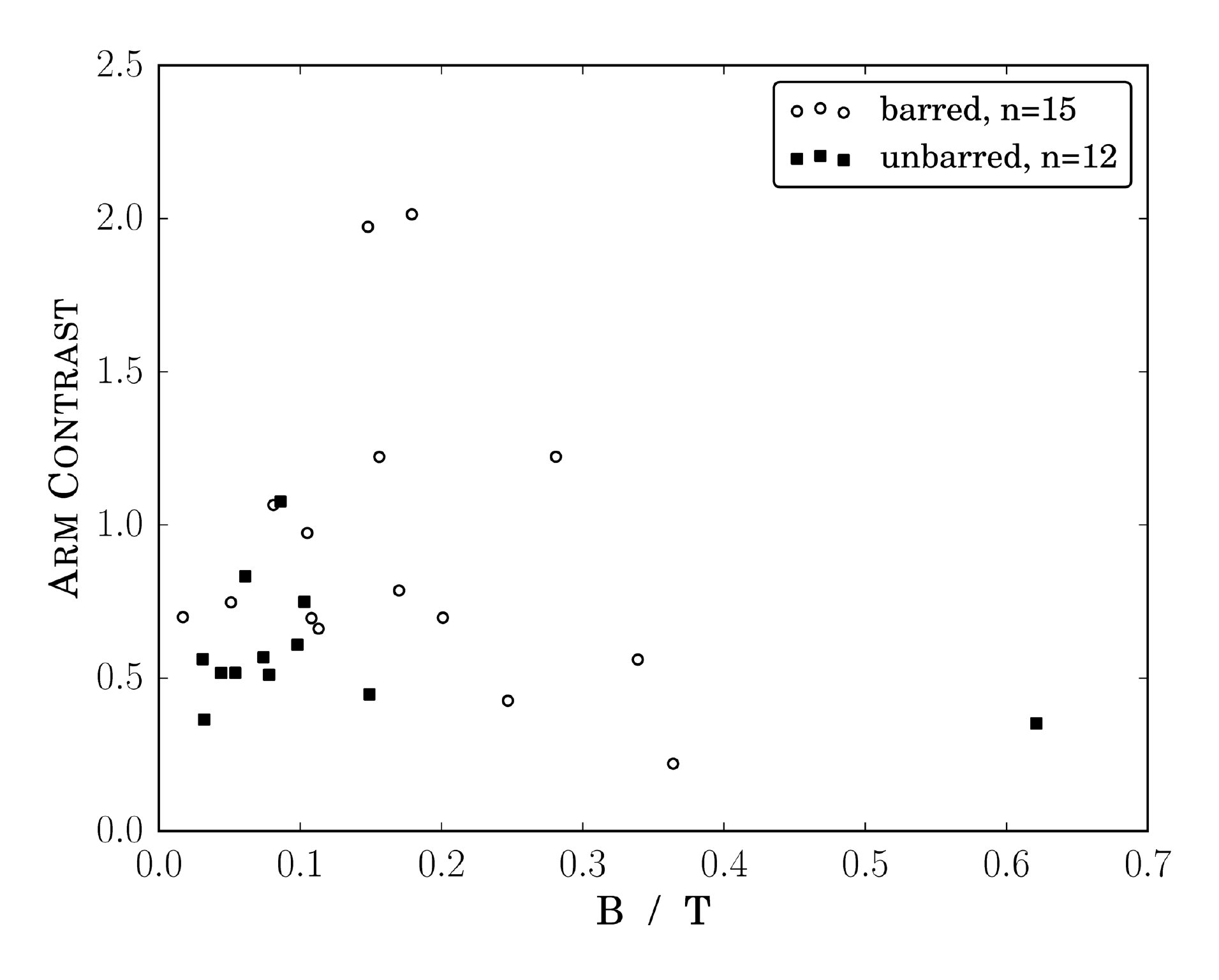}
	\caption{
		Arm contrast as a function of the bulge-to-total luminosity ratio of
		grand-design spirals split for barred and unbarred galaxies. The
		Spearman Rank correlation coefficients are $-0.26$ for barred and
		$-0.03$ for unbarred galaxies. Considering only unbarred galaxies with
		a bulge-to-total ratio lower than $0.1$, the correlation coefficient
		becomes $0.53$. 
	}
	\label{fig:bt-armcontrast_barredunbarred}
\end{figure}

We point out that there are a number of caveats concerning the presented
connection between bulges and persistent spiral wave modes.  Firstly, 25/41 of
the unbarred grand-design galaxies in our sample actually have bulges with a
\sersic index less than \rev{2}, and are thus presumably not classical bulges. This
means that the theory above does not apply for the majority of the unbarred
grand-design galaxies in our sample. Therefore, a reflection of the density
wave off the bulge seems only possible for 16 of the 41 unbarred grand-design
galaxies in our sample. To present a theory that may account for the majority
of the grand-design spiral arms is beyond the scope of this paper.  Secondly,
if a reflection of the density wave on the classical bulge were indeed the
explanation, one could reasonably expect to find a correlation between the
bulge-to-total ratio and the arm-interarm contrast (see Sect.
\ref{sec:contrast_measurements}) for galaxies with grand-design spirals, and
particularly for unbarred galaxies, in order for the reflection to be possible
and efficient.  We do find a trend in this direction in Fig.
\ref{fig:bt-armcontrast_barredunbarred} for bulge-to-total ratios lower than
$\sim 0.1$, but then higher bulge-to-total ratios have a scatter in their
spiral arm contrasts. More measurements are clearly necessary to verify these
results.  We speculate that perhaps galaxies with the highest bulge-to-total
ratios, which are early-type galaxies, also have hotter discs and less gas so
they cannot amplify the reflected spirals very well. This could account for the
drop in the arm-interarm contrast at high bulge-to-total ratios. 
Furthermore,  as will be discussed in Sect. \ref{subsec:results_contrast}, 
no correlation  between arm-interarm contrast and either the bulge \sersic
index or its effective radius \rev{is found}. 

Thus a second, very straightforward explanation is proposed by another subgroup
of the authors of this paper. As we showed in Figs. \ref{fig:tdist_class} and
\ref{fig:mass_distribution}, galaxies with grand-design spirals are in general
more massive and of earlier types than the flocculent galaxies. It is, however,
well known that early type massive spirals have generally strong classical
bulges \citep[e.g.][and references therein]{gadotti2009,salo2015}. Thus
grand-design spirals should be found more often in galaxies with strong
classical bulges, than in galaxies with no such bulges, as we indeed showed in
Fig. \ref{fig:bulge}.  In this very straightforward explanation the link
between the classical bulge and the arm class is not a link of cause to effect.
It is simply due to the fact that both are found preferably in massive early
type disc galaxies, i.e. they are due to the mass of the parent galaxy and
independent of the theory that explains the spiral structure.  Indeed, the mass
of a galaxy is often understood to be a crucial parameter, determining many of
the galaxy properties.

\section{Spiral arm and bar contrast measurements}
\label{sec:contrast_measurements}
In the following we quantify the visual classification of galaxies in the
spiral arm classes. 
\rev{
The strength of spiral arms can be parametrized by the arm-interarm contrast or
by the relative Fourier intensity amplitudes, which give comparable results
\citep{ee2011}. However, in this study we focus on measurements of the
arm-interarm contrast. 
Therefore the intensity in the spiral arms as
well as in the interarm regions is measured and the ratio of these two values computed. The
same procedure is applied to measure the bar-interbar contrast. }

\rev{A detailed description of this measurement is provided below. }
The automated
procedure is applied to a suitable subsample of \ssg\ as described in Sect.
\ref{sec:data_sample}.  Eventually, our pipeline produces images of all
galaxies in polar coordinates. In those images the positions of the spiral arms
and the measurement parameters are highlighted. Furthermore, we provide
contrast profiles as a function of the radius as well as measurements of the
bar and arm contrast. 
These data products and all input and output parameters of the measurement are
publicly available. 

\subsection{Performing the contrast measurements}
\label{subsec:contrast_measurement}
In this study, we make use of the \SI{3.6}{\microns} images provided by
pipeline 4 (P4) \citep{salo2015} of \ssg. The images have a pixel scale of
\ang{;;0.75} and a PSF FWHM of \ang{;;1.7}. The flux in this wavelength arises
mainly from old stars and thus highlights the old stellar component of spiral
galaxies.  Nevertheless, 10 - 30\% of the flux at \SI{3.6}{\microns} still has
its origin in dust emissions \citep{querejeta2015}.  Therefore we also examine
the stellar mass maps provided by pipeline 5 (P5) \citep{querejeta2015} in
which the light of the dust emissions is removed.  A comparison of our
measurements using both the P4 and P5 data indicates that the P4 data produces
\rev{approximately 10\% higher contrasts. This effect might be connected to 
emission from warm dust.}
\rev{No other systematic differences between the two datasets were found.} 
Since the P4-sample is substantially
larger, we chose to conduct the measurements on this dataset. 

In the first step the images are converted into polar coordinates using an IRAF
script of F.R. Chromey (1991). The x-axis of the resulting $r-\theta$ images
shows the azimuthal angle with one pixel referring to one degree whereas the
y-axis displays the radius in linear steps and keeps the pixel scale of
\ang{;;0.75}.  The contrast is measured in dependency of the radius. We use
radial steps of 4 pixels corresponding to approximately twice the FWHM of the
point spread function. 
For every radial step we obtain the intensity as a function of the azimuthal
angle.  In order to reliably detect the spiral arms, \rev{it is necessary} to exclude any
outliers in the intensity curve, e.g. globular clusters or areas with \rev{dust
emission}. To do so, we use two different methods. The first method (V1)
replaces outliers in the intensity curve with the local median whereas the
second method (V2) calculates the median in rectangles with a width of 10
azimuthal degrees and a height of 4 pixels.  Here the width of 10
azimuthal degrees \rev{is chosen} because no major structural changes \rev{are expected} within the
galaxy in this order of magnitude.  A comparison of both versions indicates
that method V2 produces approximately 10\% higher values of the final
contrasts. Apart from this, no systematic differences between the two methods
are found. Due to the better physical justification of method V2, 
\rev{the measurements are conducted with this method. }

The resulting intensity profile as a function of the azimuthal angle is
smoothed using a Savitzky-Golay filter \rev{and} used
to determine the local extrema numerically. Local maxima indicate spiral arms
whereas local minima indicate inter-arm regions.  For the calculation of the
arm-interarm contrast we compute the average of all detected maxima and minima,
respectively. The resulting contrast is converted to a magnitude with the
equation 
\begin{eqnarray}
	C(R)  =  2.5 \times \log \left( \dfrac{I_{\mathrm{max}}}{I_{\mathrm{min}}} \right)
	\label{eqn:contrast}
\end{eqnarray}
as used by \citet{ee2011}.

Based on the data of \citet{salo2015} we distinguish between the bar and arm
contrast.  Here the bar-interbar contrast is measured from the bulge effective
radius $\rbulge$ up to the bar semi-major axis $\rbar$ and the arm-interarm
contrast from the bar radius up to the maximum radius $\rmax$. 

It is essential for the quality of the measurement to exclude the background,
since a measurement of the background does not contain any physical
information.
\rev{  
The background levels were measured in pipeline 3 of \ssg. Two adjacent annuli
were used which surround the galaxy and are split in 45 boxes with 1000 unmasked
pixels each. The median sky level and the local pixel-to-pixel noise was
measured in each box \citep{jc2015}. 
}
Moreover, the small pixel values of the background region would
cause unreasonably high values of the contrast if a measurement of background
against background or galaxy against background would be done.  In order to
derive a reasonable maximum radius $\rmax$ for our arm contrast measurements we
use two different approaches. Firstly, the maximum radius \rev{is chosen} to be a
multiple of the disc exponential scale length. Since this method often
includes a too large background region, a strong bias of highly inclined
galaxies producing high arm contrasts is introduced. 
In the second method the measurement \rev{is stopped} at the radius $\rmax$ at which
10\% of the pixel values are below 3 times the local pixel-to-pixel noise.  We
overplot all galaxy images with the corresponding maximum radius and check them
visually. In the majority of the cases this estimate matches the size of the
galaxy \rev{well}. Nevertheless, for 11 galaxies \rev{it is necessary to} increase the maximum
radius to 2 times the local noise in order to improve the conformity between
maximum radius and galaxy size. 
This method of the determination of the maximum radius has the advantage that it
is related to the size of the galaxy itself, its inclination as well as the
quality of the data. Only a very weak bias towards high inclination galaxies
producing high arm-interarm contrasts \rev{is detected}. Furthermore the
contrast profile \rev{is expected} to be smooth within the galaxy. Therefore the
mean and median of the contrast are expected to be similar. The second method is
in consistency with these prospects. Therefore we conclude that it is more
reasonable to use the latter method for the determination of the maximum radius.
Brighter bars \rev{are also expected} to produce higher bar-interbar contrasts.
In order to check the reliability of the measurement, the bar contrast \rev{is
plotted} as a function of the bar-to-total luminosity ratio. The plot (not shown
here) indicates a clear correlation between both quantities and confirms thereby
the functionality of the code. The Spearman Rank correlation coefficients are
$\rho = 0.29$, $0.60$ and $0.69$ for the flocculent, multi-armed and
grand-design galaxies, respectively. 

\begin{figure*}
	\subfloat
	{
		\includegraphics[width=0.33\hsize]{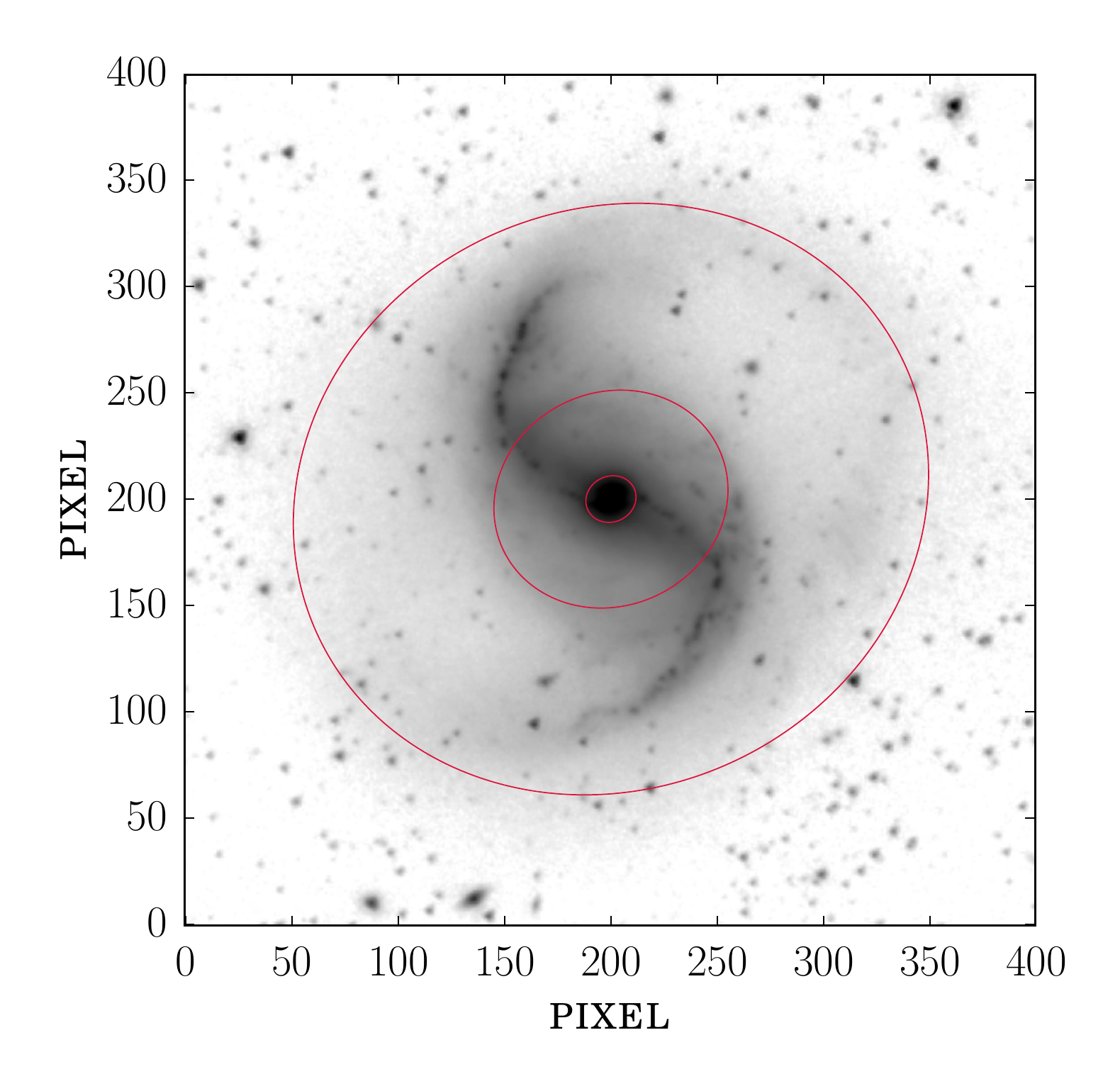}
	}
	\subfloat
	{
		\includegraphics[width=0.33\hsize]{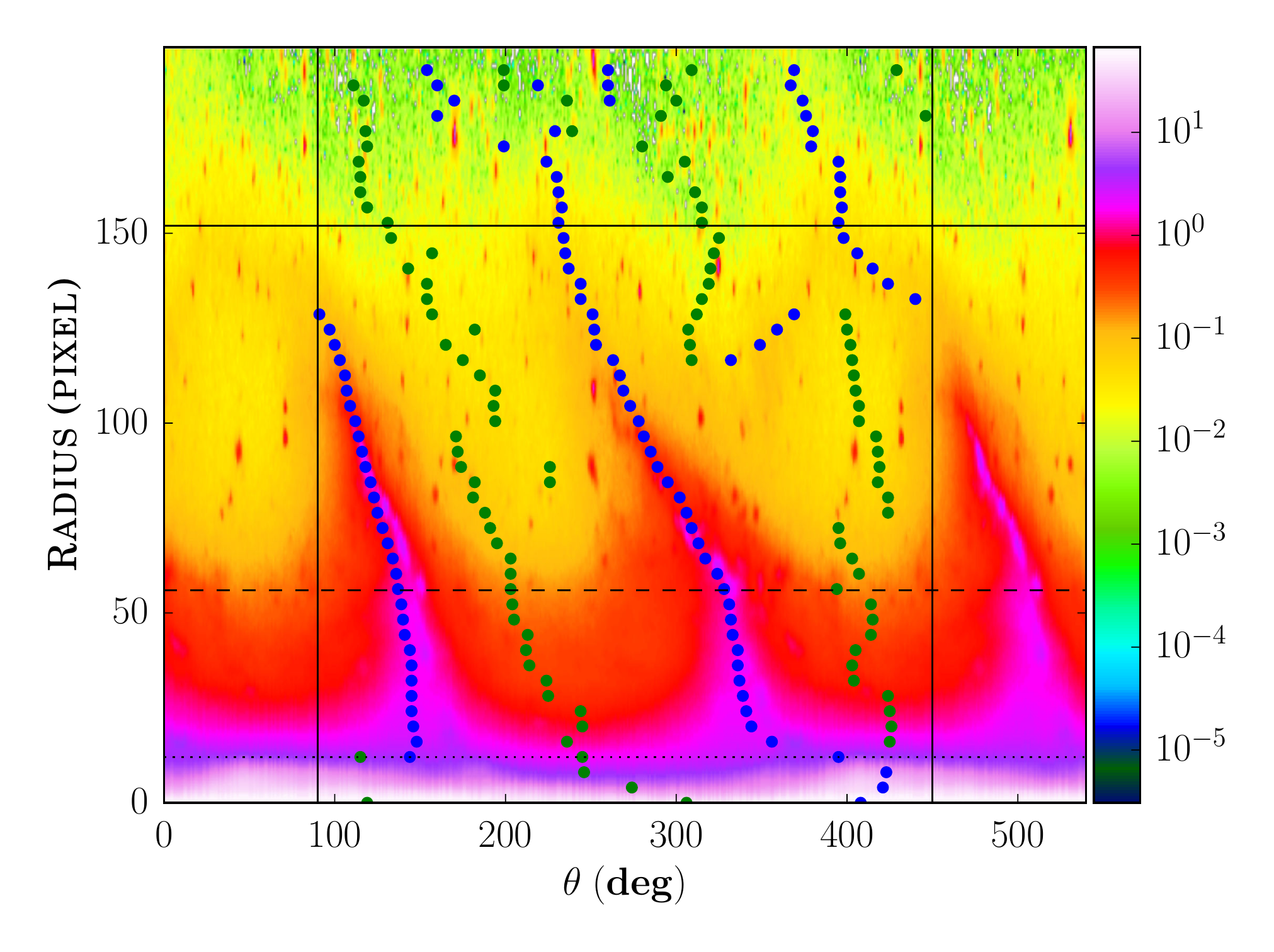}
		\label{subfig:rt_NGC0986}
	}	
	\subfloat
	{
		\includegraphics[width=0.33\hsize]{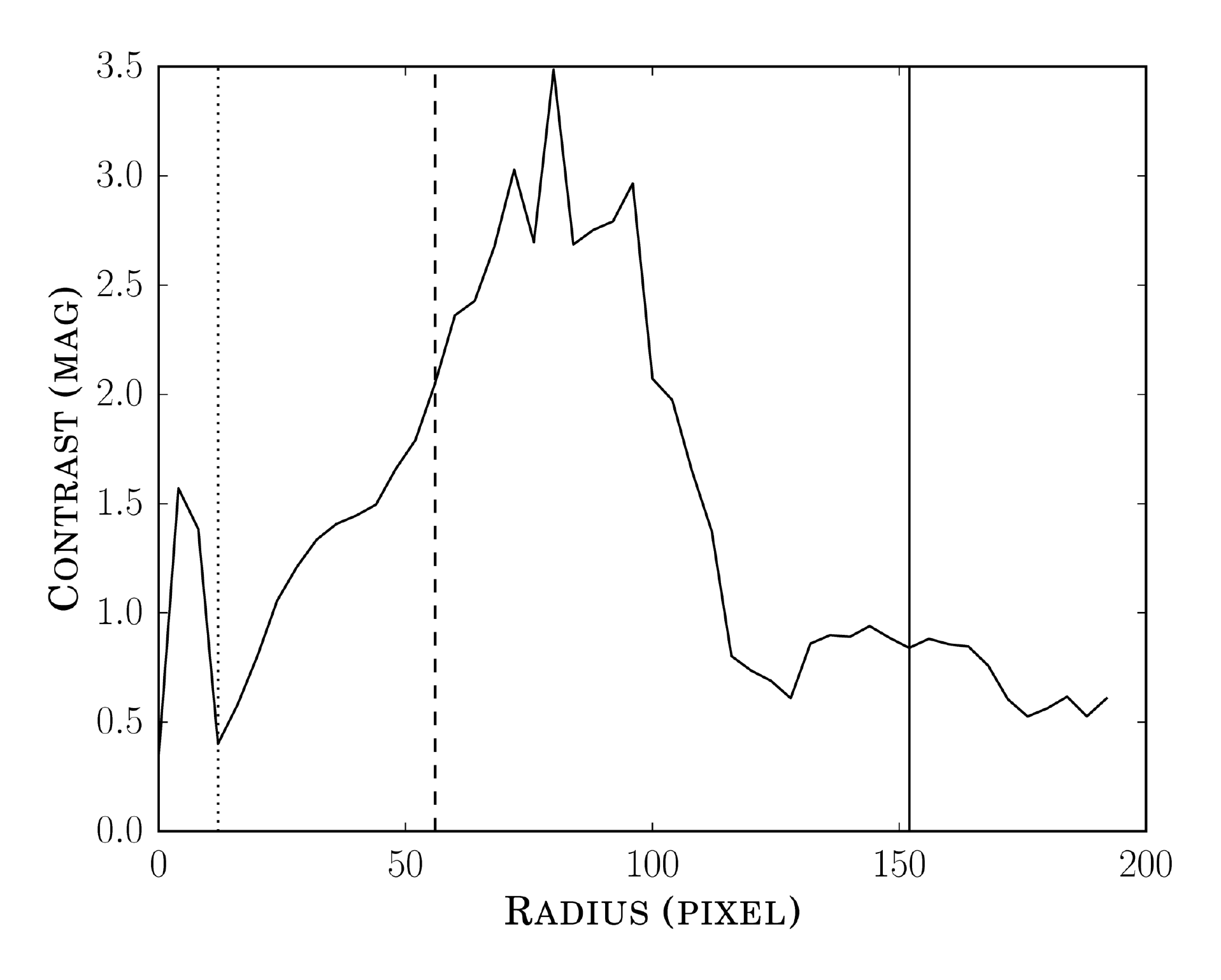}
		\label{subfig:acr_NGC0986}
	}
	\caption{
		Left: Image of the galaxy NGC 986 as processed by \ssg\ pipeline 4.
		Marked is the radius of bulge, bar and maximum radius $\rmax$ of the
		measurement. 
		Centre: The $r-\theta$ image of the same galaxy. The vertical lines
		define a range of $360^{\circ}$ for the azimuthal angle from an
		arbitrary starting point. The dotted and dashed horizontal lines refer
		to the effective radius of the bulge $\rbulge$ and the bar semi-major
		axis $\rbar$, respectively.  The solid horizontal line displays
		$\rmax$.  Blue and green points mark the positions of the maxima and
		minima for each radial step.  
		Right: The contrast as a function of the radius for the same galaxy.
		Vertical lines mark the bulge effective radius, the bar semi-major axis
		and the maximum radius of the measurement.
	}
	\label{fig:measurements}
\end{figure*}

Figure \ref{fig:measurements} shows an $r-\theta$ image of the galaxy NGC 986
and helps to illustrate how the contrast measurement is done. In addition, it
shows the resulting contrast as a function of the radius. In the $r-\theta$
image, the vertical lines define a range of $360^{\circ}$ for the azimuthal
angle from an arbitrary starting point. The dotted and dashed horizontal lines
refer to the effective radius of the bulge $\rbulge$ and the bar semi-major axis
$\rbar$, respectively.  The solid horizontal line displays the maximum radius
$\rmax$ of the measurement. Blue and green points mark the positions of the
maxima and minima, \rev{respectively,} and therewith the positions of the spiral
arms as well as the interarm regions.  

\subsection{Comparison with previous measurements}
\label{subsec:comparison}
\begin{figure}
	\centering
	\includegraphics[width=\hsize]{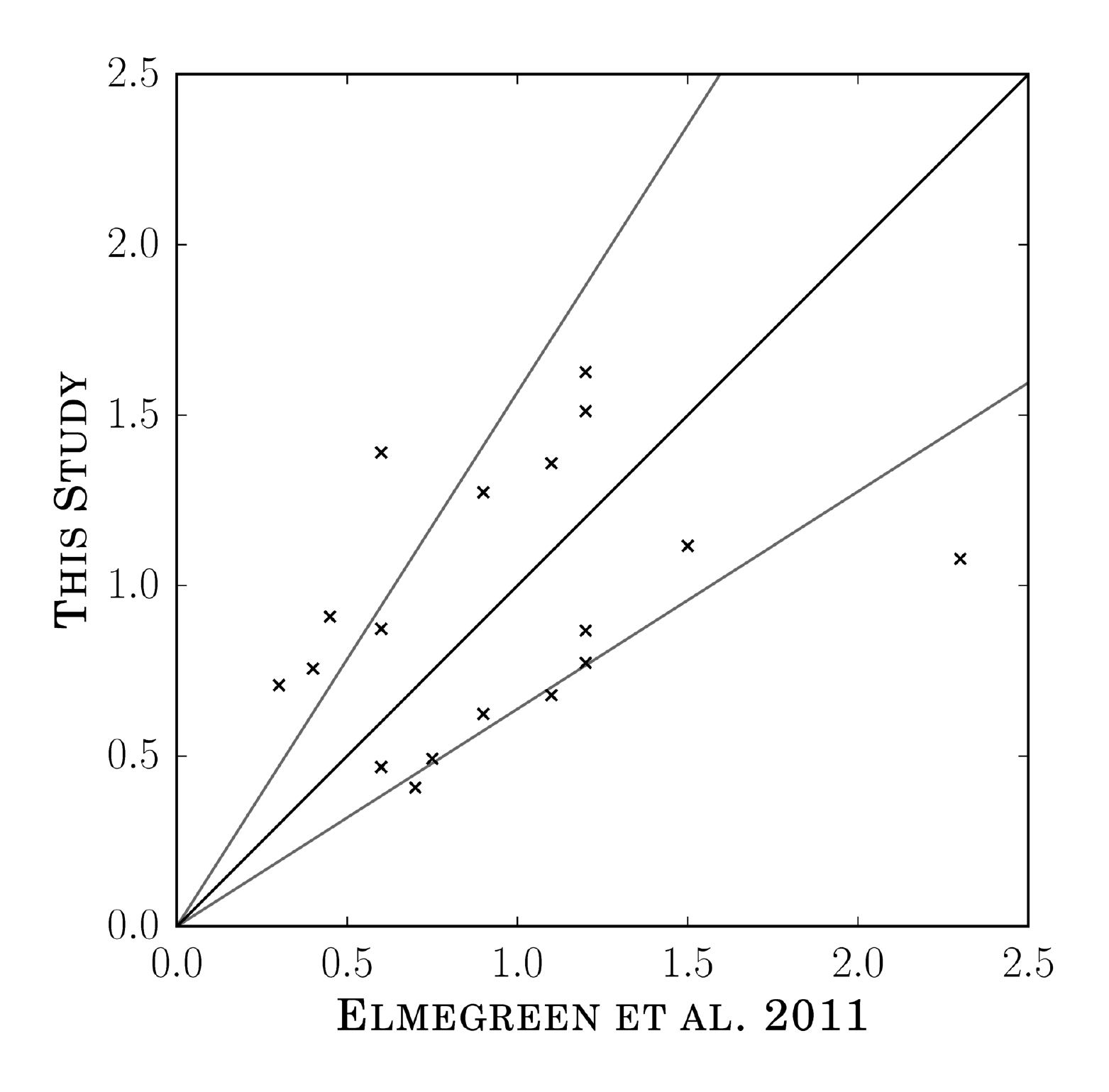}
	\caption{
		Comparison of the arm-interarm contrast as measured in this study and
		\citet{ee2011}. The black line represents a one-to-one correspondence, whereas
		grey line refers to a deviation of 25\%.
	}
	\label{fig:comparison_ee2011_bittner}
\end{figure}

Our results \rev{are compared} to previous measurements of the arm-interarm contrast.
As a reference we use the measurements of \citet{ee2011} who investigated
spiral arm properties of 46 galaxies based on the \SI{3.6}{\microns} images of
\ssg. The two datasets have 18 galaxies in common.  The comparison of these two
studies is displayed in Fig. \ref{fig:comparison_ee2011_bittner}.
The black line represents a one-to-one correspondence whereas the grey lines
refer to a deviation of 25\%.  Since the measurements strongly depend on the
chosen set of maximum and minimum radius, we recalculate for this plot the
arm-interarm contrast using the radii \rev{from \citet{ee2011}, 
to be specific one bulge effective radius and $1.0 \riso$ as minimum and maximum radius,
respectively. }

The plot shows typical deviations of $\sim 25\%$ between the two studies. This discrepancy can easily be explained by
the different measurement methods.  Firstly, \citet{ee2011} measures the contrast in 1 pixel wide cuts in radial steps
of $0.05 \riso$. In contrast, \rev{this study uses radial steps of four pixels
	and computes the median of this four-pixel-wide
cut.} Secondly, \citet{ee2011} determined the maximum and minimum radii individually by hand whereas we connect the
maximum radius to the background noise and the minimum radius to one bulge effective radius (for unbarred galaxies) or
the bar semi-major axis (for barred galaxies), as determined by \citet{salo2015}. Since both studies conclude that the
measurement strongly depends on the chosen radii, this produces differences in the measurements. 
In addition, the contrast profiles \rev{are compared} to those of \citet{ee2011}. For the majority of the galaxies these
contrast profiles are similar in shape and the same features can easily be detected in the profiles of both studies.
The main difference is that the arm-interarm contrasts, as measured by \citet{ee2011}, have on average higher values.
This shift can be explained by some differences in the measurement methods. Since \citet{ee2011} did all measurements by
hand, they also detected and avoided foreground stars manually.  In contrast, we compute the median in bins of 4 pixels
in radius and 10 azimuthal degrees in order to exclude any outliers in the intensity curve.  Furthermore, for every
radial step a Savitzky-Golay filter \rev{is applied} to improve the detection of spiral arms and interarm regions.  This
may slightly wash out any maxima and minima in the intensity profile and therefore lead to systematically lower values
of the contrast. 
Considering the reasonable scatter in the plot, the different sets of maximum and minimum radii as well as the
similarities of the contrast profiles, we conclude that our measurements are in agreement with the results of
\citet{ee2011}.

\subsection{Results}
\label{subsec:results_contrast}
\begin{figure*}
	\subfloat
	{
		\includegraphics[width=0.5\hsize]{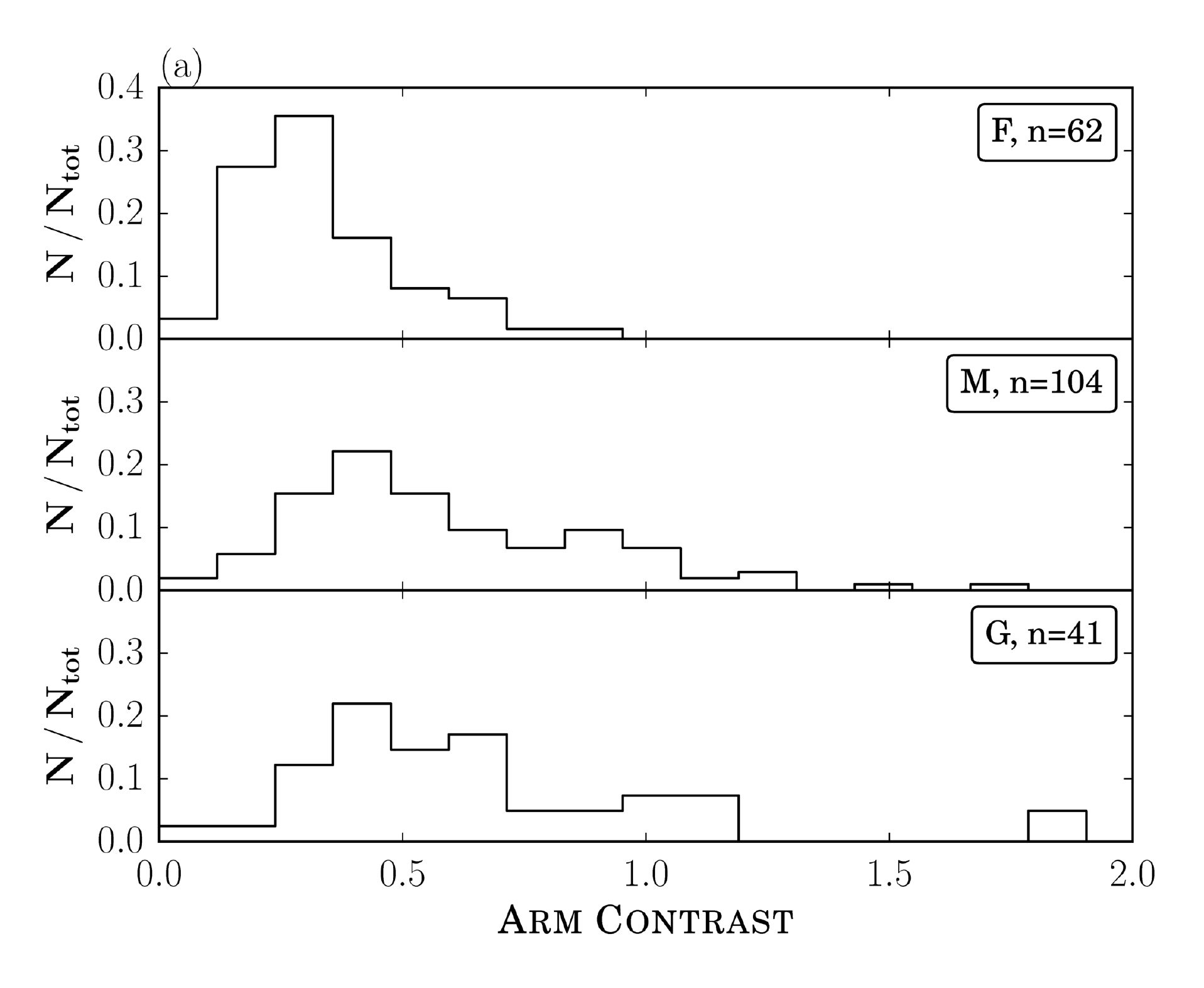}
		\label{subfig:armstrength_dist_regular}
	}
	\subfloat
	{
		\includegraphics[width=0.5\hsize]{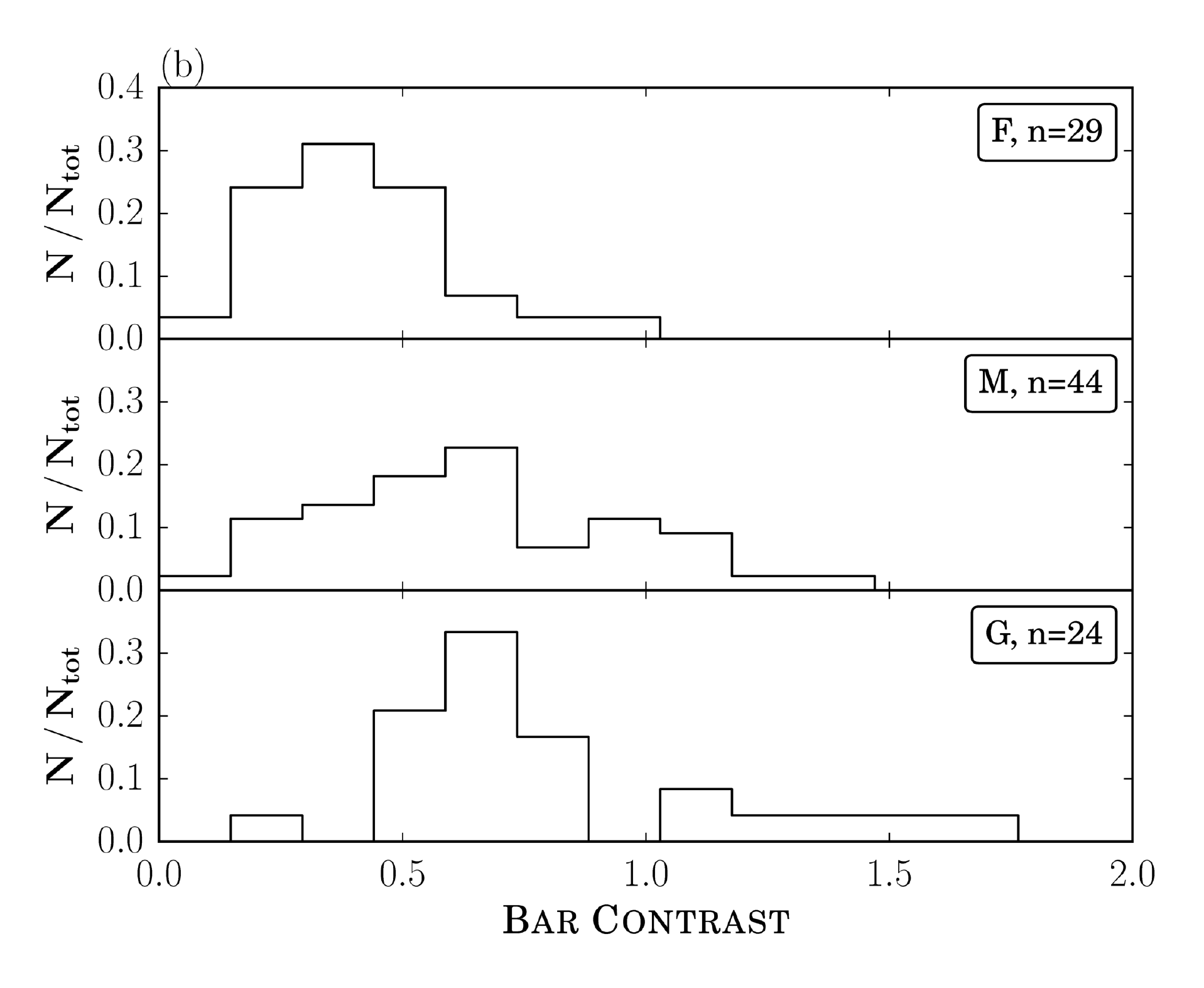}
		\label{subfig:barstrength_dist}
	}
	\caption{
		Distributions of the median of the contrast profile in the range (a)
		$\rbar < R < \rmax$ and (b) $\rbulge < R < \rbar$. The distributions of
		both contrasts increase from flocculent to grand-design galaxies. It is
		remarkable that the increase of the contrast is more striking for the
		bar.
		}
	\label{fig:strength_dist}
\end{figure*}

\begin{table}
	\centering
	\begin{tabular}{lllcl}
\toprule
\toprule
Name												&	Bar 				& Arm Cl.	 	&	T-statistic			&	P-value					\\
\midrule
Arm Cont.											&						& 	F - M	 	& 	$-6.8		$ 		&$ 	1.7\e{-10}$ 	 		\\ 
(Fig. \ref{subfig:armstrength_dist_regular})		&						& 	F - G	 	& 	$-5.2		$ 		&$ 	2.8\e{-6} $		 		\\ 
													&						& 	M - G	 	& 	$-5.8\e{-1}	$ 		&$ 	5.7\e{-1} $		 		\vspace{0.1cm} \\ 

Bar Cont. 											&						& 	F - M	 	& 	$-3.9$	 			& 	$2.4\e{-4}$ 	 		\\ 
(Fig. \ref{subfig:barstrength_dist})				&						& 	F - G	 	& 	$-4.8$	 			& 	$2.6\e{-5}$ 	 		\\ 
													&						& 	M - G	 	& 	$-1.9$	 			& 	$6.6\e{-2}$ 	 		\vspace{0.1cm} \\ 

Arm Cont.	 	 									&	SB					& 	F - M	 	& 	$-5.5$	 			& 	$8.2\e{-7}$ 	 		\\ 
(Fig. \ref{fig:armstrength_dist_bar})   			&   	 				& 	F - G	 	& 	$-5.0$	 			& 	$3.2\e{-5}$ 	 		\\ 
													&	 					& 	M - G	 	& 	$-1.3$	 			& 	$2.0\e{-1}$ 	 		\vspace{0.1cm} \\ 

													&	SA					& 	F - M	 	& 	$-4.1		$ 		& $	9.1\e{-5}$ 	 			\\ 
													&	 					& 	F - G	 	& 	$-2.4		$ 		& $	2.1\e{-2}$ 	 			\\ 
													&	 					& 	M - G	 	& 	$9.2\e{-1}	$ 		& $	3.6\e{-1}$ 	 			\vspace{0.1cm} \\ 

													&	SB				 	& 	F - F	 	& 	$-1.5	$ 			&$ 	1.5\e{-1} $	 			\\ 
													&	\,\,$\updownarrow$	& 	M - M	 	& 	$1.0	$ 			&$ 	3.0\e{-1} $	 			\\ 
													&	SA					& 	G - G	 	& 	$2.5	$ 			&$ 	1.6\e{-2} $	 			\\ 
\bottomrule
	\end{tabular}
	\caption{
		Overview of the results of Student t-tests for all plots presented in
		Sec. \ref{sec:contrast_measurements}. The terms ``SA'' and ``SB'' refer
		to unbarred and barred galaxies, respectively. 
	}
	\label{tab:ttest_contrast_measurement}
\end{table}
\rev{The distributions of the arm and bar contrast separated by arm class are shown 
in Fig. \ref{fig:strength_dist}. A Student's t-test indicates for both the arm
and bar contrast that flocculents have significantly lower contrasts whereas the 
distributions of multi-armed and grand-design galaxies are similar. }
Interestingly, the increase of the contrast from flocculent to grand-design
galaxies is more striking for the bar-interbar contrast.

\begin{figure*}
	\subfloat
	{
		\includegraphics[width=0.5\hsize]{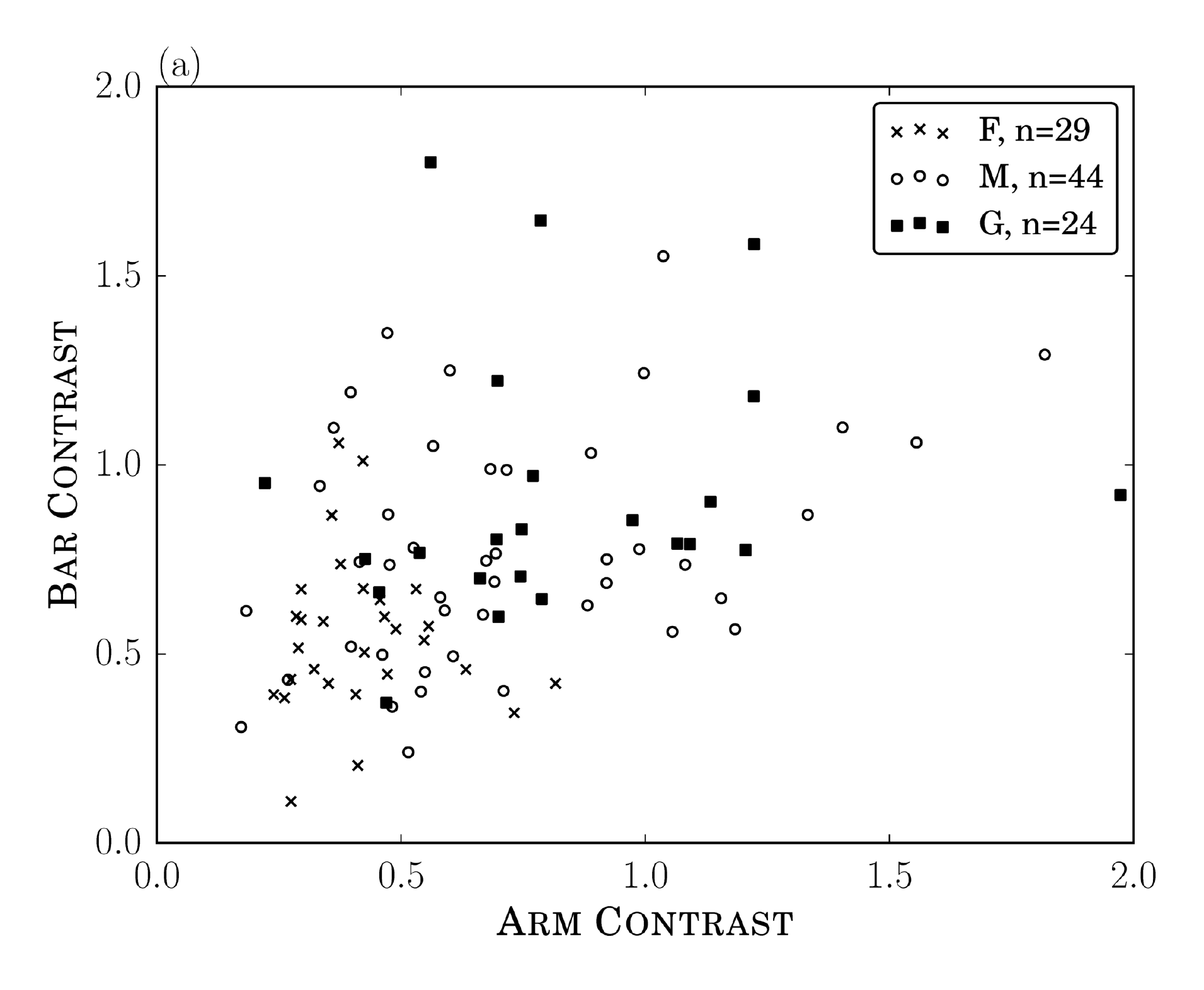}
		\label{fig:arm-bar}
	}
	\subfloat
	{
		\includegraphics[width=0.5\hsize]{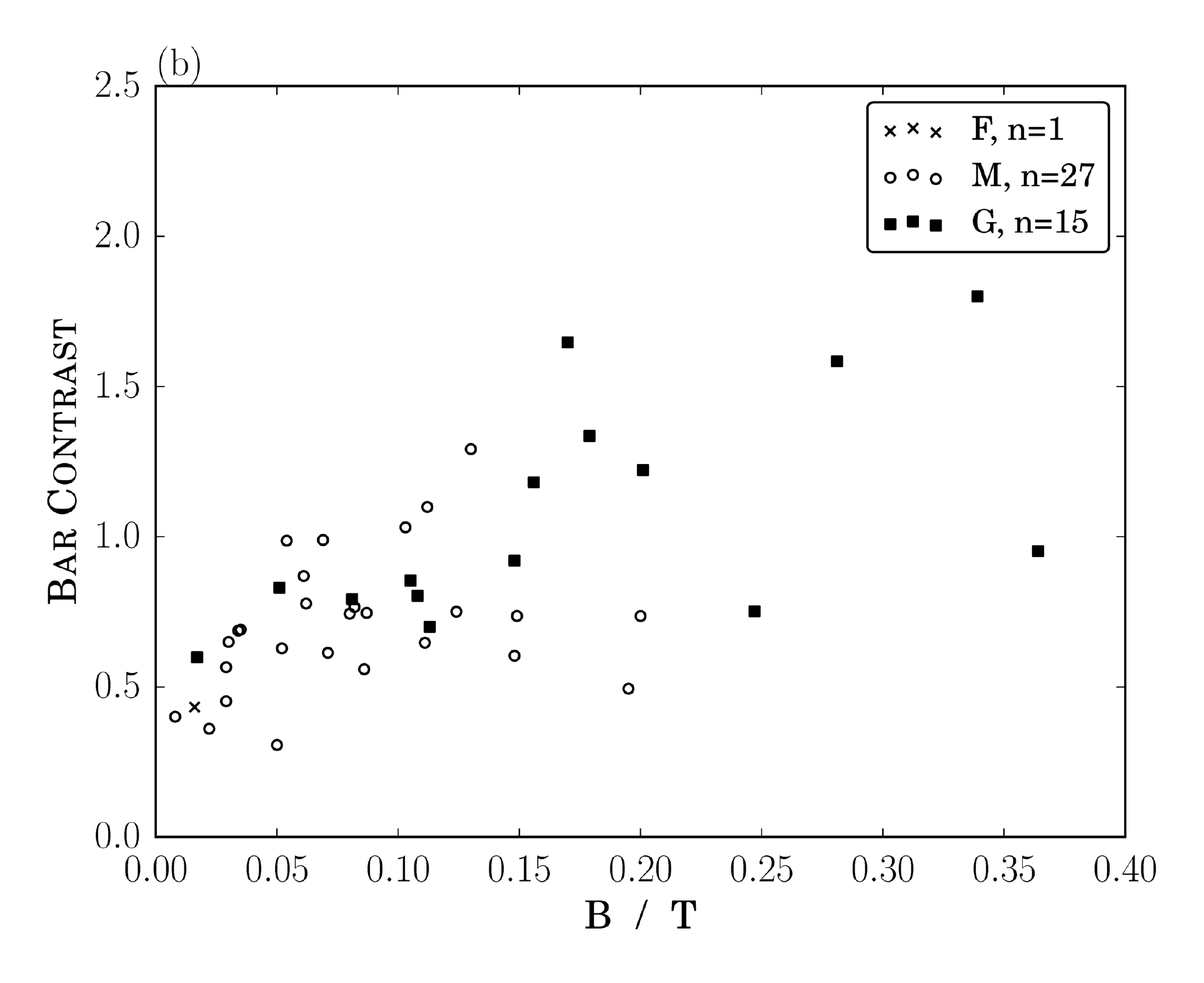}
		\label{fig:frel_B-barstrength}
	}
	\caption{
		(a) Bar contrast as a function of the arm contrast. Galaxies with high
		bar contrasts tend to have higher arm contrasts as well. The Spearman
		Rank correlation coefficients are $\rho_F = 0.10$, $\rho_M = 0.29$ and
		$\rho_G = 0.39$. 
		(b) Bar contrast as a function of the bulge-to-total luminosity ratio.
		Galaxies with high bulge-to-total ratios tend to have high bar
		contrasts, too. The Spearman Rank correlation coefficients are $\rho_M
		= 0.40$ and $\rho_G = 0.65$.  Since this plot includes only one
		flocculent galaxy, \rev{ a correlation coefficient for this
		spiral arm class cannot be computed.}
	}
	\label{fig:arm-bar_frelB-barstrength}
\end{figure*}

\rev{In Fig. \ref{fig:arm-bar}  we examine the bar contrast as a function of the
arm contrast. Galaxies with high bar contrasts tend to have high arm contrasts
as well, but the correlation is quite weak with Spearman Rank correlation
coefficients of $0.10$, $0.29$ and $0.39$ for flocculent, multi-armed and
grand-design galaxies, respectively.}

\begin{figure}
	\centering
	\includegraphics[width=\hsize]{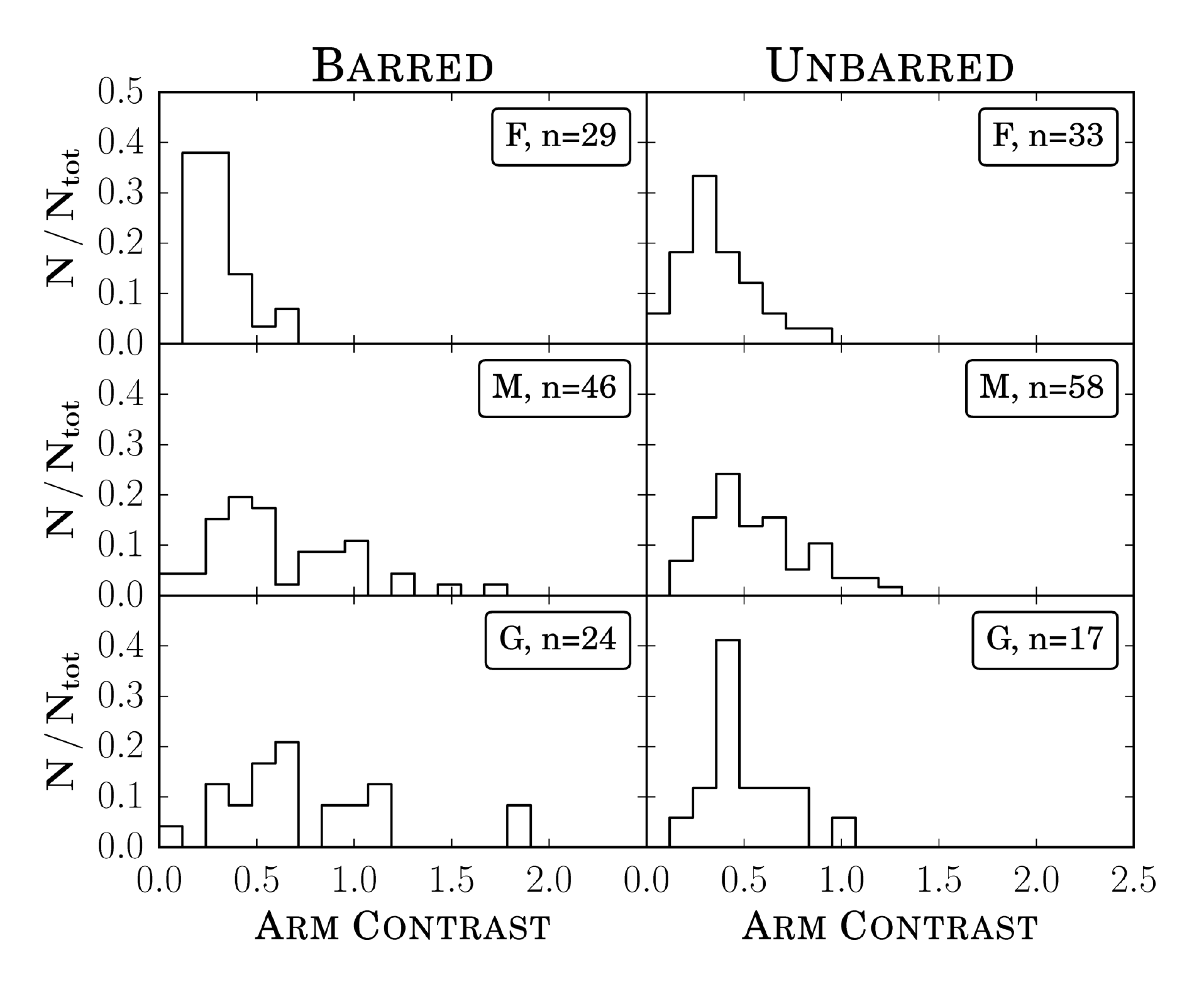}
	\caption{
		Distributions of the arm-interarm contrast for both barred (left
		panels) and unbarred galaxies (right panels).  A Student's t-test
		indicates significantly lower arm contrasts of flocculents for both
		barred and unbarred galaxies. The plot does not show a difference of
		the arm contrasts between barred and unbarred galaxies. 
	}
	\label{fig:armstrength_dist_bar}
\end{figure}

In Fig. \ref{fig:frel_B-barstrength} the bar contrast as a function of
the bulge-to-total luminosity ratio \rev{is presented}. \rev{Galaxies with a high
bulge-to-total ratio tend to have higher bar contrasts with Spearman Rank
correlation coefficients for multi-armed and grand-design galaxies of $\rho
= 0.40$ and $\rho = 0.65$, respectively.} Since this plot includes only one
flocculent galaxy, a correlation coefficient for this spiral
arm class cannot be calculated. This figure contains a substantially lower number of galaxies
because multiple additional constraints apply to this plot (in contrast to e.g.
Figs.  \ref{fig:strength_dist}, \ref{fig:arm-bar} and
\ref{fig:armstrength_dist_bar}).  In order to plot the bar contrast as a
function of the bulge-to-total ratio, \rev{the plot is} limited to galaxies \rev{which}
have contrast measurements and require a bulge as well as a bar component in
the decompositions. 
In line with the correlations between the bar contrast and the bulge-to-total
and bar-to-total ratios, an inverse correlation between the bar
contrast and the disc-to-total luminosity ratio \rev{is found} (not shown here).

\rev{ We also compared arm and bar contrast to other fundamental galaxy properties. }
These included the bulge \sersic index, bulge effective radius, the bar
semi-major axis and axial ratio, the bar-to-total ratio, the disc-to-total
ratio, the stellar mass surface density and nucleus-to-total ratio.  For none
of them a clear connection to the arm or bar contrast \rev{is found}. 

\rev{In Fig. \ref{fig:armstrength_dist_bar} we investigate the distribution of the 
arm-interarm contrast for both barred (left panels) and unbarred galaxies (right panels). }
Flocculent galaxies have
significantly lower arm contrasts for both barred and unbarred galaxies \rev{whereas} the
distributions of multi-armed and grand-design galaxies are similar.  
\rev{In addition, no significant differences between the distributions of barred and
unbarred galaxies are found.}

\subsection{Discussion}
\label{subsec:discussion_contrast}
\rev{The results of the arm and bar contrast measurements are presented in Fig.
\ref{fig:strength_dist} which indicates} that both the arm and bar contrasts of
flocculent galaxies are lower compared to multi-armed and grand-design
galaxies. This is in good agreement with the visual appearance of the different
spiral arm classes.  Nevertheless, we expected to find a more significant
difference between the contrasts of multi-armed and grand-design galaxies,
based on their visual appearance. 
In Sect. \ref{subsubsec:spiral_waves} we \rev{discussed} that the spiral arms of
grand-design galaxies possibly arise through a standing spiral wave mode.
However, the corresponding theory does not make predictions about the amplitude
of the spiral pattern, which implies that the arm-interarm contrast of
grand-design galaxies does not necessarily need to be higher as compared to
multi-armed galaxies.  Instead, since the $m = 2$ mode of the spiral wave is in
this theoretical framework amplified most strongly in grand-design galaxies,
grand-design galaxies \rev{are expected} to show a higher degree of symmetry than
multi-armed galaxies. This is in good agreement with our findings and the
visual appearance of the different spiral arm classes. Nevertheless, we remind
the reader of the caveats discussed in the second half of Sect.
\ref{subsubsec:spiral_waves} with respect to this interpretation \rev{and of the
existence of an alternative, simpler explanation.}

The lack of a strong arm contrast difference between grand-design and
multi-armed galaxies could be related to the \rev{differences in the techniques used to classify} 
galaxies into the spiral arm classes. The \rev{visual} impression of a galaxy is
certainly connected to multiple factors beyond the amplitude of the spiral arms
and the interarm regions.  These aspects could include e.g. the length of the
spiral arms and their symmetry.  However, these factors are not considered in
our measurement.  Therewith they could generate \rev{differences} between the
\rev{visual classification} of the galaxy and our quantitative measurements. 

Furthermore, the plots indicate that the increase of the bar-interbar contrast
from flocculent to grand-design galaxies is more striking compared to the
increase of the arm-interarm contrast.  This finding could indicate that bars
are more important for the secular evolution of the galaxy and the development
of spiral structure than the spiral arms themselves.  Consistent with this idea
is the fact that the bar contrasts are \rev{approximately 0.11} higher than the arm
contrasts.  This explains why linear theories can properly describe spiral arms
but not bars. 
\rev{In order to analyse the relation of bars and spiral arms in more detail, Fig. 
\ref{fig:arm-bar} shows the bar contrast as a function of the arm contrast.}
Multi-armed and grand-design galaxies with high
bar contrasts tend to have high arm contrasts as well. This result is
consistent with previous findings of e.g. \citet{ee1985,ann1987,
elmegreen2007,buta2009,salo2010}. However, this does not hold true for
flocculent galaxies, as indicated by a correlation coefficient of $\rho_F =
0.10$, and indicates that their spiral structure is generated by local
gravitational instabilities instead of being driven by a bar or a companion. 

In connection to this result, the distributions of the arm contrast
for barred and unbarred galaxies \rev{are analysed} separately (see Fig.
\ref{fig:armstrength_dist_bar}).  Significant differences
of the arm contrast \rev{are not obvious,} regardless of the existence of a bar.  This result points
out that spiral arms, whether or not they arise from standing spiral wave \rev{modes},
local gravitational instabilities or are triggered by tidal interactions or
bars, have the capacity to reach similar arm-interarm contrasts.  However, an
existing bar supports spiral arms in the sense that stronger bars trigger
spiral arms with higher amplitudes. 
\rev{This might also be true if the pattern speed of the spiral arms is lower than
that of the bar itself, even if bar and spiral arms break and reconnect
continuously \citep[see][]{tagger1987,sellwood1988}. }

\rev{Furthermore, Fig. \ref{fig:frel_B-barstrength} indicates a clear
correlation between the bulge-to-total luminosity ratio and the bar contrast for
multi-armed and grand-design galaxies.}  It is remarkable that no other
correlations of fundamental galaxy parameters with the bar contrast are obvious.
This could indicate a strong connection between classical bulges and large bars
in early-type disc galaxies.  In Sect.  \ref{subsec:general_discussion} we found
that the bulges of multi-armed and grand-design galaxies are mostly classical
bulges.  Within this framework the classical bulge forms through a merger of
individual galaxies or of clumps in a protogalaxy. After the formation of a disc
and a bar, the bulge supports the growth of the bar through the transfer of
angular momentum from the bar \rev{to} the bulge, as discussed by
\citet{lia2002} and \citet{lia2003}.  Thus, the existence of a classical bulge
supports the \revv{growth} of a strong bar.

\section{disc breaks}
\label{sec:disk_breaks}
An important property of disc galaxies is their radial surface brightness
profile.  Different observations suggest that the majority of all disc galaxies
have double exponential profiles. In this section, we explore the properties of
these disc breaks in galaxies with different arm classes. In addition, we
connect disc breaks to the contrast measurement of Sect.
\ref{sec:contrast_measurements}.  For this analysis only barred galaxies
\rev{are considered}, as the number of unbarred galaxies in the sample
\rev{used} here is small (see Sect. \ref{sec:data_sample}).

\subsection{Results}
\label{subsec:results_diskbreaks}
\begin{figure}
	\centering
	\includegraphics[width=\hsize]{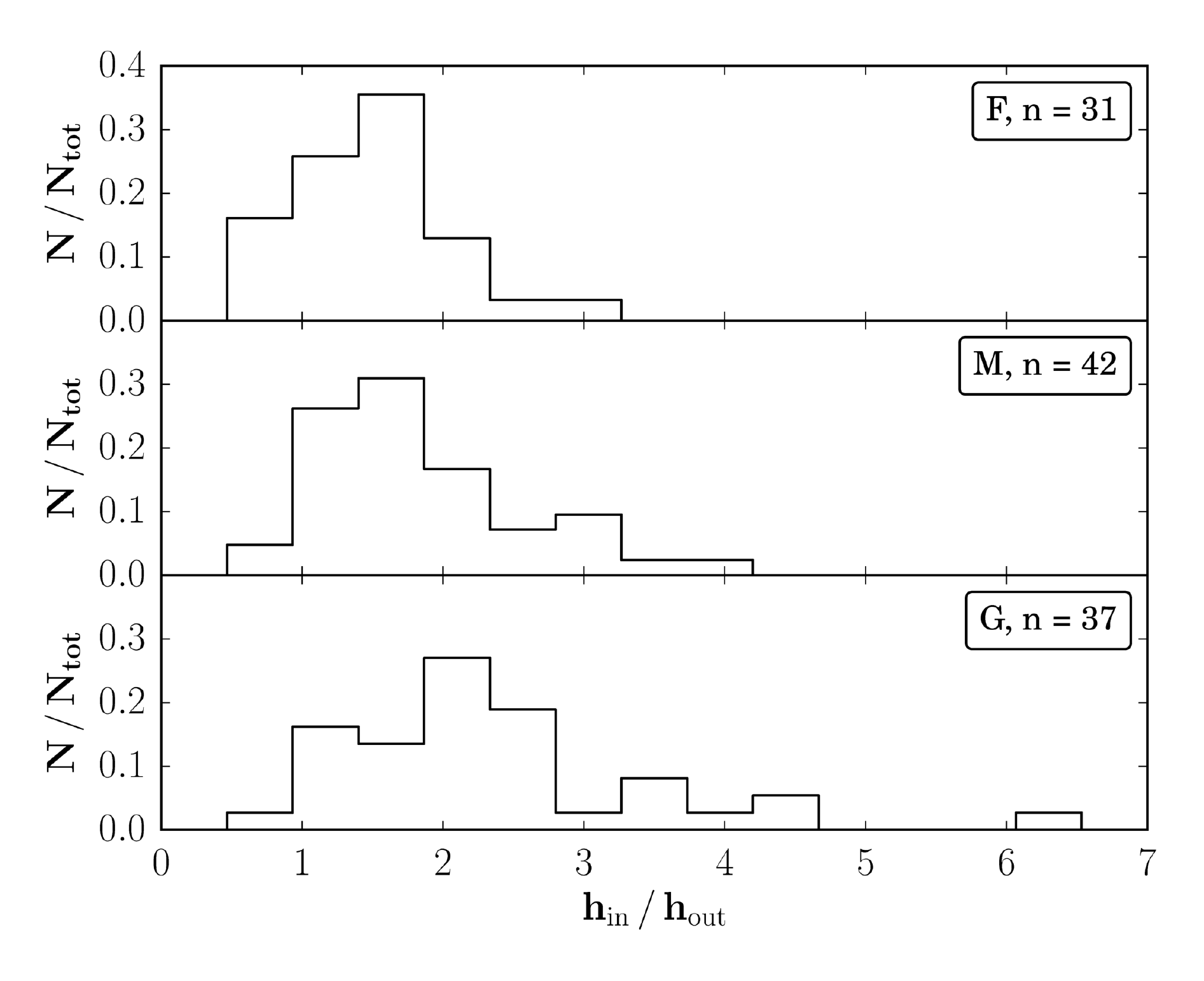}
	\caption{
		Distributions of the ratio of disc inner to outer scale length.
		Grand-design galaxies have higher values compared to flocculents.
		Multi-armed galaxies seem to be an intermediate case. 
	}
	\label{fig:hin_hout}
\end{figure}

\begin{figure*}
	\subfloat
	{
		\includegraphics[width=0.5\hsize]{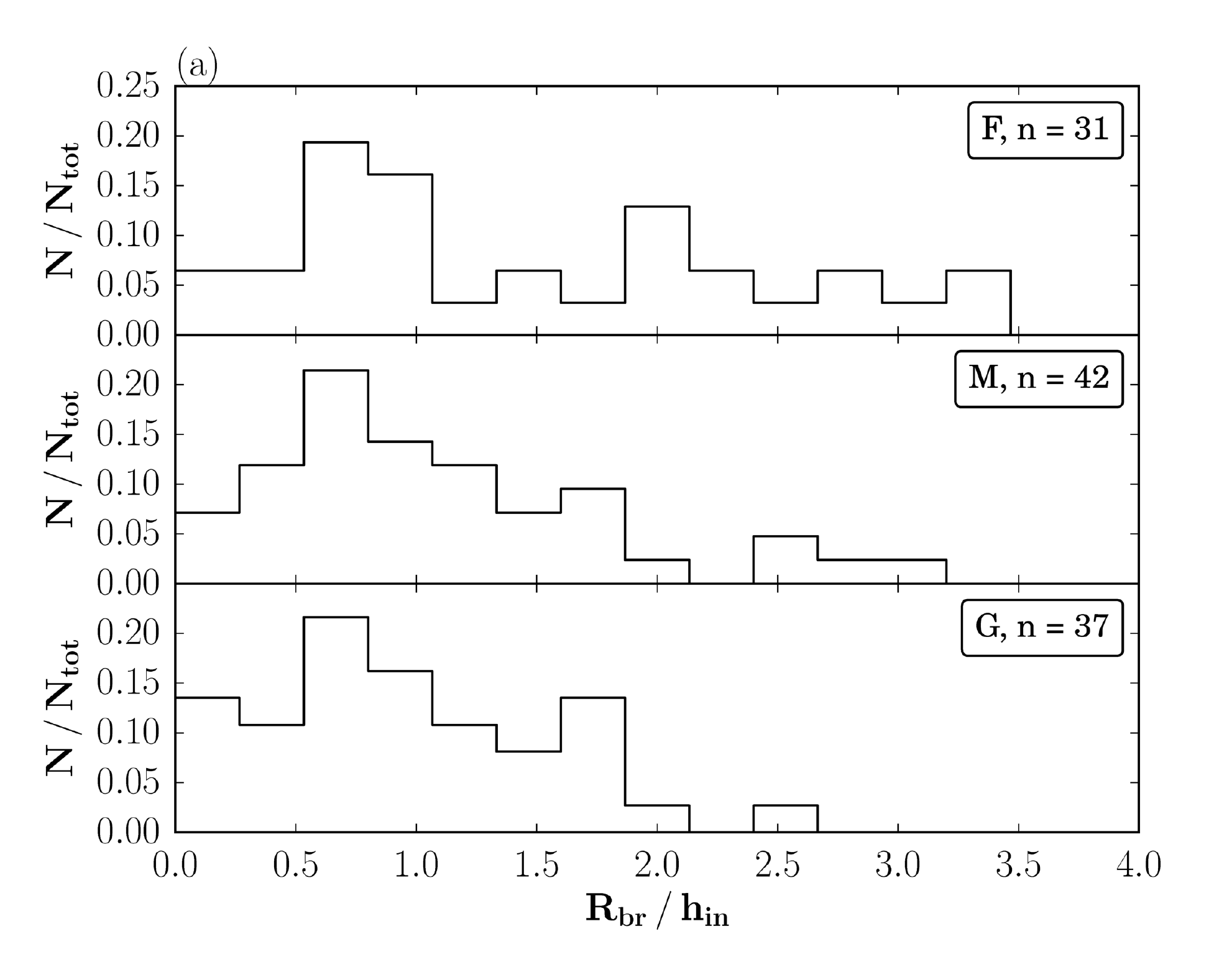}
		\label{subfig:rbr_hin}
	}
	\subfloat
	{
		\includegraphics[width=0.5\hsize]{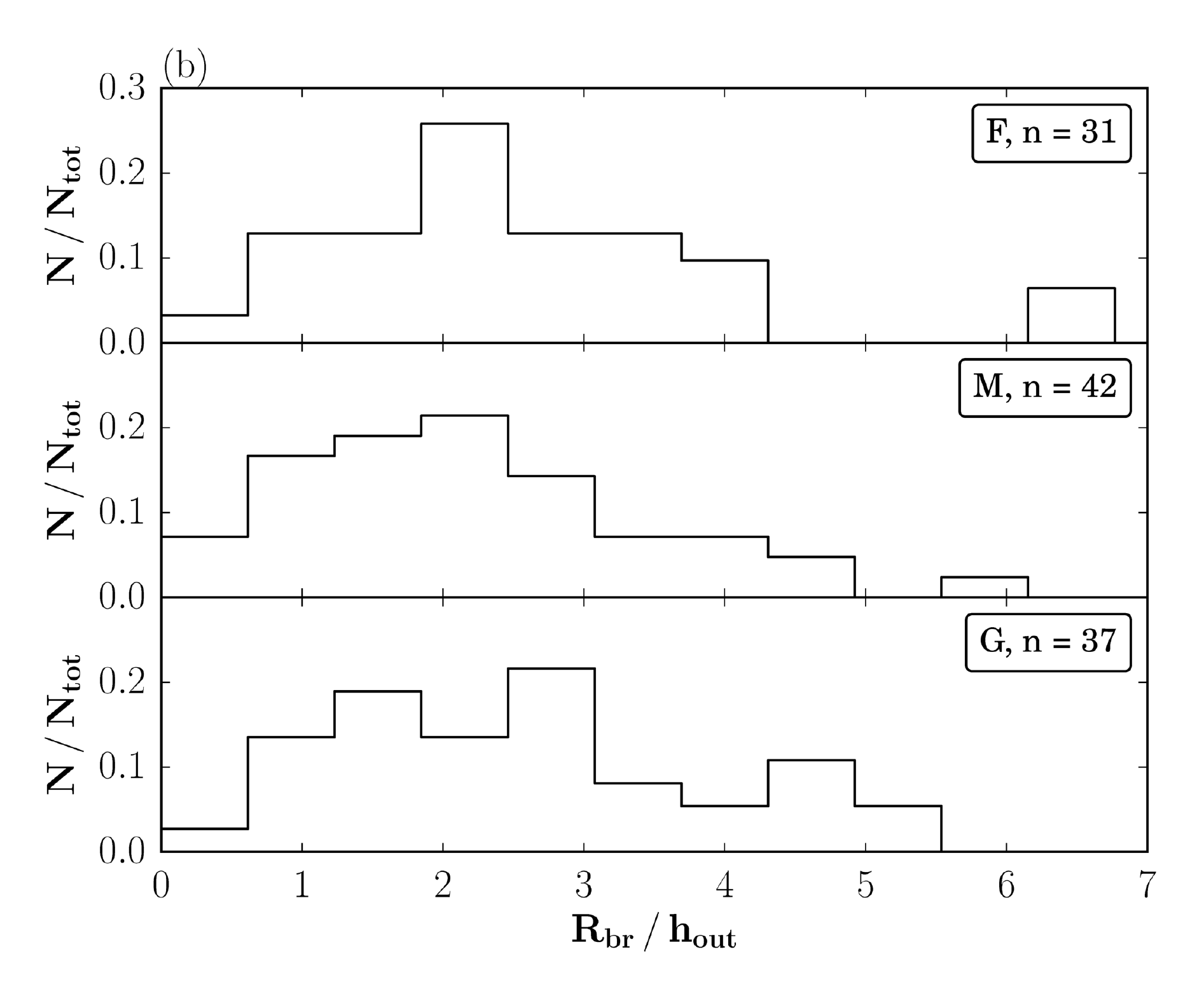}
		\label{subfig:rbr_hout}
	}
	\caption{
		Distributions of the break radius normalized by (a) the disc inner and
		(b) outer scale length. All distributions are similar regardless of the
		spiral arm class. 
	}
\end{figure*}

\begin{table}
	\centering
	\begin{tabular}{llcl}
\toprule
\toprule
Name								& Arm Class		&	T-statistic		&	P-value				\\
\midrule
$\hin / \hout$						& 	F - M	 	& 	$-2.3     $	 	& 	$2.3\e{-2}$	 		\\ 
(Fig. \ref{fig:hin_hout})           & 	F - G	 	& 	$-4.1     $	 	& 	$1.6\e{-4}$	 		\\ 
									& 	M - G	 	& 	$-2.4     $	 	& 	$2.1\e{-2}$	 		\vspace{0.1cm}\\ 

$\rbr / \hin$						& 	F - M	 	& 	$2.0     $   	& 	$4.9\e{-2}$	 		\\ 
(Fig. \ref{subfig:rbr_hin})   		& 	F - G	 	& 	$2.7     $   	& 	$1.1\e{-2}$	 		\\ 
                        			& 	M - G	 	& 	$7.2\e{-1 }$	& 	$4.8\e{-1}$	 		\vspace{0.1cm}\\ 

$\rbr / \hout$						& 	F - M	 	& 	$1.2	$   	& 	$2.4\e{-1}$	 		\\ 
(Fig. \ref{subfig:rbr_hout})   		& 	F - G	 	& 	$1.1\e{-1}$		& 	$9.1\e{-1}$	 		\\ 
                          			& 	M - G	 	& 	$-1.2     $		& 	$2.3\e{-1}$	 		\\ 
\bottomrule
	\end{tabular}
	\caption{
		Overview of the results of Student t-tests for all plots presented
		in Sec. \ref{sec:disk_breaks} 
	}
	\label{tab:ttest_disk_breaks}
\end{table}
In the first step, the distributions of the disc inner $\hin$ and outer scale
length $\hout$ for the three different arm classes \rev{are investigated} (not
shown here).  The disc inner scale length increases from flocculent to
grand-design galaxies.  Considering the disc outer scale length, multi-armed and
grand-design galaxies have similar distributions whereas flocculent galaxies
have lower disc outer scale lengths.  
In addition, the ratio of disc inner to outer scale length \rev{is used} as an
indicator for the strength of the break and this break strength is, on average,
higher for grand-design galaxies than for flocculents. Multi-armed galaxies are
an intermediate case (see Fig. \ref{fig:hin_hout}). 
Figures \ref{subfig:rbr_hin} and \ref{subfig:rbr_hout} present the break radius
$\rbr$ normalized by the disc inner and outer scale length.  \rev{There are no}
differences of the distributions of the spiral arm classes regardless of the
used normalization parameter. 

\begin{figure}
	\centering
	\includegraphics[width=\hsize]{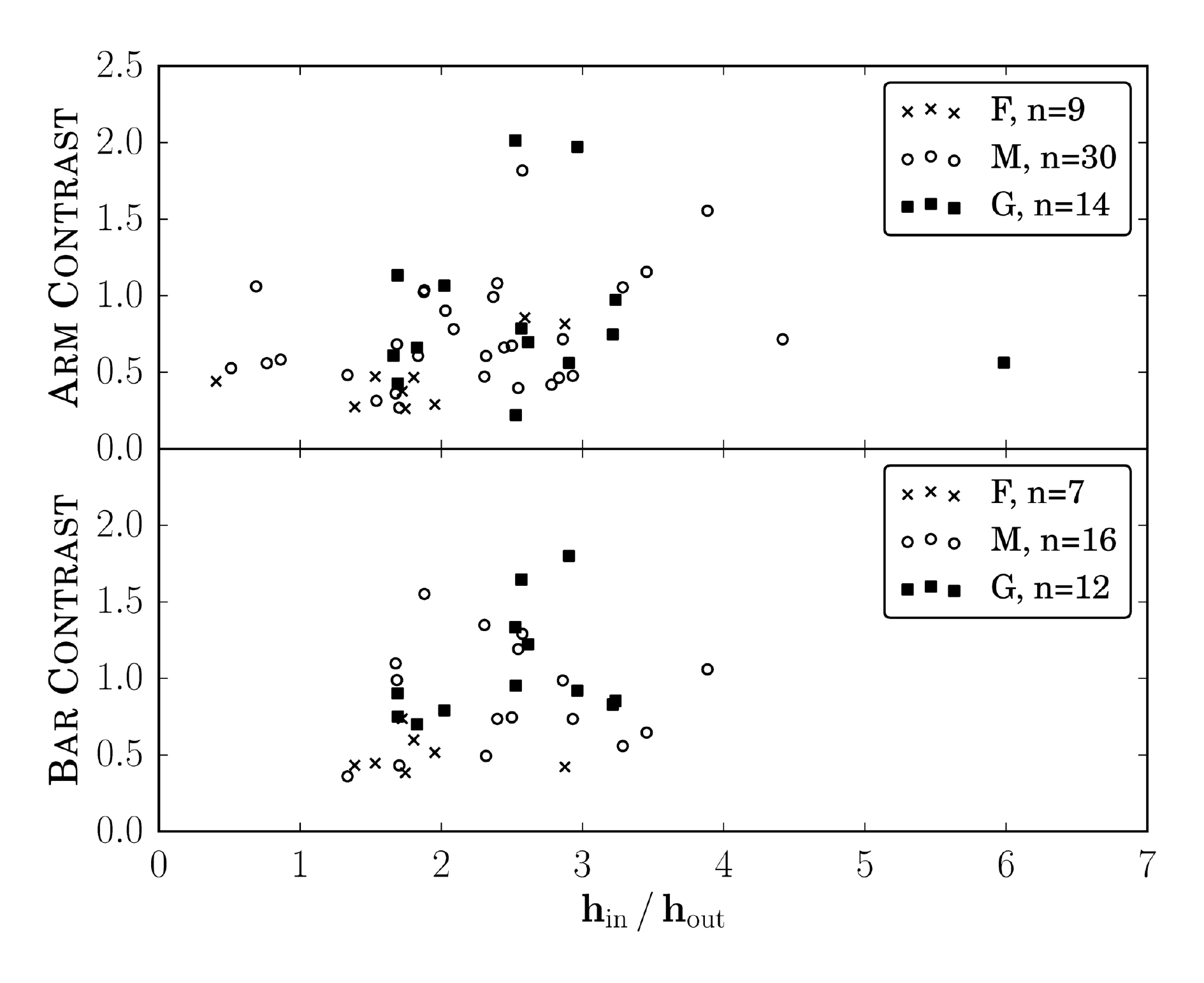}
	\caption{
		Bar and arm contrast as a function of the break strength. 
		\rev{A correlation with the arm contrast is not obvious.}
		However, grand-design
		galaxies with a strong break tend to have higher bar-interbar
		contrasts.  The correlations coefficients are $\rho_F = -0.07$, $\rho_M
		= 9 \e{-3}$ and $\rho_G = 0.33$, respectively.
	}
	\label{fig:hin_hout_asbs}
\end{figure}

\begin{figure*}
	\subfloat
	{
		\includegraphics[width=0.5\hsize]{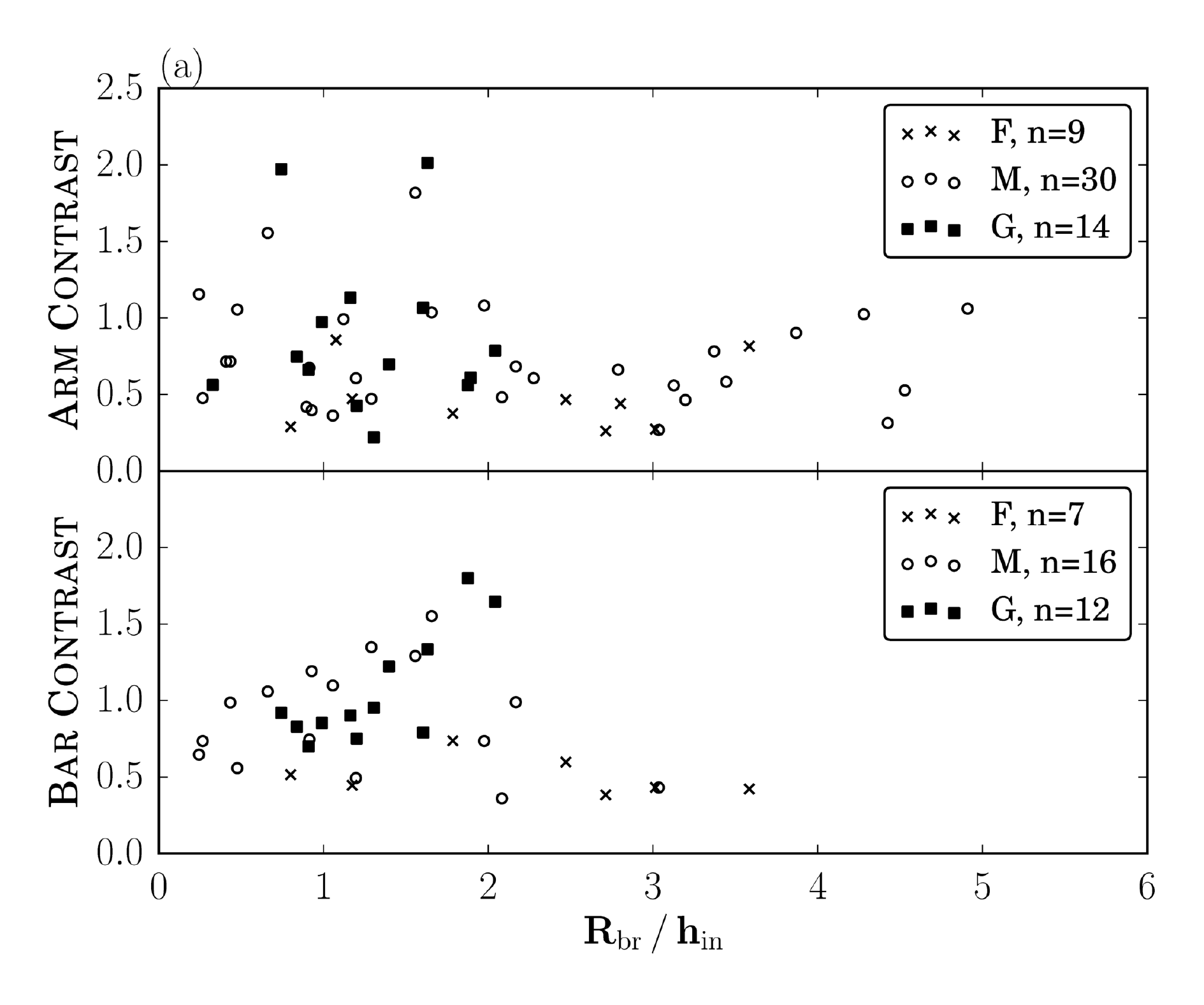}
		\label{subfig:rbr_hin_asbs}
	}
	\subfloat
	{
		\includegraphics[width=0.5\hsize]{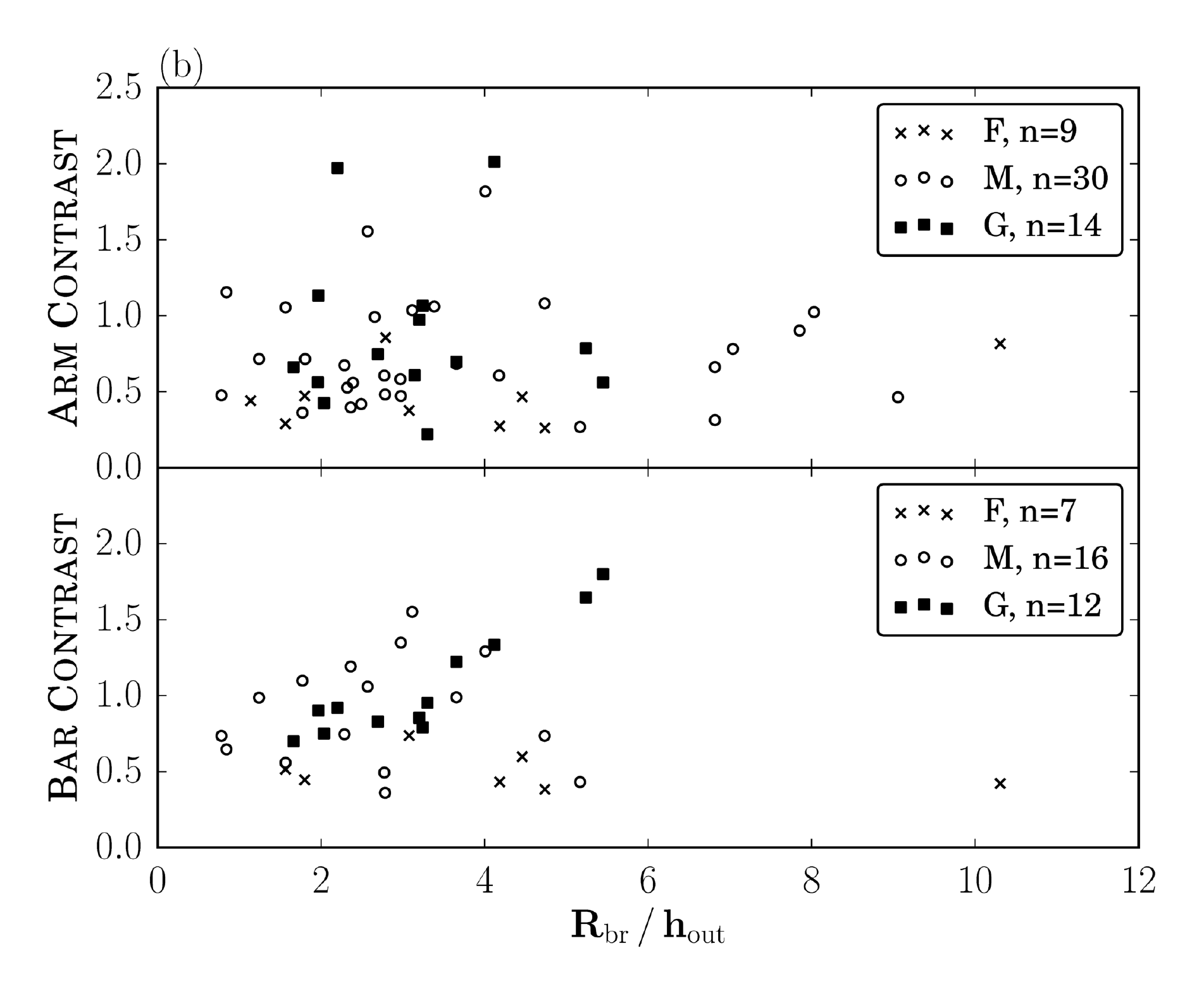}
		\label{subfig:rbr_hout_asbs}
	}
	\caption{
		Bar and arm contrast as a function of the break radius in units of (a)
		the disc inner and (b) disc outer scale length. \rev{Correlations with the 
		arm contrast are not obvious.} The plots indicate a positive
		correlation between the bar contrast and the break radius for
		grand-design galaxies. The Spearman Rank correlation coefficients are
		$\rho = 0.64$ (left panel) and $0.85$ (right panel).  It is remarkable
		that the bar contrast of flocculents shows an inverse correlation with
		the break radius. Their correlation coefficients are $\rho = -0.61$
		(left panel) and $-0.57$ (right panel).  Multi-armed galaxies do not
		show any correlation as indicated by correlation coefficients of $\rho
		= 0.01$ (left panel) and $0.14$ (right panel).  
	}
	\label{fig:rbr_asbs}
\end{figure*}

In the following we make use of our contrast measurements of Sect.
\ref{sec:contrast_measurements} and plot the arm and bar contrast as a function
of the break strength (see Fig.  \ref{fig:hin_hout_asbs}). \rev{There are no}
correlations between the arm contrast and the break strength.  However,
grand-design galaxies with a strong break have a weak tendency to have higher
bar-interbar contrasts.  The correlation coefficients for the spiral arm classes
are $\rho_F = -0.07$, $\rho_M = 9 \e{-3}$ and $\rho_G = 0.33$.  Furthermore, the
scatter is large and this \rev{result} must be confirmed with a larger sample.
In addition, \rev{Fig. \ref{fig:rbr_asbs} shows} arm and bar contrast as a
function of the break radius normalized by the disc inner and outer scale
length. Since the following results are based on a small number of galaxies,
\rev{their statistical significance is weak.}
Figure \ref{fig:rbr_asbs} does not indicate any correlation between break radius
and arm contrast.  However, for both normalization parameters a correlation of
the break radius with the bar contrast \rev{is detected}. Grand-design galaxies
with high bar contrasts have break radii that are larger relative to the disc.
The Spearman Rank correlation coefficients are $\rho_G = 0.64$ and $0.85$ for
the break radius normalized by the disc inner and outer scale length,
respectively.  It is remarkable that flocculent galaxies show an inverse
correlation with correlation coefficients of $\rho_F = -0.61$ and $-0.57$. In
fact, flocculents seem to have similar bar contrasts for their full range of
break radii in the plot. Thus, the comparably high correlation coefficients may
be caused by the small number of galaxies.  Multi-armed galaxies do not show any
correlation between these two parameters.  

\subsection{Discussion}
\label{subsec:discussion_diskbreak}
As discussed in Sect.  \ref{sec:general_properties} and
\ref{sec:contrast_measurements}, grand-design galaxies have, on average,
stronger bars.  Furthermore, the bars of multi-armed and
grand-design galaxies have a flat radial profile (see Fig.
\ref{subfig:bar_sersic}).  By comparing their observations with the simulation
results of \citet{lia2002}, and particularly those of \citet{lia2003}, the
previous studies by \citet{gadotti2007} and \citet{lia2009} were able to
indicate that the angular momentum redistribution of flat bars is more
efficient than of exponential bars.  Therefore we expect the angular momentum
redistribution in grand-design galaxies to be stronger. 
As a result of this, the inner discs should develop faster and lead to flatter
profiles.  This is consistent with the increase of the disc inner scale length
from flocculent to grand-design galaxies as well as with the idea that bars
drive disc breaks \citep[see][and references therein]{jc2013}. 
Another result of this mechanism is that grand-design galaxies have a stronger
disc break, as indicated by the ratio of disc inner to outer scale length (see
Fig. \ref{fig:hin_hout}).  Multi-armed galaxies seem to be an intermediate case
consistent with their bar properties (see Fig. \ref{fig:bar}), as compared to
the other spiral arm classes. 

Possible correlations between the disc break and the bar as well as arm contrast
\rev{are investigated as well} (see Fig. \ref{fig:hin_hout_asbs} and
\ref{fig:rbr_asbs}).  Clear connections between the break radius and the bar
contrast \rev{are found}.  However, this does not hold true for the arm
contrast. Thus, the spiral structure of galaxies does not seem to influence the
development or evolution of disc breaks in barred galaxies. At the same time
this points out the importance of bars for disc breaks. 
The correlations between bar contrast and break radius depend strongly on the
spiral arm class of the galaxy.  Considering grand-design galaxies, our plots
indicate a positive correlation between the bar contrast and the break radius
normalized by the disc inner as well as outer scale length. Grand-design
galaxies have flat bars which are known to efficiently redistribute the angular
momentum of a galaxy \citep{gadotti2007,lia2009}. \rev{As a result,} the bar strongly
influences the distribution of matter in the inner disc and the break radius
itself. Since previous studies (see \citet{kim2016} for the observations and
\citet{lia2013} for a review of the theoretical work) indicate that the bar
grows longer and stronger with time, the break radius \rev{is expected} to evolve
with time.  Thus, high break radii occur in galaxies with long and strong bars,
\rev{which} is in good agreement with our results. 
In contrast, the Spearman Rank correlation coefficients of flocculent galaxies
indicate an inverse correlation between break radius and bar contrast. However,
the plot seems to show similar values of the bar contrast for the full span of
break radii. This indicates that disc breaks of flocculent galaxies are not
connected to the properties of the bar. Hence, disc breaks of flocculents may
be generated by other mechanisms than in grand-design galaxies. 

\section{Summary and Conclusions}
\label{sec:conclusions}
We investigated how fundamental galaxy properties, including the properties of
bars and disc breaks, are related to the properties of spiral arms.  Using $3.6
\micron$ images from the Spitzer Survey of Stellar Structure in Galaxies
(\ssg), we performed measurements of arm-interarm and bar-interbar contrasts
and also considered previously published measurements of fundamental galaxy
parameters, including visual classification into the three different arm
classes (flocculent, multi-armed and grand-design). 
The main results from this study can be summarised as follows:
\begin{enumerate}
	\item Our measurements of the arm-interarm contrast compare well with the
		results of a previous study, with typical differences of approximately
		25\%. We discuss differences in the measurement methods and point out
		striking similarities in the resulting radial contrast profiles. Thus
		we conclude that the measurements of both studies are in reasonable
		agreement (see Sect. \ref{subsec:comparison}). 
	\item The arm contrasts of flocculent galaxies are significantly lower as
		compared to the other spiral arm classes. However, the arm
		contrasts of multi-armed and grand-design galaxies are more similar
		than expected from a visual classification.  Interestingly, the bar
		contrast, and its increase from flocculent to grand-design galaxies, is
		systematically more significant as compared to the arm contrast (see
		Fig. \ref{fig:strength_dist}). 
	\item Flocculent galaxies are clearly distinguished from the other spiral
		arm classes, in particular by their lower total stellar masses and
		surface densities. In contrast, multi-armed and grand-design galaxies
		share many fundamental parameters, excluding some bar properties and
		the bulge-to-total luminosity ratio.  In particular, 
		almost all flocculent galaxies either have no bulge or have extended,
		less massive (possibly disc-like) bulges \rev{whereas} grand-design and
		multi-armed galaxies tend to have more classical-type bulges with
		slightly more massive bulges in the grand-design spirals (see Sect.
		\ref{sec:general_properties}). 
	\item Considering multi-armed and grand-design galaxies, a strong
		correlation between the bulge-to-total ratio and the bar contrast
		\rev{is found} and \rev{we} conclude that the existence of a classical
		bulge could enhance bar evolution.  In addition, a weaker correlation
		between arm and bar contrast is found which corroborates the findings of
		previous studies (see Fig. \ref{fig:arm-bar_frelB-barstrength}). 
	\item Similar arm-interarm contrasts are detected in both barred and
		unbarred galaxies (see Fig. \ref{fig:armstrength_dist_bar}). This
		indicates that spiral arms have the capacity to reach similar arm
		contrasts regardless of which mechanism triggers the spiral structure.
		However, the highest arm contrasts are found exclusively in barred
		galaxies. 
	\item We show that the bar contrast of grand-design galaxies correlates
		with the disc break radius, reinforcing previous conclusions on the
		connection between bars and disc breaks. However, such correlation is
		absent for the arm contrast or the other spiral arm classes (see Sect.
		\ref{subsec:discussion_diskbreak}). 
\end{enumerate}

Our measurements of the arm and bar contrasts as well as the corresponding radial \rev{contrast} profiles
are available to the community at \url{http://homepages.physik.uni-muenchen.de/~a.bittner/projects/arm_contrasts/overview.html}. 

\section*{Acknowledgements}
\label{sec:acknowledgments}
We thank an anonymous referee for very helpful suggestions. 
EA and AB thank the CNES (Centre National d'Etudes Spatiales, France) for
financial support.  This work is based on observations made with the Spitzer
Space Telescope and made use of the NASA/IPAC Extragalactic Database (NED),
which are operated by the Jet Propulsion Laboratory, California Institute of
Technology, under a contract with the National Aeronautics and Space
Administration (NASA).  We also acknowledge the usage of the HyperLeda database
(\url{http://leda.univ-lyon1.fr}).  This research has made use of NASA's
Astrophysics Data System Bibliographic Services.


\bibliographystyle{mnras}
\bibliography{literature}

\bsp	
\label{lastpage}
\end{document}